\documentclass[a4paper,11pt]{article}
\pdfoutput=1 

\usepackage{jheppub} 
\usepackage{etoolbox}                     
\makeatletter
\patchcmd{\maketitle}{\@fpheader}{Prepared for submission to JHEP}{}{}
\makeatother
\usepackage[T1]{fontenc}
\usepackage{amsmath,amssymb,mathtools,slashed}
\usepackage{color,xcolor}
\definecolor{blue}{rgb}{0,0,0.5}
\usepackage[utf8]{inputenc}
\usepackage{enumerate}
\usepackage[printonlyused]{acronym}
\usepackage{footnote}
\makesavenoteenv{tabular} 
\makesavenoteenv{table} 
\usepackage{subcaption}
\usepackage{booktabs}
\usepackage{hhline}
\usepackage{multirow}
\usepackage{cancel}
\usepackage[normalem]{ulem}
\usepackage{appendix}
\usepackage{slashed}
\usepackage{rotating}
\usepackage{ifthen}

\definecolor{darkgreen}{RGB}{0,175,10}
\definecolor{brown}{RGB}{150,50,0}
\definecolor{AK}{rgb}{0,0,1.0}
\definecolor{}{rgb}{0.1,0.6,0.1}

\newcommand{\ba}{\begin{eqnarray}}
\newcommand{\ea}{\end{eqnarray}}
\newcommand{\be}{\begin{equation}}
\newcommand{\ee}{\end{equation}}
\newcommand{\Dplpill}{D^+\to \pi^+\ell^+\ell^-}

\renewcommand{\text}[1]{%
\ifthenelse{\equal{#1}{p2}}{p^2}{}%
\ifthenelse{\equal{#1}{q2}}{q^2}{}%
\ifthenelse{\equal{#1}{pplq2}}{s}{}%
\ifthenelse{\equal{#1}{Icdd}}{I_{cdd}}{}%
\ifthenelse{\equal{#1}{Icd}}{I_{cd}}{}%
\ifthenelse{\equal{#1}{Icdfq}}{I_{cdfq}}{}%
\ifthenelse{\equal{#1}{Icdc}}{I_{cdc}}{}%
\ifthenelse{\equal{#1}{Icdfp}}{I_{cdfp}}{}%
\ifthenelse{\equal{#1}{mc2}}{m_c^2}{}%
\ifthenelse{\equal{#1}{mpi}}{m_\pi}{}%
\ifthenelse{\equal{#1}{P2}}{P^2}{}%
}

\title{
$D\to P \ell^+\ell^-$  decays assisted by QCD light-cone sum rules     
}

\author{Anshika Bansal,}
\author{Alexander Khodjamirian and}
\author{Thomas Mannel}
\affiliation{Center for Particle Physics Siegen (CPPS), Theoretische Physik 1, 
\\ Universit\"at Siegen, D-57068 Siegen, Germany}

\emailAdd{anshika.bansal@uni-siegen.de,  khodjamirian@physik.uni-siegen.de, 
mannel@physik.uni-siegen.de}

\preprint{SI-HEP-2025-12,  P3H-25-033}

\abstract{
We suggest a new method to analyse the rare $D_{(s)}\to P \ell^+\ell^-$
 decays ($P=\pi,K$), combining QCD light-cone sum rules (LCSR) with 
 hadronic dispersion relations. As our main study case, we consider
 the $D^+\to\pi^+\ell^+\ell^-$ mode
 which attracts much of interest from the point of view of GIM 
 cancellation and potential new sources of the FCNC 
 $c\to u$ transitions. The  hadronic amplitude of this decay is dominated by the combination of  
 weak annihilation with the emission of a virtual photon.
 This  $D^+\to \pi^+\gamma^*$ amplitude is calculated
 at spacelike photon virtualities, $-q^2 \gg \Lambda_{QCD}^2$, 
 using LCSR with pion distribution amplitudes. The LCSR results are then fitted to the hadronic dispersion 
 relation in the variable $q^2$. The fit allows us 
 to determine relative phases of the $\rho$-,$\omega$- and 
 $\phi$-meson terms and 
 to estimate the contributions of heavier hadronic states. The latter are 
 parameterized in two different ways: first, with a $z$-expansion and second, using a model 
 with excited vector-meson resonances.
 The resulting $D^+\to\pi^+\ell^+\ell^-$ width obtained from 
 the LCSR-assisted dispersion relation is not much smaller than
 the current upper bound measured by LHCb collaboration. 
 In comparison with our  result for the dominant 
 $D^+\to \pi^+\gamma^*$ transition, the contribution induced by the short distance 
$c\to u \ell^+\ell^-$  quark transition generated  by the effective $O_9$ operator in Standard Model turns out to be practically invisible. To detect its effect, one readily
 needs a enhancement of the coefficient $C_9$ from non-standard effects by many orders of magnitude.
 We also establish new $U$-spin symmetry relations 
 between the amplitudes of $D\to P \ell^+\ell^-$  decays
 and apply our method to the Cabibbo-favoured modes 
 $D^+_s\to\pi^+\ell^+\ell^-$ and $D^0\to\bar{K}^0\ell^+\ell^-$ which 
 proceed via the same weak annihilation mechanism.
}

\begin{document}

\maketitle
\flushbottom
\section{Introduction}
\label{sext:intto}

Flavour-changing neutral current (FCNC) transitions play an important role in the 
search for effects from physics beyond the Standard Model (BSM). The underlying quark transitions 
are absent at tree level in the Standard Model  and thus can proceed only through loops, which turn out to be finite due to the so called GIM mechanism. At the perturbative level, 
this leads -- in addition to CKM suppressions -- to the usual loop-factor of $1/(16 \pi^2)$, 
rendering these contributions in SM small. In turn, there could be additional contributions from heavy new particles. This makes the FCNC transitions highly sensitive to BSM effects.  
However, these quark processes are relevant only at short distances, where QCD perturbation theory 
is valid. For the corresponding hadronic processes, this simple picture is distorted by the long distance effects, which are genuinely non-perturbative and thus difficult to evaluate.

For bottom quarks, the short-distance FCNC $b \to s$ and $b \to d$ transitions  are enhanced by the large top-quark mass, and the $b \to s$ transition does not even suffer from a CKM suppression. As an example, the short-distance transition $b \to s \ell^+ \ell^- $ contributes significantly to the process $B \to K^{(*)} \ell^+ \ell^-$. 
However, there are long-distance effects from the $b \to s \bar{c} c$ quark transitions which  
generate the same final state, since the intermediate charm-anticharm states can annihilate into 
the lepton pair. These contributions are sizable and still are one of the 
major sources of uncertainty. 

For charm quarks, it is way more difficult to get a handle on the FCNC $c \to u$  transition,  
since it suffers from a factor $m_b^2 / M_W^2$ and from a severe CKM suppression due to the small 
CKM elements, linking the third generation to the first and the second ones. Consequently, the desired 
short-distance contributions are deeply buried under large long-distance effects, making a search for 
BSM particles, applying the same strategy as in the bottom sector, almost impossible. 

The $D^+\to \pi^+ \ell^+\ell^-$ decay is the simplest  process involving the FCNC $c\to u \ell^+\ell^-$ quark
transition. Dominant long-distance effects  in this decay involve an ordinary weak $D^+ \to \pi^+$ transition, combined with an electromagnetic (e.m.) production of the lepton pair. 
A simple way to realize this mechanism
is to combine the $D^+ \to \pi^+ V $ singly-Cabibbo suppressed (SCS) decay with the $V\to \ell^+\ell^-$ e.m. process,
where $V=\rho^0,\omega,\phi$ is a neutral vector meson. This has been implemented in many previous
studies of the $D\to \pi \ell^+\ell^-$ decays (see e.g., \cite{Fajfer:2001sa,Burdman:2001tf, Fajfer:2005ke, deBoer:2015boa}), treating the vector meson resonances with a simple Breit-Wigner shape. These long-distance 
contributions turn out to be many orders of magnitude larger than the contribution induced by the 
short distance $c\to u \ell^+\ell^-$ quark transition, making it hard to separate one from the other.

Currently, the most restrictive experimental upper bound, $BR(D^+\!\!\to \pi^+ \mu^+\mu^-)<6.7\cdot 10^{-8}$, was
obtained by LHCb collaboration \cite{LHCb:2020car}, including a veto on the $\rho,\omega,\phi$ 
resonance region of the lepton pair invariant mass. However, even outside the vetoed region the main
contribution is still given by shoulders of these resonances as well as by the contributions of excited 
vector mesons. Hence, any sensitivity to the desired 
short-distance contribution depends crucially on the proper description of the long-distance effects, also 
outside the resonance regions. Thus 
it is timely to revisit the $D^+\to \pi^+ \ell^+\ell^-$ and similar decays with refined theoretical methods. 

On the theory side, a necessary prerequisite to
study the $c\to u\ell^+\ell^-$ transitions 
was the accurate calculation \cite{deBoer:2016dcg}  of the Wilson coefficients in the 
relevant effective Hamiltonian. An approach largely based on QCD factorization (QCDf) was then suggested in \cite{Feldmann:2017izn}
for the decay channel $D\to \rho\ell^+\ell^-$, where the contributions of various quark  topologies were 
obtained using the QCDf analysis  \cite{Beneke:2001at} of $b\to s\ell^+\ell^-$ transitions.
An important new element in  \cite{Feldmann:2017izn} was the treatment of resonance contributions employing a duality representation of the loop diagram in terms of an infinite tower of resonances suggested in \cite{Blok:1997hs}.
For the $D^+ \to \pi^+ \ell^+\ell^-$ decay, 
a similar QCDf-based method was then used in \cite{Bharucha:2020eup}.
Certain important questions still remain open. First, a description 
of the  $D^+ \to \pi^+ \ell^+\ell^-$ and similar decays in terms of QCDf diagrams is,
strictly speaking, only possible at a spacelike momentum transfer to the lepton pair, 
far from the intermediate vector-meson resonances. Second, the QCDf method employs 
the $m_c\to \infty$ limit, which may receive significant $1/m_c$ corrections. 

In this paper, we suggest and use a different new approach to these processes, 
taking as a study case the $D^+\to \pi^+ \ell^+\ell^-$ decay. 
We focus on the dominant  contribution to the decay amplitude, 
stemming from the  combination of the weak and e.m. interactions. 
This contribution is obtained from the amplitude of the $D\to\pi\gamma^*$ transition, followed 
by the conversion of the virtual photon into a lepton pair. The amplitude for 
$D\to\pi\gamma^*$ is given by a nonlocal matrix element of the weak effective Hamiltonian and 
an electromagnetic current.

We will use QCD light-cone sum rule (LCSR) to 
calculate the $D\to\pi\gamma^*$ amplitude at spacelike photon virtuality, $q^2<0$, far below the light hadronic 
thresholds in the virtual photon channel, and at finite $c$-quark mass, i.e we assume that $-q^2\gg \Lambda_{\rm QCD}^2$, having in mind that 
also $m_c^2\gg \Lambda_{\rm QCD}^2$.
The LCSR  is obtained employing  the pion distribution amplitudes (DAs) and a $D$-meson interpolating 
current. The underlying correlation function is computed perturbatively and to the leading twist. The main contribution is given by the weak annihilation diagrams combined with the virtual photon emission. To obtain the decay amplitude in the physical (timelike) region $ 4 m_\ell^2 < q^2 < (m_D - m_\pi)^2$ of  the 
 $D^+\to \pi^+ \ell^+\ell^-$ decay, we  employ a dispersion relation in the variable $q^2$, where we model the 
hadronic spectral function, taking into account, apart from the lowest light vector mesons, also 
the contributions of heavier states. The calculation in the unphysical region of $q^2$ is then used
to determine the unknown parameters of the dispersion relation, in particular, the relative phases of the 
vector meson terms and the integral over the contributions of heavier states. 
Our main phenomenological result is the lepton-pair invariant mass spectrum 
of the $D^+\to \pi^+ \ell^+\ell^-$ decay.
In addition, to assess the role of the "genuine" short-distance FCNC effects, 
we separately calculate the contribution of the point-like $c\to u\ell^+\ell^- $ operator $O_9$
to the amplitude of $D^+\to \pi^+ \ell^+\ell^-$. In SM, this contribution is heavily suppressed by a CKM coefficient $V_{ub}V_{cb}^*$. 

The method set up in this paper can easily be extended to other
$D_{(s)}\to P \ell^+\ell^-$ modes, where $P=\pi,K,\eta,\eta'$.
Their complete list, given in Table~\ref{tab:modes},
also includes the Cabibbo-favoured (CF) and doubly-Cabibbo-suppressed (DCS) modes,
which, as far as we know, were not given much attention in the literature, except in the early paper \cite{Fajfer:2001sa},
see also \cite{Sanchez:2022nsh}.  
The CF and DCS modes share 
common hadronic dynamics with the SCS decays, all of them being dominated by the annihilation 
mechanism. As an additional application of our approach, we obtain the decay width of the CF modes $D_s^+\to\pi^+\ell^+\ell^-$ and $D^0\to\bar{K}^0\ell^+\ell^-$, taking into account the $s$-quark
mass in LCSR and thus, quantitatively assess
the flavour $SU(3)$ violation in these decays. 
\begin{table}
\centering
\begin{tabular}{|l|c|l|}
\hline
Decay mode & Cabibbo hierarchy   & BR,  upper limit \cite{PDG}\\
\hline
&&\\[-4mm]
$D^+\to \pi^+ \ell^+\ell^-$    & SCS & $1.1\times 10^{-6}\, (\ell = e)$\\
&  &$6.7\times 10^{-8}\, (\ell = \mu)$ (*)\\
\hline
 $D^+\to K^+ \ell^+\ell^-$  & DCS   &$8.5\times 10^{-7}\, (\ell = e)$ (*)\\ 
&    &$5.4\times 10^{-8}\, (\ell = \mu)$ (*)\\ 
 \hline\hline
&&\\[-4mm]
$D^0\to \bar{K}^0\ell^+\ell^-$  & CF  & $2.4\times 10^{-5}\, (\ell = e)$\\ 
&     &$2.6\times 10^{-4}\, (\ell = \mu)$\\
\hline
&&\\[-4mm]
$D^0\to \pi^0\ell^+\ell^-$    & SCS  &$4\times 10^{-6} \,(\ell = e)$\\
&  &$1.8\times 10^{-4}\, (\ell = \mu)$\\
\hline
&&\\[-4mm]
$D^0\to \eta \ell^+\ell^-$    & SCS & $3\times 10^{-6}\, (\ell = e)$\\
&   &$5.3\times 10^{-4}\, (\ell = \mu)$\\
\hline
&&\\[-4mm]
$D^0\to \eta' \ell^+\ell^-$    & SCS & -\\
\hline
&&\\[-4mm]
$D^0\to K^0 \ell^+\ell^-$    & DCS  & - \\
\hline
\hline
&&\\[-4mm]
$D_s^+\to \pi^+ \ell^+\ell^-$    &  CF    &$5.5\times 10^{-6}\, (\ell = e)$ (*)\\ 
&    &$1.8\times 10^{-7} \,(\ell = \mu)$ (*)\\
\hline
&&\\[-4mm]
$D_s^+\to K^+ \ell^+\ell^-$    & SCS &$3.7\times 10^{-6} \,(\ell = e)$\\
& &$1.4\times 10^{-7}\, (\ell = \mu)$ (*)\\
\hline
\end{tabular}
 \caption{ The $D_{(s)}\to P\ell^+\ell^-$ decays and the experimental 
 upper limits of their branching fractions. The limits marked by (*) result from the latest LHCb measurements \cite{LHCb:2020car}, in which the region of lepton-pair invariant
 mass between 525 MeV and 1250 MeV was vetoed.
  }
\label{tab:modes}
 \end{table}

The plan of this paper is the following. In Section~\ref{sect:oper}, we consider the effective Hamiltonian relevant 
for $D_{(s)}\to P\ell^+\ell^-$ decays and specify the dominant contributions to the $D^+\to\pi^+\ell^+\ell^-$ decay, governed by the  $D^+\to\pi^+\gamma^*$ transition amplitude.  
In Section~\ref{sect:topol}, we discuss the quark topologies contributing  to this amplitude.
In Section~\ref{sect:uspin}, considering the limit of 
$SU(3)_{fl}$ symmetry, we obtain
new relations between different $D_{(s)}\to P\ell^+\ell^-$ amplitudes 
following from the $U$-spin flavour symmetry.
Section~\ref{sect:lcsr} is devoted to the derivation of LCSR with the pion DAs
for the  $D^+\to\pi^+\gamma^*$ amplitude.
Having this amplitude at spacelike  $q^2$ calculated from LCSR, in Section~\ref{sect:hadrdisp}, we match it to the hadronic dispersion relation in terms of neutral vector mesons and their excited states. 
Our numerical analysis of the 
$D^+\to \pi^+\gamma^*$ amplitude and our predictions for the differential 
decay width of $D^+\to \pi^+\ell^+\ell^-$ decay are presented in Section~\ref{sect:num}, together with analogous results
for CF decays $D_s^+ \to \pi^+\ell^+\ell^-$ and $D^0\to \bar{K}^0\ell^+\ell^-$. The paper contains four appendices: 
In Appendices \ref{app:diagA} and \ref{app:diagL}, the detailed computations of OPE diagrams and their spectral densities with, respectively, annihilation and loop topologies are presented. 
In Appendix~\ref{app:CF}, the details of annihilation diagrams for CF transitions are discussed. The Appendix~\ref{app:Vexc} presents our 
estimates of the input parameters for 
excited vector mesons.

\section{ Effective operators and the $D_{(s)}\to P\ell^+\ell^-$ amplitude}
\label{sect:oper}
For the effective Hamiltonian generating the $c\to u \ell^+\ell^-  $ strangeness-conserving $(\Delta S=0)$ transitions,  we use the most up-to-date  analysis of its operators and their
Wilson coefficients from \cite{deBoer:2016dcg}. The general expression is 
\begin{eqnarray}
H^{(\Delta S=0)} _{eff}=\frac{4G_F}{\sqrt{2}}\left[\sum_{{\cal D}=d,s}\lambda_{\cal D} 
\Big[C_1(\mu)O_1^{\cal D}+C_2(\mu)O_2^{\cal D}\Big] -\lambda_b\sum_{i=3}^{10}C_i(\mu)O_i \right] \,,
\label{eq:Heff}
\end{eqnarray}
where $\lambda_{{\cal D}}= V_{u{\cal D}} V^*_{c{\cal D}} ~~({\cal D}=d,s)$ and $\lambda_b = V_{ub} V^*_{cb}$
is a short-hand notation
for the product of CKM matrix elements. 
We isolate the dominant current-current 
operators~\footnote{Note that we use the numbering of operators 
usually adopted for the effective Hamiltonian of nonleptonic decays, 
so that our $O_{1(2)}$ corresponds to $O_{2(1)}$ in \cite{deBoer:2016dcg}. }
\begin{eqnarray}
O_1^{\cal D}=\left(\bar{u}_L\gamma_\mu {\cal D}_L\right)\left(\bar{{\cal D}}_L\gamma^\mu c_L\right)\,,~~
O_2^{\cal D}=\left(\bar{u}_L\gamma_\mu t^a{\cal D}_L\right)
\left(\bar{{\cal D}}_L\gamma^\mu t^a c_L\right)\,,  ~~~({\cal D}=d,s)\,,
\label{eq:O2}
\end{eqnarray}
with the left-handed quarks denoted as e.g., $u_L= \frac{1}{2}(1-\gamma_5)u$ and with colour matrices 
$t^a\equiv \lambda^a/2$.
The effective operators $O_{3-9}$ are absent at the renormalization scale $\mu$ larger than $m_b$, 
and their Wilson coefficients at  $\mu\sim m_c$, relevant for the decays we consider, are 
much smaller than $C_{1,2}$. Moreover, the contributions of these operators to the decay amplitudes are, in addition, strongly suppressed by the very small CKM factor
$|\lambda_b|\sim 10^{-4}$. 
Throughout this paper, we put $\lambda_b=0$ in the effective Hamiltonian. 
In this approximation, which we call the GIM limit, the operators $O_{3-9}$ do not contribute to the $D_{(s)}\to P\ell^+\ell^-$ decay amplitudes. Combined with  CKM unitarity,  the GIM limit
yields $\lambda_s=-\lambda_d $, so that  the effective Hamiltonian (\ref{eq:Heff}) becomes:
\begin{eqnarray}
H^{(\Delta S=0,\,\lambda_b=0)}_{eff}&=&\frac{4G_F}{\sqrt{2}}\lambda_d
\Big[C_1(\mu) \big[\left(\bar{u}_L\gamma_\mu d_L\right)\left(\bar{d}_L\gamma^\mu c_L\right)
-\left(\bar{u}_L\gamma_\mu s_L\right)\left(\bar{s}_L\gamma^\mu c_L\right)\big]
\nonumber \\
&+&
C_2(\mu) \big[\left(\bar{u}_L\gamma_\mu t^a d_L\right)
\left(\bar{d}_L\gamma^\mu t^a c_L\right)
-\left(\bar{u}_L\gamma_\mu t^a s_L\right)
\left(\bar{s}_L\gamma^\mu t^a c_L\right)
\big] \Big]\,.
\label{eq:HeffS0}
\end{eqnarray}
where  the  current-current operators $O_{1,2}^{\cal D}$
are now spelled out explicitly.
The corresponding Hamiltonians for the CF and 
DCS decays with $\Delta S = \pm 1$ are, respectively:
\begin{eqnarray}
H^{(\Delta S=-1)}_{eff}=\frac{4G_F}{\sqrt{2}}  V_{ud} V_{cs}^*
\Big[C_1(\mu) \left(\bar{u}_L\gamma_\mu d_L\right)\left(\bar{s}_L\gamma^\mu c_L\right)+
C_2(\mu) \left(\bar{u}_L\gamma_\mu t^ad_L\right)
\left(\bar{s}_L\gamma^\mu t^a c_L\right)
\Big] \,,
\label{eq:HeffS1}
\end{eqnarray}
\begin{eqnarray}
H^{(\Delta S=+1)}_{eff}=\frac{4G_F}{\sqrt{2}}  V_{us} V_{cd}^*
\Big[C_1(\mu) \left(\bar{u}_L\gamma_\mu s_L\right)\left(\bar{d}_L\gamma^\mu c_L\right)+
C_2(\mu) \left(\bar{u}_L\gamma_\mu t^a s_L\right)
\left(\bar{d}_L\gamma^\mu t^ac_L\right)   
\Big] \,.
\label{eq:HeffS2}
\end{eqnarray}
The leading contribution to the decays $D_{(s)}\to P \ell^+\ell^-$ proceeds via an 
additional electromagnetic (e.m.) interaction, generated by the e.m. quark  current 

$$
j_\mu^{em}=\sum_{q=u,d,s,c} Q_q\bar{q}\gamma_\mu q, ~~~~~Q_{u,c}=2/3, ~~Q_{d,s}=-1/3\,.
$$
In what follows, to estimate a typical deviation from the GIM limit, we will also need the operator
 \begin{equation}
 O_9=\frac{\alpha_{em}}{4\pi} 
 \left(\bar{u}_L\gamma_\mu c_L\right)\left(\bar{\ell}\gamma^\mu \ell\right)
 \label{eq:O9}
 \end{equation}
which  has the largest Wilson coefficient  among the operators $O_{3-9}$. Note that  the Wilson coefficient 
for $O_{10}$ in the charmed sector vanishes \cite{deBoer:2016dcg}.  

The $D_{(s)}\to P \ell^+\ell^-$ 
decay amplitude, in the most general form, is 
given by the matrix element of the effective Hamiltonian 
 between the initial and final states:
\begin{equation}
A(D_{(s)}\to P  \ell^+\ell^-)=
\langle \ell^+\ell^- P(p) | H_{eff} | D_{(s)}(p+q)\rangle\,,
\label{eq:ampl1}
\end{equation}
where $\ell=e,\mu$,  and  $q$ denotes the four-momentum of the lepton pair, 
so that the kinematically allowed region is $4m_\ell^2<q^2<(m_{D_{(s)}}-m_P)^2$. The  other two
kinematical invariants are fixed by the on-shell conditions: $(p+q)^2=m_D^2$, $p^2=m_P^2$.
For a decay mode with $\Delta S=0$ ,  $\Delta S=-1$ or 
$\Delta S=1$,  we take for the effective Hamiltonian in (\ref{eq:ampl1}) 
the  expression (\ref{eq:HeffS0}), (\ref{eq:HeffS1}) or (\ref{eq:HeffS2}),
respectively.\\
Retaining in $H_{eff}$ only the quark current-current operators, one needs to insert in (\ref{eq:ampl1}) additional
e.m. interactions to create the lepton pair in the final state. 
It is straightforward to factorize out in (\ref{eq:ampl1}) the leptonic part 
together with the intermediate photon propagator. 
For the $\Dplpill$ decay mode the amplitude becomes: 
\begin{eqnarray}
A(D^+\to \pi^+\ell^+\ell^-)=\left(\frac{16\pi\alpha_{em}G_F}{\sqrt{2}}\right)
\lambda_{d} \frac{\bar{u}_\ell(q_-)\gamma^\mu v_\ell(q_+)}{q^2}
{\cal A}_{\mu\, (D^+\to \pi^+\gamma^*)}(p,q)\,,
\label{eq:ampl2}
\end{eqnarray}
 where  $q_{-(+)}$ is the momentum of the lepton (antilepton),
 represented by bispinors, so that $q_-+q_+=q$, and
\begin{eqnarray}
&&{\cal A}_{\mu}^{(D^+\to \pi^+\gamma^*)}(p,q)=i\int d^4xe^{iq\cdot x}  \langle \pi^+(p)|T\bigg\{ j_\mu^{em}(x),
\bigg(C_1(\mu)\big[O_1^{d}(0)- O_1^{ s}(0)\big] \nonumber\\
&&+C_2(\mu)\big[O_2^{d}(0)-O_2^{s}(0)\big]\bigg)\bigg\}|D^+(p+q)\rangle=
\Big[(p\cdot q)q_\mu -q^2p_\mu\Big] {\cal A}_{(D^+ \pi^+\gamma^*)}(q^2)\,, 
\label{eq:HME}
\end{eqnarray}
is the hadronic matrix element  of the $D^+\to \pi^+\gamma^*$ transition.
In this matrix element, a nonlocal product of quark operators is sandwiched between the $D^+$ and $\pi^+$ states. Kinematically, a $D\to P\gamma^*$ transition is similar to 
the $D\to PV$ nonleptonic decay. In both cases, due to the 
total angular momentum conservation, the final state particles are in the $P$-wave. Hence, the $D^+\to\pi^+\gamma^*$ transition is determined by a single invariant amplitude (nonlocal form factor) depending on $q^2$. The form of the structure multiplying this amplitude in (\ref{eq:HME}) is dictated by the conservation of e.m. current. Note that at $q^2=0$, if we multiply 
(\ref{eq:HME}) by a real 
(transverse) photon polarization vector, the $D^+\to\pi^+\gamma$ amplitude vanishes as expected.

The dependence of Wilson coefficients $C_{1,2}$ 
on the renormalization scale indicated in (\ref{eq:HME}) also deserves a comment.
Generally, this dependence is expected to be compensated by  scale dependence
emerging in the perturbative part of the calculated hadronic matrix element. 
In the LCSRs used in this paper, the object calculated in QCD is the
correlation function containing a virtual $c$-quark, 
and the actual calculation is done at leading, zeroth order in $\alpha_s$. 
In the absence of perturbative gluon corrections, as it is usually done 
for LCSRs for other $D$-meson form factors, it is conceivable to adopt an optimal scale $\mu\sim m_c$ 
for all renormalizable parameters. Hence, we will assume the same scale
also for the Wilson coefficients.

Inserting the decomposition (\ref{eq:HME}) in (\ref{eq:ampl2}), yields for the $D^+ \to \pi^+\ell^+\ell^-$ decay amplitude in the GIM limit: 
\begin{eqnarray}
A(D^+\to \pi^+\ell^+\ell^-)=-\left(\frac{16\pi\alpha_{em}G_F}{\sqrt{2}}\right)
\lambda_{d} \left[\bar{u}_\ell(q_-)\gamma_\mu v_\ell(q_+)\right]p^\mu
{\cal A}_{(D^+\pi^+\gamma^*)}(q^2)\,,
\label{eq:amplA}
\end{eqnarray}
determined by the nonlocal
form factor ${\cal A}_{(D^+\pi^+\gamma^*)}(q^2)$
in the timelike region $4m_\ell^2<q^2<(m_D-m_\pi)^2$ of the momentum transfer. The differential branching fraction of this decay is then:
\begin{eqnarray}
\frac{dBR(D^+\to \pi^+\ell^+\ell^-)}{dq^2}&=&\frac{4\alpha_{em}^2 G_F^2\lambda_d^2}{3\pi}\big|{\cal A}_{(D^+\pi^+\gamma^*)}(q^2)\big|^2
\nonumber\\
&\times&\sqrt{1-\frac{4m_\ell^2}{q^2}}\left(
1+\frac{2m_\ell^2}{q^2}\right)\big(p_{(D^+\pi^+\gamma^*)}(q^2)\big)^3\,\tau_{D^+}\,,
\label{eq:dBR}
\end{eqnarray}
where the $q^2$-dependent kinematical factor, $p_{(D\pi\gamma^*)}(q^2)=[(m_D^2+m_\pi^2-q^2)^2/(4m_D^2)-m_\pi^2]^{1/2}$
is equal to the 3-momentum of pion in the rest frame of the $D$ meson, and $\tau_{D^+}$ is the lifetime of $D^+$. 

Definitions analogous to (\ref{eq:ampl2}) and (\ref{eq:HME}) for all other $D_{(s)}\to P \ell^+\ell^-$ decays are easily introduced, replacing  the 
flavour content of the initial and final states, the effective operators and CKM factors correspondingly. 
Note that the amplitude (\ref{eq:HME}) is similar to the one parameterizing the so-called "charm-loop" effect in 
$B\to K \ell^+\ell^-$ decays (see e.g., \cite{Khodjamirian:2010vf,Khodjamirian:2012rm}).

Physics beyond the SM can manifest itself by an enhancement of the Wilson coefficient 
$C_9$, and so it is instructive to estimate 
the contribution to  $\Dplpill$ induced by the 
operator $O_9$ which appears if we restore $\lambda_b$ in the effective Hamiltonian (\ref{eq:Heff}). 
In the SM, the latter contribution 
has a simple hadronic structure, 
because the effective operator $O_9$ involves a pointlike emission of the lepton pair.
Using the definition (\ref{eq:O9}),
we obtain
\begin{eqnarray}
A^{(O_9)}(D^+\to \pi^+\ell^+\ell^-)&\equiv&
-\frac{4G_F}{\sqrt{2}}\lambda_b C_9(\mu)
\langle \ell^+\ell^- \pi^+(p) | O_9 | D^+(p+q)\rangle  
\nonumber\\
&=&  - \frac{\alpha_{em}G_F}{\sqrt{2}\pi}\lambda_b C_9 (\mu)
\left[\bar{u}_\ell(q_-)\gamma_\mu v_\ell(q_+)\right]p^\mu f^+_{D\pi}(q^2)\,,
\label{eq:ampl3}
\end{eqnarray}
where a standard definition of the $D^+\to \pi^+$ form factors is used:
\begin{eqnarray}
\langle \pi^+ (p)| \bar{u}\gamma_\rho c |D^+(p+q)\rangle = 
(2p_\rho + q_\rho)    f^+_{D\pi}(q^2) +q_\rho f^-_{D\pi}(q^2)\, . 
\label{eq:DpiFF}
\end{eqnarray}
Note that only the vector form factor $f^+_{D\pi}$ contributes to 
(\ref{eq:ampl3}), and the terms 
proportional to $q_\rho$ vanish due to lepton current conservation. Furthermore, this part of the amplitude depends on the renormalization scale $\mu$, which is 
compensated by a corresponding dependence of the additional contributions from $O_1$ and $O_2$  to the decay amplitude, 
which appear at $\lambda_b \neq 0$
and we denote them as $A^{(O_1,O_2)} (q^2)$. 
The resulting 
 $O(\lambda_b)$   amplitude, 
$A^{(O_1,O_2)} (q^2)  + A^{(O_9)} (q^2)$,
becomes $\mu$ independent. 
In order to make a statement on the size of the $O_9$ contribution in the SM, we take $C_9$ at $\mu\sim m_c$ and add (\ref{eq:ampl3}) to (\ref{eq:amplA}), replacing in the latter:
\begin{equation}
{\cal A}_{(D^+\pi^+\gamma^*)}(q^2) \to 
{\cal A}_{(D^+\pi^+\gamma^*)}(q^2)\ + \frac{\lambda_b}{\lambda_d}\frac{C_9 (\mu\sim m_c)}{16\pi^2}f^+_{D\pi}(q^2)\,.
\label{eq:O9add}
\end{equation}
Below, we use the LCSR approach with pion DAs to obtain the $D^+ \to\pi^+ \gamma^*$ amplitude, whereas the  $D\to\pi$ form factor is already well known from lattice QCD or LCSR calculations. 
As we shall see, in the SM, the part of the amplitude in (\ref{eq:O9add}) proportional to $\lambda_b C_9$ is negligibly small. 
 Hence, we refrain from defining an effective $C_9$ coefficient, which absorbs the dominant part proportional to 
 $\lambda_d C_{1,2}$, as it is frequently done in the literature.

\section{Quark topologies}
\label{sect:topol}
\begin{figure}[t]\centering 
\includegraphics[scale=0.2]{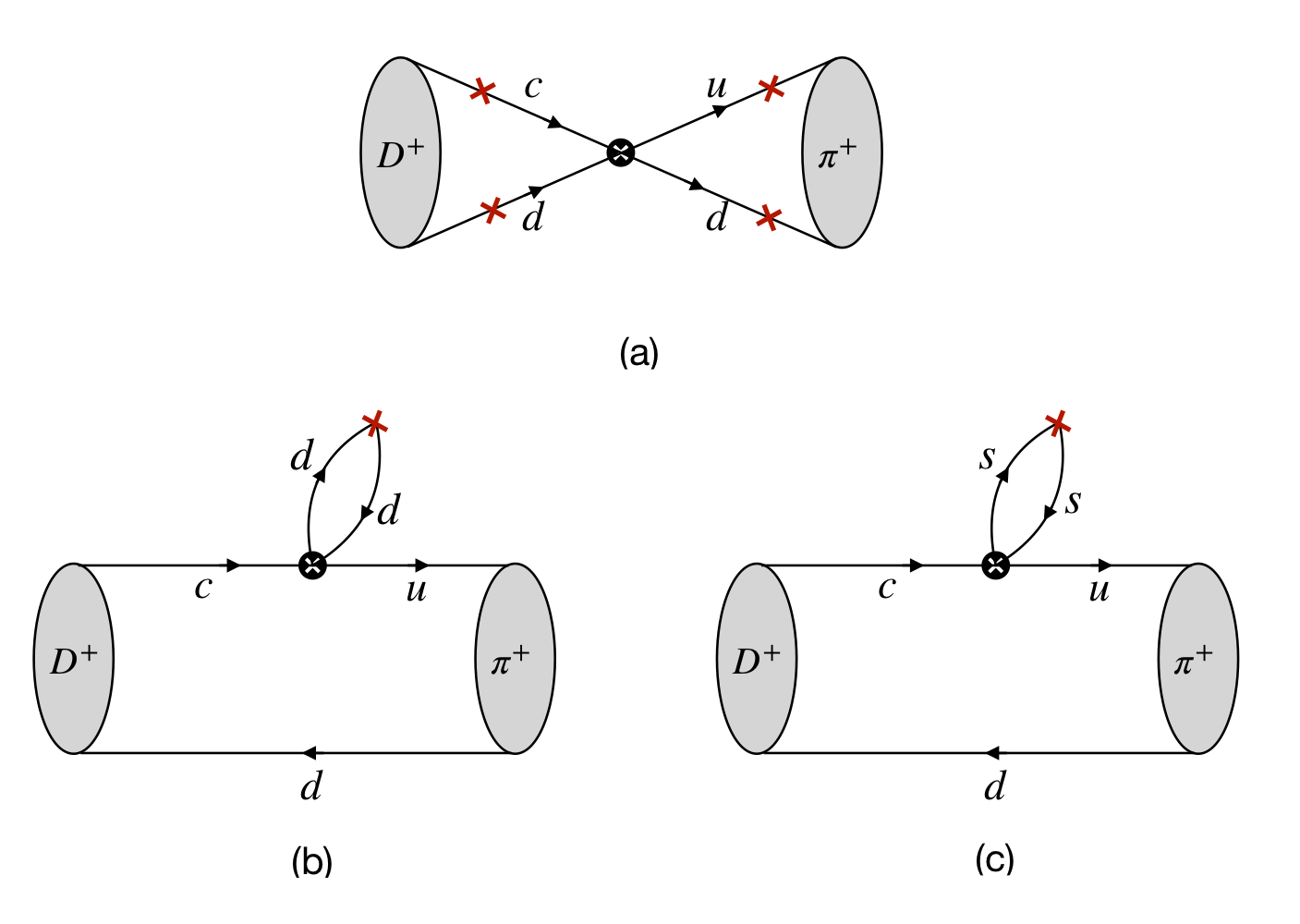}~~
\caption{Diagrams with: 
(a) the annihilation topology, (b) the $d$-quark and (c) the $s$-quark loop  topologies for the $D^+\to \pi^+\gamma^*$ transition. The virtual photon emission is shown with a cross.}
\label{fig:Dpill_LO}
\end{figure}

Quark diagrams contributing to the  $D_{(s)}\to P \gamma^*$ transitions 
have either annihilation~\footnote{
In our nomenclature, this topology includes both the annihilation of the initial quark-antiquark pair via $W$-boson and the scattering of the quark and antiquark via $W$-exchange.}
or  loop topology.   
Under quark topologies, we understand, as usual,  all possible  ways to connect quark fields  of the effective operators with the 
valence quarks and antiquarks of the initial $D_{(s)}$-meson and 
final $P$-meson. Note that the loop topologies with 
both $s$ and $d$ flavours, are
possible only in the  SCS transitions, whereas the CF and DCS 
$D_{(s)}\to P \gamma^*$ modes  have  only annihilation topology.\\
In Fig.~\ref{fig:Dpill_LO}(a), the annihilation topology for the $D^+\to \pi^+\gamma^*$ hadronic matrix element is shown, where a virtual photon can be emitted from any of the  quark lines, generating a set of four diagrams. 
In Fig.~\ref{fig:Dpill_LO}(b) and \ref{fig:Dpill_LO}(c), the $d$- and $s$-quark loop topologies
are presented, respectively.

In what follows, the $D\to \pi\gamma^*$ hadronic matrix elements  with both annihilation and loop topologies are computed
at spacelike photon virtualities. To this end, we
use LCSRs, where soft  quark-gluon interactions are implicitly absorbed in the nonperturbative objects, such as the pion DAs  and $D\to\pi$
form factors. In this paper, we will confine ourselves to the leading order (LO) of this computation, neglecting all perturbative gluon exchanges. At LO, the 
$d$- or $s$-loop topology 
factorizes, and is reduced to  a single diagram where the  virtual photon is emitted from the loop.

For the loop topology,
a complete GIM cancellation between the $d$- and $s$-loop
diagrams takes place only in the $SU(3)_{fl}$ limit of QCD, that is,
neglecting the  quark mass difference $m_s-m_d$. 
Note that it is  an additional
condition which is independent of the GIM limit $\lambda_b=0$ in the effective Hamiltonian
(\ref{eq:HeffS0}). In LCSRs,  we shall take the nonvanishing light quark masses into account, and, hence,  will be able to quantitatively assess the $SU(3)_{fl}$ violation.
On the other hand, the annihilation topology in  the SCS $D_{(s)}\to P\ell^+\ell^-$ decays 
has no relation to the GIM mechanism, because for the majority of  decay modes  
either $d$-flavour or $s$-flavour part of the current-current 
operators $O^{ d,s}_{1,2}$  contributes, cf. Fig.~\ref{fig:Dpill_LO}(a).
The only exceptions are the $D^0\to \eta\ell^+\ell^-$ and $D^0\to \eta'\ell^+\ell^-$
decays where both $d$ and $s$ parts contribute.
 \begin{figure}[h]\centering 
\includegraphics[scale=0.12]{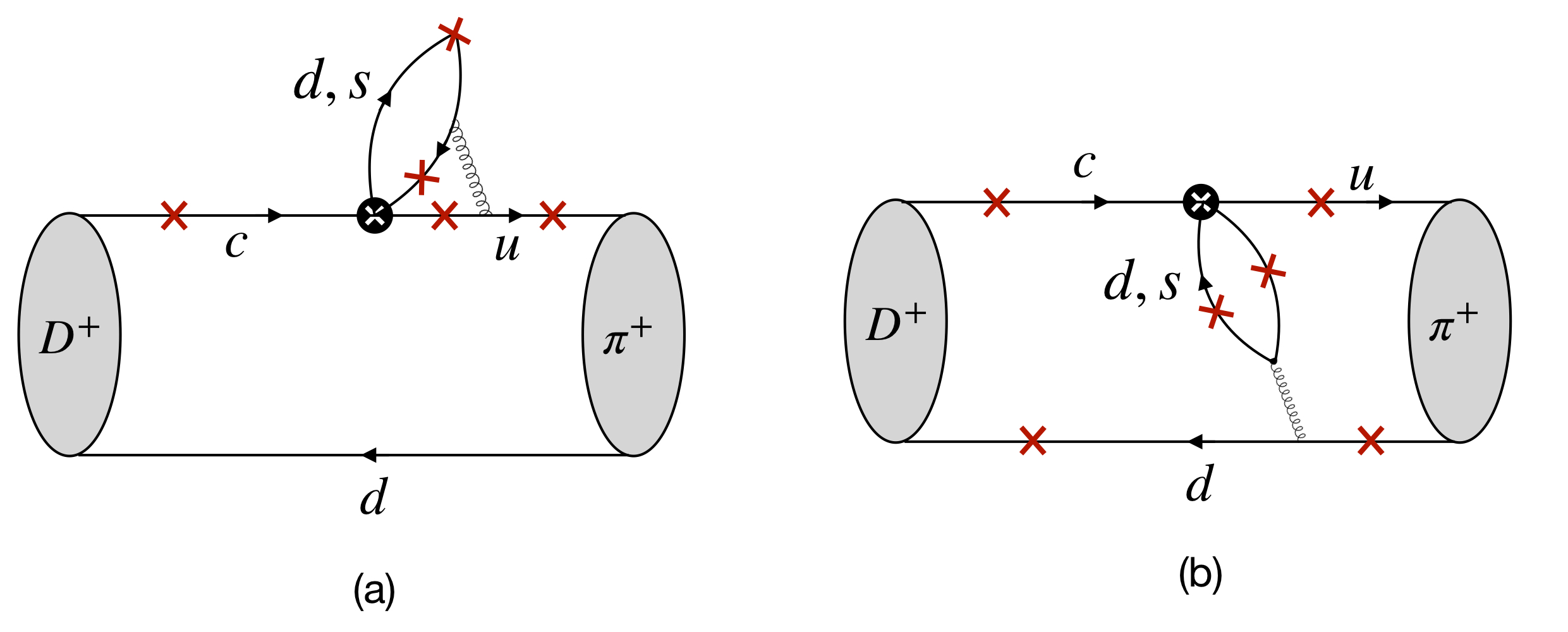}~~
\caption{ The $d,s$ loop topologies in $D^+\to \pi^+\ell^+\ell^-$ decay
at NLO, including the hard gluon exchange.
Crosses indicate possible positions of a virtual photon emission.
 }
\label{fig:diagsNLO}
\end{figure}

In this paper, we neglect the next-to-leading order (NLO) effects, 
generated by perturbative gluon exchanges in the quark diagrams contributing to LCSRs. 
Note that in the NLO diagrams with the loop topology, if the loop is connected by a gluon exchange with the rest of the 
diagram (see for illustration, Fig.~\ref{fig:diagsNLO}), a virtual photon emission is possible not only from the loop quarks.
We expect these effects to be inessential, because as we shall see, the sum of the LO $d$- and $s$-loop contributions that we calculate from LCSRs  is already small,  and there should be an additional $O(\alpha_s)$ suppression at NLO.
In the case of annihilation topology, NLO effects 
would certainly improve the accuracy of LCSRs, but they
involve two-loop diagrams, which demand technically difficult calculations that are out of our scope here.

Going beyond the GIM limit and restoring $\lambda_b$, that is, using instead of  $\lambda_s= -\lambda_d$ 
 the full CKM unitarity relation $\lambda_s= -\lambda_d-\lambda_b$\,,
 generates additional $s$-loop topology contributions of the operators $O^s_{1,2}$ to $D^+\to \pi^+\ell^+\ell^-$. 
 Other effects proportional to $\lambda_b$, with both the loop and annihilation topologies, stem from the 
 operators $O_{3-6}$, and are important, e.g., 
 for assessing a tiny direct CP-violation in $D\to \pi\ell^+\ell^-$.
 In this paper, we neglect all these suppressed contributions.  Still, just to have an idea of 
 how small is the contribution  of the short-distance $c\to u \ell^+\ell^-$ transition in
 $D^+ \to \pi^+\ell^+\ell^-$ compared to the dominant long-distance amplitude, we will assess the effect of the operator $O_9$ in the hadronic matrix element of this decay using Eq.~\ref{eq:O9add}. 
 
\section{Use of $U$-spin symmetry}
\label{sect:uspin}

It is instructive to use the $U$-spin subgroup of the approximate $SU(3)_{fl}$ symmetry in the analysis of $D_{(s)}\to P \ell^+\ell^-$
decays. To our knowledge, this was not done before, while  there are several recent applications of the $U$-spin
to nonleptonic charm decays, see e.g., \cite{Grossman:2019xcj,Gavrilova:2022hbx}.

Our conventions  for $U$-spin multiplets  are the same as in 
\cite{Jung:2009pb} (except the sign choice)
where the $B$-meson  decays 
were analysed using $U$-spin symmetry. Here we just need to replace the $b$ quark by $c$-quark.
The $d$ and $s$ quarks and antiquarks form $U$-doublets: 
\begin{eqnarray}
\left( \begin{array}{c}
|d\rangle\\
 |s\rangle
\end{array} \right)=
\left( \begin{array}{c}
 |1/2,+1/2\rangle   \\
 |1/2,-1/2\rangle
\end{array} \right)\,,
~~\left( \begin{array}{c}
|\bar{s}\rangle\\
 |\bar{d}\rangle
\end{array} \right)=
\left( \begin{array}{ll}
 &|1/2,+1/2\rangle   \\
 -&|1/2,-1/2\rangle
\end{array} \right)\,,
\label{eq:Udoubl}
\end{eqnarray}
whereas all other quark flavours are $U$-spin singlets. Combining quarks with antiquarks to form various  
meson states, we see that the charmed mesons with $C=+1$ 
are subdivided into a $U$-doublet and a singlet:
\begin{eqnarray}
|D^+_{(U=1/2)}\rangle\equiv\left( \begin{array}{c}
|D^+_s\rangle= |c\bar{s}\rangle\\
 |D^+\rangle=|c\bar{d}\rangle
\end{array} \right)=
\left( \begin{array}{ll}
 &|1/2,+1/2\rangle   \\
 -&|1/2,-1/2\rangle
\end{array} \right)\,,~~~
|D^0\rangle= |c\bar{u}\rangle=|0,0\rangle\,. 
\end{eqnarray}
The same multiplets are formed by their charge conjugates
with $C=-1$:
\begin{eqnarray}
|D^-_{(U=1/2)}\rangle\equiv\left( \begin{array}{c}
|D^-\rangle= |d\bar{c}\rangle\\
 |D^-_s\rangle=|s\bar{c}\rangle
\end{array} \right)=
\left( \begin{array}{ll}
 &|1/2,+1/2\rangle   \\
 &|1/2,-1/2\rangle
\end{array} \right)\,,~~~
|\bar{D}^0\rangle= |u\bar{c}\rangle=|0,0\rangle\,. 
\end{eqnarray}
Furthermore, the light pseudoscalar mesons form two doublets,
\begin{eqnarray}
&&|P^+_{(U=1/2)}\rangle\equiv\left( \begin{array}{c}
|K^+\rangle= |u\bar{s}\rangle\\
 |\pi^+\rangle=|u\bar{d}\rangle
\end{array} \right)=
\left( \begin{array}{c}
 |1/2,+1/2\rangle   \\
 -|1/2,-1/2\rangle
\end{array} \right)\,,
\nonumber\\
&&|P^-_{(U=1/2)}\rangle\equiv
\left( \begin{array}{c}
|\pi^-\rangle= |d\bar{u}\rangle\\
 |K^-\rangle=|s\bar{u}\rangle
\end{array} \right)=
\left( \begin{array}{c}
 |1/2,+1/2\rangle   \\
 |1/2,-1/2\rangle
\end{array} \right)\,,
\label{eq:Udoublpseud}
\end{eqnarray}
and one triplet, 
\begin{eqnarray}
|P^0_{(U=1)}\rangle\equiv \left( \begin{array}{c}
|K^0\rangle= |d\bar{s} \rangle\\[1mm]
\frac{\sqrt{3}}{2}|\eta_8\rangle
-\frac{1}{2} |\pi^0\rangle
= \frac{1}{\sqrt{2}} |d\bar{d}-s\bar{s}\rangle\\[1.5mm]
|\bar{K}^0\rangle= |s\bar{d}\rangle
\end{array} \right)=
\left( \begin{array}{cl}
 &|1,+1\rangle   \\
 -&|1,0\rangle\\
 -&|1,-1\rangle 
\end{array} \right)\,,
\label{eq:U1}
\end{eqnarray}
where $|\eta_8\rangle=|\bar{u}u+\bar{d}d-2\bar{s}s\rangle/
\sqrt{6}$ is the dominant, $SU(3)_{fl}$-octet
component of $\eta$ meson. There are in addition, two
$U$-singlet states: 
$\eta_u = -\frac{1}{2}|\eta_8\rangle-
\frac{\sqrt{3}}{2}|\pi^0\rangle$ orthogonal to the $U_3=0$ state in the above triplet , and $|\eta'\rangle=|\bar{u}u+\bar{d}d+\bar{s}s\rangle/
\sqrt{3}$.\\
Furthermore, looking at the flavour structure of
the  three effective Hamiltonians  (\ref{eq:HeffS0}), (\ref{eq:HeffS1}) and (\ref{eq:HeffS2}), we notice that the current-current operators $O_{1}$ and $O_2$ entering them also  
 form  a $U$-triplet:
\begin{eqnarray}
O_1^{(U=1)}\equiv\left( \begin{array}{c}
\left(\bar{u}_L\gamma_\mu s_L\right)\left(\bar{d}_L\gamma^\mu c_L\right)
\\[2mm]
1/\sqrt{2}\left [\left(\bar{u}_L\gamma_\mu d_L\right)\left(\bar{d}_L\gamma^\mu c_L\right)-
\left(\bar{u}_L\gamma_\mu s_L\right)\left(\bar{s}_L\gamma^\mu c_L\right)\right]
\\[2mm]
\left(\bar{u}_L\gamma_\mu d_L\right)\left(\bar{s}_L\gamma^\mu c_L\right)
\end{array} \right)=
\left( \begin{array}{cl}
 &|1,+1\rangle   \\
 -&|1,0\rangle\\
 -&|1,-1\rangle 
\end{array} \right)\,,
\end{eqnarray}
and the triplet $O_2^{(U=1)}$ is defined similarly.
Importantly,  the quark e.m. current is a $U$-singlet. Hence, combining it 
with the linear combination $C_1O_1^{(U=1)}+C_2O_2^{(U=1)}$, 
does not alter the $U$-spin properties of the effective operators.
As a result, the overlap
of weak and e.m. interactions generating the $D_{(s)}\to P \gamma^*$ transitions represents a $U$-triplet operator: 
\begin{equation}
\hat{O}^{(U=1)}_\mu(q)=i\int d^4xe^{iq\cdot x}\, T\big\{ j_\mu^{em}(x),
\big[C_1(\mu)O_1^{(U=1)}(0)+C_2(\mu)O_2^{(U=1)}(0)\big]\big\}\,.
\label{eq:OU1}
\end{equation}
Sandwiching this operator between the $U$-multiplets of 
the initial-state and final-state mesons, we form the two possible $U$-invariant 
amplitudes, represented by the matrix elements:
\begin{eqnarray}
{\cal A}_\mu^{(D^+_{1/2}\to P^+_{1/2}\gamma^*)}
\equiv\langle P^+_{(U=1/2)}| \hat{O}_\mu^{(U=1)}|D^+_{(U=1/2)}\rangle\,,~~ ~~~
{\cal A}_\mu^{(D^0\to P^0_{1}\gamma^*)}
\equiv \langle P^0_{(U=1)}| \hat{O}_\mu^{(U=1)}|D^0\rangle \,,
\label{eq:HME3}
\end{eqnarray}
where we omit meson momenta for brevity. In addition, the 
two amplitudes of  the $U$-singlet $D^0$ transitions to the 
$U$-singlet pseudoscalar states vanish:  
\begin{equation}
\langle \eta_u| \hat{O}_\mu^{(U=1)}|D^0\rangle = 
\langle \eta'| \hat{O}_\mu^{(U=1)}|D^0\rangle = 0\,,
\label{eq:Usingl}
\end{equation}
simply because they 
cannot be transformed from one to another via
$U$-triplet interaction.

Expressing the amplitude of each decay mode in terms
of the $U$-invariant amplitudes (\ref{eq:HME3}), and using the additional 
conditions (\ref{eq:Usingl}), we obtain altogether seven relations:
\begin{eqnarray}
&&{\cal A}_{(D^+ \pi^+\gamma^*)}(q^2)= 
-{\cal A}_{(D_s^+K^+\gamma^*)}(q^2)
={\cal A}_{(D^+_s \pi^+\gamma^*)}(q^2)=
{\cal A}_{(D^+ K^+\gamma^*)}(q^2)
\,,
\label{eq:Urel1}\\
&&{\cal A}_{(D^0 \bar{K}^0\gamma^*)}(q^2)= {\cal A}_{(D^0 K^0\gamma^*)}(q^2)
=-1/2 {\cal A}_{(D^0 \pi^0\gamma^*)}(q^2) +\sqrt{3}/2{\cal A}_{(D^0 \eta_8\gamma^*)}(q^2)
\,,
\label{eq:Urel2}\\
&&{\cal A}_{(D^0 \eta_8\gamma^*)}(q^2)=-\sqrt{3}\,{\cal A}_{(D^0 \pi^0\gamma^*)}(q^2) \,,
\label{eq:Urel3}\\
&&{\cal A}_{(D^0 \eta'\gamma^*)}(q^2)=0\,,
\label{eq:Urel4}
\end{eqnarray}
written for the 
Lorentz-invariant amplitudes, defined similar to (\ref{eq:HME}).
The same relations are valid for the amplitudes of charge-conjugated decay modes. 
Note also that the sum of $d$ and $s$-loop topology contributions  (Fig.1 (b),(c)) to SCS 
amplitudes vanishes  
in the $U$-spin symmetry limit. Hence, only the annihilation topology contributes to all  
decay amplitudes entering the above relations.

The $U$-spin relations  can also be obtained from a simple comparison 
between annihilation topology diagrams, replacing the $s$-quark with $d$-quark and vice versa. This  comparison makes  clear that the 
$D^0\to P^0\gamma^*$ and $D^{\pm}_{s}\to P^{\pm} \gamma^*$ amplitudes differ from each other 
mainly because a photon emission distinguishes the light $u$-quark 
in $D^0$ from the $d(s)$ quark in $D^{\pm}_{(s)}$ by their different electric charges. 
In addition, we notice that the  relations (\ref{eq:Urel3}) and (\ref{eq:Urel4}) are in fact also the consequences of GIM cancellation, this time for the annihilation diagrams. Indeed, the $\bar{s}s$ and $\bar{d}d$ valence 
components in $\eta_u$ and $\eta'$
are equal and have the same sign, whereas the diagrams with these quark pairs in the final state have opposite signs.

All amplitude relations written down above, except the triangle relation (\ref{eq:Urel2})
for $D^0$ decays, yield the following simple equalities between differential decay widths of the SCS, CF and DCS decays: 
\begin{eqnarray}
 &&\frac{d\Gamma(D^+\to \pi^+\ell^+\ell^-)}{dq^2}=\frac{d\Gamma(D_s^+\to K^+\ell^+\ell^-)}{dq^2}
 \nonumber\\
 =
&&\frac{|V_{cd}|^2}{|V_{cs}|^2} \frac{d\Gamma(D_s^+\to \pi^+\ell^+\ell^-)}{dq^2}=
\frac{|V_{ud}|^2}{|V_{cd}|^2} \frac{d\Gamma(D^+\to K^+\ell^+\ell^-)}{dq^2}\,,
 \label{eq:rel_widths}   
\end{eqnarray}
where the ratios of phase space factors, 
reflecting kinematical corrections, 
which cancel in the $U$-spin symmetry limit, can be easily restored
and are not shown. On the other hand, the equality
\begin{equation}
\frac{d\Gamma(D^0\to \bar{K}^0\ell^+\ell^-)}{dq^2}=
\frac{|V_{ud}V^*_{cs}|^2}{|V_{us}V^*_{cd}|^2} \frac{d\Gamma(D^0\to K^0\ell^+\ell^-)}{dq^2}
\end{equation}
is evidently free from kinematical corrections.
However, it is clear that these relations cannot be taken point by point, since the exact shape 
of the spectrum is dominated by resonances and thus suffers from significant $U$-spin breaking, and so we 
expect that these relations hold for the integrated rates at the level of the usual $U$-spin breaking of about 20\%.

Turning to the short-distance $c\to u \ell^+\ell^-$ transition generated by the effective $O_9$ operator in the SM, we emphasize that this operator is a 
$U$-singlet, hence its contributions to $D_{(s)}\to P \ell^+\ell^-$ decays are 
violating the above relations.
However, the $U$-spin violation stemming from the $m_s-m_d$ difference is apparently a larger effect. 
In what follows, we will predict from LCSRs the 
$D^+\to \pi^+\ell^+\ell^-$ and $ D^+_s\to \pi^+\ell^+\ell^-$
decay amplitudes, with $m_s\neq 0$ for the $D_s$ mode, thus, assessing the accuracy of the $U$-spin symmetry relations
obtained above.

\section{ Light-Cone Sum Rule for the $D^+\to \pi^+ \gamma^*$ amplitude}
\label{sect:lcsr}
Here, we derive the LCSR~\footnote{ The  method of QCD light-cone sum rules (LCSRs) was originally developed in \cite{Balitsky:1986st,Balitsky:1989ry,Chernyak:1990ag}.  }
for the 
$D^+\to \pi^+\gamma^*$ transition amplitude (nonlocal form factor) defined in 
(\ref{eq:HME}).  Since we deal with a pion in the final state, it is natural to use the version of LCSRs with the pion DAs.  
The starting point is the vacuum-to-pion correlation function
\begin{eqnarray}
&&{\cal F}_\mu(p,q,k)=-\int \! d^4x \, e^{iq\cdot x}
\!\!\int \! d^4y \, e^{-i(p+q)\cdot y}
\nonumber\\
&&\times \langle \pi^+(p-k)|T\Big\{j_\mu^{em}(x)
\sum\limits_{i=1,2} C_i\big[O_i^{d}(0)- O_i^{ s}(0)\big] 
j_5^D(y)\Big\}|0\rangle\,,
\label{eq:corr}
\end{eqnarray}
of the e.m. quark current, the operators from 
$H_{eff}^{(\Delta S=0,\lambda_b=0)}$ and the $D$-meson interpolating current 
$$
j_5^D=(m_c+m_d)\bar{c}i\gamma_5d\,. 
$$ 
The four-momenta related to the 
currents are, respectively, $q$ and $p+q$, and the 
pion is on-shell, so that $(p-k)^2=m_\pi^2$. 

In order to be able to set up a dispersion relation in the variable
$(p+q)^2$ to be used in the LCSR analysis, in (\ref{eq:corr}) we introduce an artificial four-momentum $k$ 
at the four-quark operator vertex. This procedure was first suggested in \cite{Khodjamirian:2000mi} and then used in other LCSRs for heavy hadron nonleptonic decays \cite{Khodjamirian:2005wn,Khodjamirian:2017zdu}.
Adding the momentum $k$\,, 
transforms the correlation function (\ref{eq:corr}) into a $2 \to 2$ 
scattering amplitude, which depends not only on
the squared external momenta: $q^2$, $(p+q)^2$, $(p-k)^2=m_\pi^2$ and $k^2$, but also on the two additional 
invariant variables $p^2$ and $(p+q-k)^2\equiv P^2$.
For simplicity and convenience, we put
\begin{equation}
k^2=0,~~~ \mbox{and} ~~~ p^2=0\,.
\label{eq:pk}
\end{equation}
The correlation function (\ref{eq:corr}) is a Lorentz vector and can be expanded into invariant amplitudes, according to 
\begin{eqnarray}
&&{\cal F}_\mu(p,q,k)
=\Big[(p\cdot q)q_\mu -q^2p_\mu\Big] F((p+q)^2,q^2,P^2)
\nonumber\\
&&+
\Big[(k\cdot q)q_\mu -q^2 k_\mu\Big]  F_{(qk)}((p+q)^2,q^2,P^2)+
\epsilon_{\mu\nu\alpha\beta}p_\nu q_\alpha k_\beta 
F_{(\epsilon)}((p+q)^2,q^2,P^2)\,,
\label{eq:expan}
\end{eqnarray}
where we already take into account the e.m. gauge invariance, which implies that $q^\mu {\cal F}_\mu(p,q,k) = 0$. 
For convenience, we define one of the Lorentz structures 
to be the same as in the decay amplitude decomposition (\ref{eq:HME}), and 
the invariant amplitude $F$ multiplying this structure 
will be used for the LCSR.

As a next step, we specify the region for  the kinematic invariants, where  OPE near the light-cone is applicable, such that the intervals $x^2\sim y^2\sim (x-y)^2\sim 0$ dominate  the coordinate integration in the
correlation function (\ref{eq:corr}). This is achieved  by choosing 
the four-momenta $(p+q)$, $q$ and $P$, spacelike and far
from the hadronic thresholds in the corresponding channels, such that:
\begin{equation}
|(p+q)^2|\sim |q^2|\sim |P^2|\gg \Lambda_{QCD}^2\,.
\label{eq:cond}
\end{equation}
Note  that, simultaneously, due to the conditions (\ref{eq:pk}), the momenta squared
$k^2$ and $p^2$  also remain sufficiently far, at a distance 
of $O(m_c^2)$, from the corresponding  hadronic thresholds.
To obtain the OPE near the light-cone, we 
contract the virtual quark fields in the correlation function (\ref{eq:corr}) into propagators
and form the pion DAs from the remaining $\bar{u}$ and $d$ quarks.
This leads to the diagrams with 
annihilation  and loop   quark topologies, similar to the ones
considered in Section~\ref{sect:topol}. One important difference
is that  the on-shell $D$-meson state is replaced 
by the interpolating current with a four-momentum  
far off shell. 

\begin{figure}[h]\centering 
\includegraphics[scale=0.25]{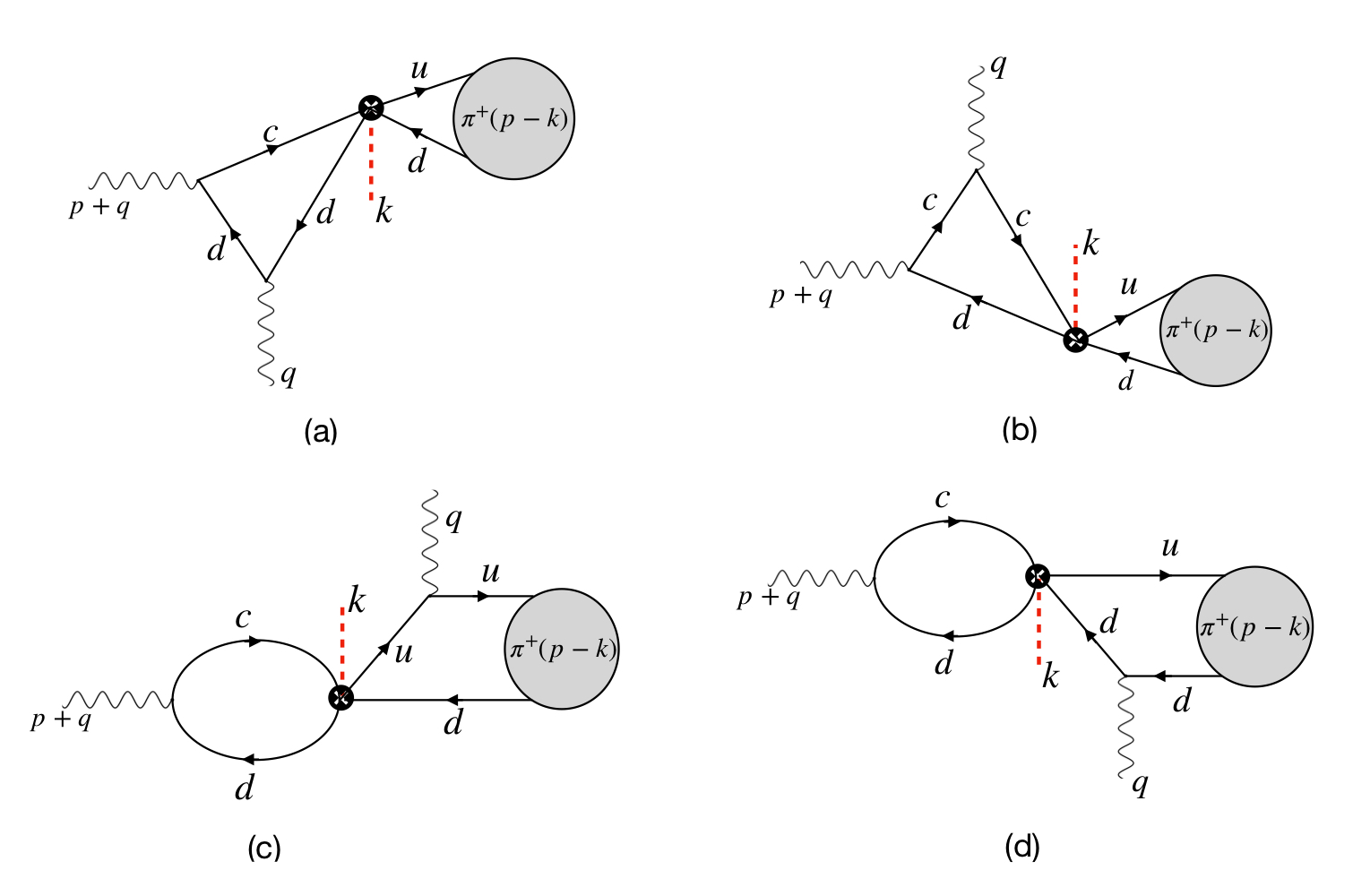}~~
\caption{ OPE diagrams for the correlation function: the annihilation topology at LO. The dashed line 
denotes the artificial momentum at the operator vertex.
}
\label{fig:lcsr_ann}
\end{figure}

At LO, there are four diagrams with the annihilation topology, depicted in Fig.~\ref{fig:lcsr_ann}. They correspond
to the four different ways to couple a virtual photon with a quark line. Note that only the operator $O_1^{d}$ contributes in the considered case of the $D^+\to \pi^+\gamma^*$ transitions. The diagrams
are factorizable and consist of a heavy-light quark loop multiplied by the vacuum-to-pion matrix element. In the first two diagrams (Fig.~\ref{fig:lcsr_ann}(a) and (b)), the pion DA is reduced to its local limit, that is, the pion decay constant. In the second pair of diagrams (Fig.~\ref{fig:lcsr_ann}(c) and (d)), the full pion DA emerges. 

\begin{figure}[h]\centering 
\includegraphics[scale=0.15]{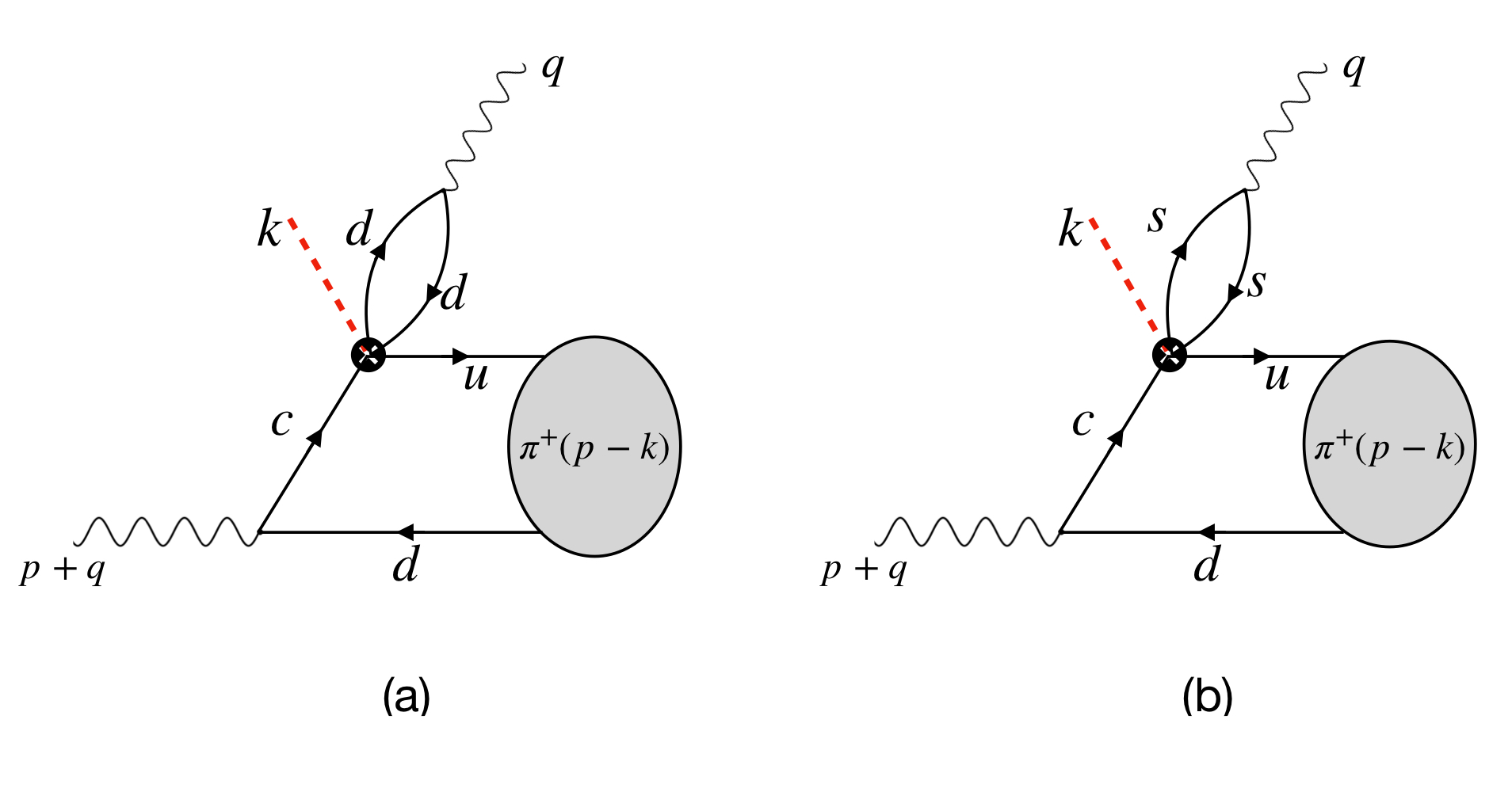}~~
\caption{ OPE diagrams for the correlation function: the loop topology at LO. }
\label{fig:lcsr_loop}
\end{figure}

The two loop-topology diagrams at LO shown in Fig.~\ref{fig:lcsr_loop}
are also factorized into a product of a two-point loop and a two-point 
correlation function with pion DAs.
As we already mentioned, the loop diagrams are present only in the case of SCS 
$D_{(s)}\to P\gamma^*$ transitions, and receive contributions from both $O_1^{d}$ and $O_1^{s}$, revealing a GIM cancellation due to the opposite sign of these operators in the  effective Hamiltonian.
Calculating these diagrams, we take into account a nonvanishing $s$-quark mass, hence, the cancellation is  incomplete.
A detailed calculation of the LO diagrams with the annihilation and loop topologies 
is presented, respectively, in Appendices~\ref{app:diagA}
and \ref{app:diagL}.

\begin{figure}[h]\centering 
\includegraphics[scale=0.28]{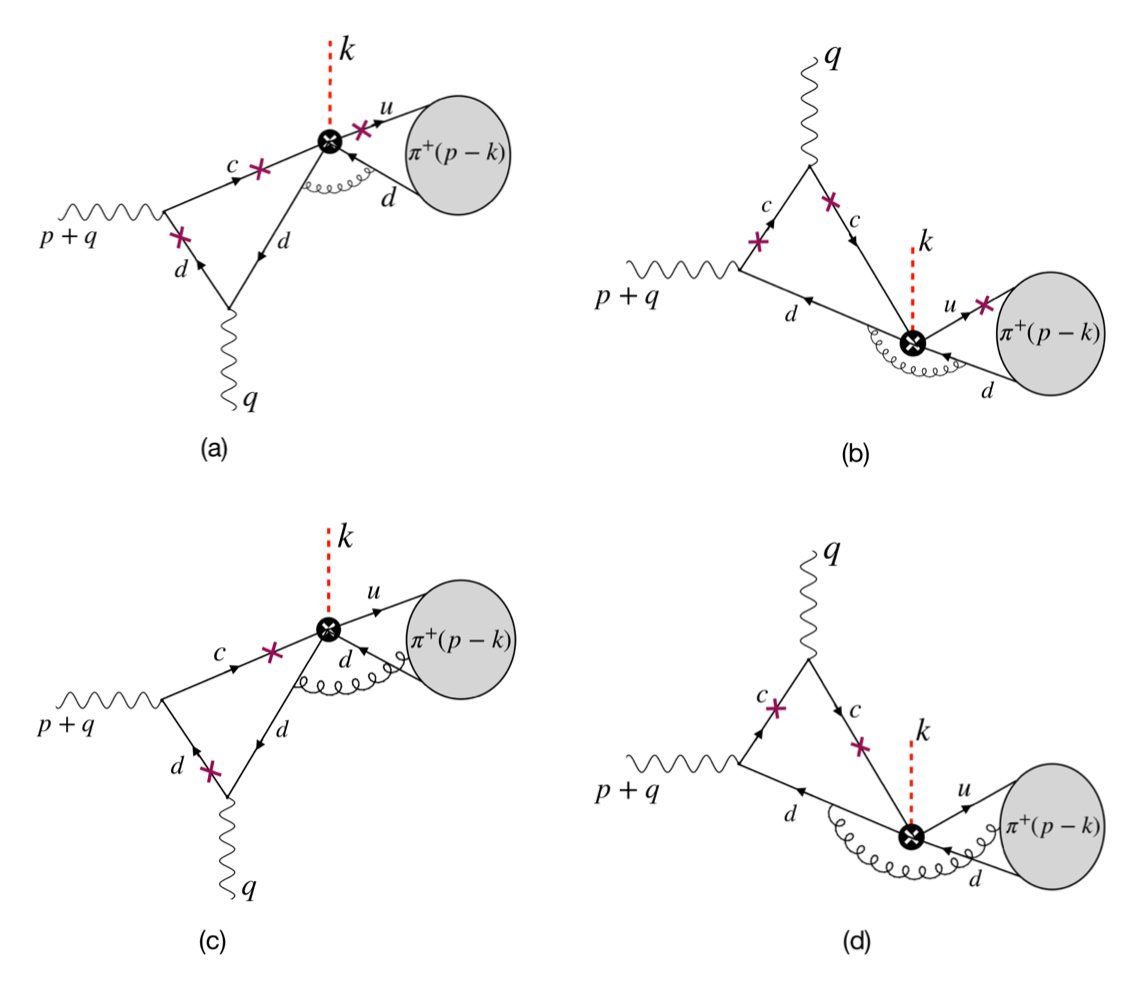}~~
\caption{ Examples of (a),(b) perturbative and (c),(d) soft-gluon corrections to the OPE 
diagrams with annihilation topology. Shown are  nonfactorizable diagrams. Crosses indicate possible ways to attach gluon lines in these diagrams.}
\label{fig:lcsr_nlo}
\end{figure}

In this paper, we confine ourselves by the LO of the correlation function.
Perturbative gluon corrections at the NLO in $\alpha_s$ are given
by two-loop  diagrams with several momentum scales. Their computation represents a technically very involved task, which is far beyond our scope.
In Fig.\ref{fig:lcsr_nlo} (a) and (b) we show, as an example, the  
nonfactorizable  NLO diagrams with annihilation topology and with the e.m. 
vertex attached to the quark lines of the $D$-meson interpolating current.
We expect that these NLO effects -- being suppressed by $O(\alpha_s(m_c))$ -- are numerically small, 
as it is the case  in the LCSRs for $D\to \pi$ semileptonic form factors
dominated by the LO contributions of the twist-2 and twist-3 pion DAs
(see e.g. \cite{Khodjamirian:2009ys}).

Soft-gluon contributions can also be included in the LCSR analysis. The corresponding nonfactorizable diagrams are
shown in Fig.~\ref{fig:lcsr_nlo}(c) and (d).
Their calculation, which is one of the possible future improvements of this work, will  involve three-particle pion DAs of higher-twist and
will induce power corrections to the LO of the correlation function.

In addition, we assume that a characteristic scale set by the virtualities in the
correlation function, in particular, the normalization scale taken for
the pion DAs, is equal to the scale $\mu \simeq m_c$ at which the Wilson coefficients multiplying the operators in (\ref{eq:corr}) are
normalized. This is a natural assumption in the absence of NLO corrections to the correlation function.

We denote the sum of all contributions taken into account in the OPE for the correlation function (\ref{eq:corr}) as  ${\cal F}_\mu^{(OPE)}(p,q,k)$. In the adopted approximation, we include 
in our calculations  the LO diagrams with annihilation and loop topologies shown
  in  Figs.~\ref{fig:lcsr_ann} and \ref{fig:lcsr_loop}. Applying the Lorentz-decomposition (\ref{eq:expan}), we isolate the 
invariant amplitude  $F^{(OPE)}((p+q)^2,q^2,P^2)$, where the momentum
variables obey  the limits (\ref{eq:cond}). 
To derive  LCSR, we need this amplitude  in the form of a dispersion
integral in the  variable $(p+q)^2$:
\begin{equation}
F^{(OPE)}((p+q)^2\!,q^2\!,P^2)= \frac{1}{\pi}\int\limits_{m_c^2}^{\infty}\! ds \,
\frac{ \mbox{Im} F^{(OPE)}(s,q^2,P^2)}{s-(p+q)^2},
\label{eq:dual}
\end{equation}
where the possible subtraction terms are ignored as they vanish after the Borel transform.
The OPE spectral density is subdivided into contributions of separate diagrams:
\begin{eqnarray}
\frac1\pi\mbox{Im} F^{(OPE)}(s,q^2,P^2)=&&
\frac1\pi \mbox{Im} F^{(a)}(s,q^2,P^2) +
\frac1\pi \mbox{Im} F^{(b)}(s,q^2,P^2) +
\frac1\pi \mbox{Im} F^{(c\, \oplus\, d)}(s,q^2,P^2) 
\nonumber\\
&&+
\frac1\pi \mbox{Im} F^{(L)}(s,q^2,P^2)\,.
\label{eq:sumIM}
\end{eqnarray}
The computation of the spectral density  for the annihilation topology 
diagrams is described in detail in Appendix~\ref{app:diagA}. 
The resulting  expressions for the imaginary parts $\mbox{Im} F^{(a),(b)}$ of the diagrams
in Fig.~\ref{fig:lcsr_ann} (a) and (b) are presented, respectively, in eqs.~(\ref{eq:ImFa}) and (\ref{eq:ImFb}). 
The corresponding expression
for the sum of diagrams in Fig.~\ref{fig:lcsr_ann} (c) and (d)  denoted
as $\mbox{Im} F^{(c\,\oplus\, d)}$ is presented in eq.~(\ref{eq:ImFcd}).
In turn, the loop-topology diagrams in Fig.~\ref{fig:lcsr_loop}
are computed in Appendix~\ref{app:diagL} in such a way that (\ref{eq:ImFL}) already contains their contribution to the final LCSR.

In the next step, following the method used for the nonleptonic $ B\to \pi \pi $ and $D \to \pi\pi$ decays 
in \cite{Khodjamirian:2000mi,Khodjamirian:2005wn,Khodjamirian:2017zdu},  we  analytically continue the OPE amplitude from spacelike $P^2<0$ to a large  fixed value 
$P^2=m_D^2$ in the timelike region, keeping the other variables  $(p+q)^2$ and $q^2$ 
spacelike. We assume that the result
of this continuation approximates the full amplitude:
\begin{equation}
F((p+q)^2,q^2,m_D^2) \simeq F^{(OPE)}((p+q)^2,q^2, P^2=m_D^2)\,,
\label{eq:dual1}
\end{equation}
which is, in essence, an assumption of the local quark-hadron duality.
An argument in favour of this approximation is that it works
rather well in the case of a two-pion final state 
for the pion e.m. form factor. Namely, the timelike pion form-factor asymptotics 
is well approximated by its QCD-based calculations at large 
spacelike momentum transfers  (see e.g., \cite{Cheng:2020vwr}).
Note also that for the $D\to\pi \gamma^*$ transition, as opposed to
LCSRs for the $D\to \pi\pi$ decay \cite{Khodjamirian:2017zdu}, we do not
need to interpolate the second hadron in the final state, 
thus avoiding additional uncertainties.  

We proceed further and represent the amplitude
(\ref{eq:dual1}) at fixed spacelike $q^2$ in a form of hadronic 
dispersion relation in the variable $(p+q)^2$. To this end, 
we employ unitarity in the channel of the $D$-meson  interpolating current 
and  isolate the ground $D$-meson contribution. We obtain:
\begin{eqnarray}
\frac{1}{\pi}\int\limits_{m_c^2}^{\infty} \!ds \,
\frac{ \mbox{Im} F^{(OPE)}(s,q^2,P^2=m_D^2)}{s-(p+q)^2}=
\frac{m_D^2 f_D 
{\cal A}_{(D^+ \pi^+\gamma^*)}(q^2) }{m_D^2-(p+q)^2} 
+\int_{s_{h_D}}\limits ^\infty \!ds\,\frac{\rho_{h_D}(s,q^2,P^2=m_D^2)}{s-(p+q)^2}\,,
\label{eq:corrdisp}
\end{eqnarray}
where the standard definition of the $D$-meson decay constant
$ \langle D^+(p+q)| j_5^D |0 \rangle=m_D^2f_D$ is used.
We remind that 
this dispersion relation is obtained for the generalized correlation function with an artificial momentum $k\neq 0$ 
and with the variable $P^2\neq (p+q)^2$. At the same time, fixing $P^2=(p+q-k)^2=m_D^2$, 
enforces $k=0$ in the $D$-meson pole term at $(p+q)^2=m_D^2$. Hence,  this term in 
(\ref{eq:corrdisp})
contains the $D\to\pi\gamma^*$ transition amplitude we need.
Note that the  equation  (\ref{eq:corrdisp})  
is valid only at large spacelike $q^2$ and $(p+q)^2$, where 
we can rely on the OPE calculation of the l.h.s.

The spectral density $\rho_{h_D}$ in the integral on r.h.s.  is a shorthand notation for the sum
over the contributions of heavier than $D$-meson hadronic states with the same quantum numbers, starting with 
the $D^*\pi$  state with the  threshold $s_{h_D}=(m_{D^*}+m_\pi)^2$: 
\begin{eqnarray}
&&\sum_{h_D=D^*\pi,....} \!\!\!\frac{1}{2\pi}d\tau_{h_D}
\langle \pi^+(p-k)\big|\int d^4x \, e^{iq\cdot x}
T\bigg\{j_\mu^{em}(x)
\sum_{i=1,2}C_i\big[O_i^{d}(0)- O_i^{ s}(0)\big] \bigg\}
\big|h_D(p+q)\rangle 
\nonumber\\
&&{}\hspace{1cm}
\times \langle h_D (p+q)\big| j_5^D(y)|0\rangle= \rho_{h_D}(s,q^2,P^2)
\Big[(p\cdot q)q_\mu -q^2p_\mu\Big] +\dots \,,
\label{eq:rhoh}
\end{eqnarray}

where $d\tau_{h_D}$ indicates the phase space integration. In the above, only
the relevant Lorentz structure is shown  (cf. (\ref{eq:expan})) and the rest is denoted by dots.

We then use the (semilocal) quark-hadron duality approximation in the $D$-meson channel and  replace the integral over the spectral density $\rho_{h_D}$ on r.h.s. of (\ref{eq:corrdisp})
by the integral over the OPE spectral density,
taken above a certain effective threshold $s_0^D$.
After that replacement, we subtract equal integrals from both sides of (\ref{eq:corrdisp}) and 
perform Borel transform. This last step brings us to the final form 
of  LCSR 
for the $D^+\to \pi^+\gamma^*$ transition amplitude (nonlocal form factor):
\begin{eqnarray}
{\cal A}^{(LCSR)}_{(D^+ \pi^+\gamma^*)}(q^2)
= \frac{1}{\pi m_D^2f_D}\int\limits_{m_c^2}^{s_0^D} \!ds\, e^{(m_D^2-s)/M^2}
\mbox{Im} F^{(OPE)}(s,q^2,m_D^2)\,,
\label{eq:lcsr}
\end{eqnarray}
with the integrand on r.h.s. given in (\ref{eq:sumIM}) and $M$ denotes the Borel parameter. This LCSR is our main new result. It is valid at $-q^2\gg \Lambda_{QCD}^2$, that is, far below hadronic thresholds in the channel of 
e.m. current.
It is instructive to establish the $q^2\to -\infty$ asymptotics of the LCSR (\ref{eq:lcsr}), determined by the limiting behaviour of $\mbox{Im} F^{(OPE)}(s,q^2\to -\infty,m_D^2)$. For the annihilation topology contributions, this limit is directly computed  
in Appendix~\ref{app:diagA} (see (\ref{eq:q2asympt_ab}) and (\ref{eq:q2asympt_cd})) yielding
$${\cal A}^{(LCSR)}_{(D^+\pi^+\gamma^*)}(q^2\to -\infty) \sim 1/q^2\,,$$ 
whereas the loop topology contribution, given by (\ref{eq:ImFL})
is suppressed by one power being  $\sim1/(q^2)^2$ at $q^2\to -\infty$.

LCSRs for all other $D_{(s)}\to P\gamma^*$ transitions can be obtained with a similar procedure, 
replacing on-shell states, DAs, effective operators and interpolation currents in the underlying correlation function (\ref{eq:corr}) accordingly.
In what follows, we will calculate the amplitudes of the CF transitions
$D_s^+\to \pi^+\gamma^*$ and $D^0\to \bar{K}^0\gamma^*$.
The corresponding correlation functions are, respectively,
\begin{eqnarray}
&&{\cal F}_\mu^{(D^+_s \pi^+\gamma^*)}(p,q,k)=-\int \! d^4x \, e^{iq\cdot x}
\!\!\int \! d^4y \, e^{-i(p+q)\cdot y}\nonumber\\
&&\times \langle \pi^+(p-k)|T\Big\{
j_\mu^{em}(x)
\Big[C_1\left(\bar{u}_L\gamma_\mu d_L\right)\left(\bar{s}_L\gamma^\mu c_L\right)+
C_2\left(\bar{u}_L\gamma_\mu t^ad_L\right)
\left(\bar{s}_L\gamma^\mu t^a c_L\right)
\Big]
j_5^{D_s}(y)\Big\}|0\rangle\,,
\label{eq:corrDs}
\end{eqnarray}
where $j_5^{D_s} =(m_c+m_s)\bar{c}i\gamma_5s$\, and 
\begin{eqnarray}
&&{\cal F}_\mu^{(D^0 \bar{K}^0\gamma^*)}(p,q,k)=-\int \! d^4x \, e^{iq\cdot x}
\!\!\int \! d^4y \, e^{-i(p+q)\cdot y}\nonumber\\
&&\times \langle \bar{K}^0(p-k)|T\Big\{
j_\mu^{em}(x)
\Big[C_1\left(\bar{u}_L\gamma_\mu d_L\right)\left(\bar{s}_L\gamma^\mu c_L\right)+
C_2\left(\bar{u}_L\gamma_\mu t^ad_L\right)
\left(\bar{s}_L\gamma^\mu t^a c_L\right)
\Big]
j_5^{D^0}(y)\Big\}|0\rangle\,,
\label{eq:corrD0}
\end{eqnarray}
where $j_5^{D^0} =(m_c+m_u)\bar{c}i\gamma_5u$\,.

As already mentioned, only annihilation topology diagrams contribute to the OPE for both correlation functions.
Compared to the diagrams  for (\ref{eq:corr}) in 
Fig.~\ref{fig:lcsr_ann}, the diagrams for (\ref{eq:corrDs}) contain a virtual $s$-quark in the 
loop instead of $d$-quark.
The diagrams for (\ref{eq:corrD0})
contain a $u$-quark in the loop instead of the $d$-quark, as well as  the kaon decay constant and DAs 
instead of, respectively, the pion decay constant and DAs.  Details of the OPE spectral density for
(\ref{eq:corrDs}) and (\ref{eq:corrD0})
 are presented in Appendix~\ref{app:CF}.

\section{Modelling the hadronic dispersion relations} 
\label{sect:hadrdisp}
In this section, we will develop a model for the hadronic matrix element of $D^+\to\pi^+\ell^+ \ell^-$ which we will fit to the LCSR calculation described in the previous section. To this end, we observe that the $D^+\to \pi^+ \gamma^*$ transition amplitude 
defined in (\ref{eq:HME}) 
obeys a dispersion relation in the variable $q^2$. This relation, derived from
analyticity  and  unitarity of the amplitude in the channel of the e.m. current~\footnote{~Intermediate steps of this derivation involve also the crossing-symmetry related amplitude 
$D^+\pi^-\to \gamma^*$.}, has the following generic form:
\begin{eqnarray}
&&{\cal A}_\mu ^{(D^+\to \pi^+\gamma^*)}(p,q)=
\frac{1}{\pi}\int\limits_{4m_\pi^2}^{\infty}\,
\frac{ds}{(s-q^2-i\epsilon)}
\nonumber\\
&&\times\Bigg[
 \frac{1}{2} \sum_{h}d\tau_h\langle 0 | j_\mu^{em} |h(q)\rangle
\langle h (q) \pi^+(p)| 
\sum\limits_{i=1,2} C_i\big[O_i^{d}- O_i^{ s}\big]
|D^+(p+q)\rangle^*\Bigg]_{(q^2=s)}\,,
\label{eq:disp0}
\end{eqnarray}
where the expression in brackets is the spectral density given by the  
sum over all intermediate hadronic states $|h\rangle$ with the quantum numbers of a virtual photon. For each state, the phase-space integration 
and summation over polarizations is implied. The lightest such state 
is a dipion $|h\rangle =|\pi^+\pi^-\rangle $ in  $P$-wave, with the threshold at $q^2=s=4m_\pi^2$. 
The dispersion relation (\ref{eq:disp0}) is valid for any $q^2$, including the spacelike region $q^2<0$, where 
it can be fitted to the LCSR result for sufficiently large negative $q^2$. Our final goal is to use a parameterized and fitted dispersive representation
of the $D^+\to \pi^+\gamma^*$ amplitude in  the physical  region $4m_\ell^2<q^2<(m_D-m_\pi)^2$  of the $D^+\to\pi^+\ell^+\ell^-$ decay. 

In order to perform such a fit, we need to model the hadronic spectral density in (\ref{eq:disp0}) in terms of a finite set of parameters. 
The models we propose take into account some of the known features of the spectral functions, 
which can be inferred from data, such as the masses and the widths of the lowest-lying resonances in the virtual photon channel. 
In the region $s\lesssim 1.0$ GeV$^2$, we approximate the hadronic spectral density in (\ref{eq:disp0}) with a sum over light neutral vector mesons, schematically: 
\begin{equation}
\Bigg[\sum_ h \dots\Bigg]_{s \lesssim 1.0 ~{\rm GeV}^2} \simeq \sum_{V= \rho^0, \,\omega, \,\phi} \cdots. 
\label{eq:rhoomegphi}
\end{equation}
 Note that in this region of the lepton-pair invariant mass, 
intermediate $\eta$ and $\eta'$ mesons also contribute to $D\to\pi\ell^+\ell^-$. However, the 
transitions $\eta\to \ell^+\ell^-$ 
and $\eta'\to \ell^+\ell^-$ proceed via two virtual photons, and hence we expect their 
contribution to be suppressed by an additional power of $\alpha_{em}$.

Turning to a more detailed description of the dispersion relation (\ref{eq:disp0}), and assuming (\ref{eq:rhoomegphi}), we define  the  decay constants of the vector mesons with respect to the e.m. current as:
\be
\langle 0 | j^{em}_\rho|V(q)\rangle = \kappa_V\epsilon^{V}_\rho m_V
f_V\,,
\label{eq:fV}
\ee
where $\epsilon^{V}$ is the polarization vector, and the coefficients: 
\begin{equation}
\kappa_{\rho^0}=1/\sqrt{2}, ~~ \kappa_\omega=1/(3\sqrt{2}),~~  \kappa_\phi=-1/3\,,
\label{eq:Vcoeff}
\end{equation}
are adjusted to the meson valence-quark content. 
The decay constants $f_V$ are determined from the $V\to e^+e^-$ decay widths.
Neglecting the electron mass, we obtain:
\begin{equation}
|f_{V}|=\frac{1}{|\kappa_V|}\sqrt{\frac{3m_V BR(V\to e^+e^-)\Gamma_{tot}^V}{4\pi\alpha_{em}^2}}\,.    
\label{eq:fVexp}
\end{equation}

In each vector-meson term entering (\ref{eq:disp0}), the decay constant $f_V$ is multiplied by  the amplitude of the weak nonleptonic $D^+\to \pi^+V $
decay, defined as 
\be
\langle \pi^+(p)V(q) |
\sum\limits_{i=1,2} C_i\big[O_i^{d}- O_i^{ s}\big]
|D^+(p+q)\rangle \equiv(\epsilon_{V}^*\cdot p)\,m_VA_{D^+\pi^+V }\,,
\label{eq:ADpiV}
\ee
and assuming that the effects proportional to $\lambda_b$ 
are negligible. We obtain the absolute value of this amplitude from the 
measured $D^+\to \pi^+ V$ width:
\be |A_{D^+ \pi^+ V}| =
\Bigg(\frac{8\pi BR(D^+\to \pi^+ V)}{\tau_{D^+} G_F^2
|\lambda_d|^2 m_{D^+}^3 \lambda ^{3/2}_{D^+ \pi^+V}}\Bigg)^{1/2}\,, 
\label{eq:DpiVampl}
\ee
where  $\lambda_{D^+\pi^+V}=\lambda(1,m_{\pi^+}^2/m_{D^+}^2,m_V^2/m_{D^+}^2)$
is the K\"all\'en function.

Substituting the hadronic matrix elements (\ref{eq:fV}) and (\ref{eq:ADpiV})  in (\ref{eq:disp0}), we isolate  the invariant amplitude defined in (\ref{eq:HME}) and obtain, motivated by the dispersion relation, an ansatz for this amplitude in the following form:
\begin{eqnarray}
{\cal A}^{(D^+ \pi^+\gamma^*)}(q^2)
&&=
 \sum_{V=\rho,\omega, \phi } \frac{r_V\, e^{i \varphi_V} }{(m_V^2-q^2-
i\sqrt{q^2}\Gamma_V(q^2))}
+\int_{s_{\rm th}}^{\infty} ds
\frac{\rho_{exc}(s)}{(s-q^2-i\epsilon)}\,,
\label{eq:dispAuns}
\end{eqnarray}
where we introduce a short-hand notation for the residues of the resonance poles,  
\begin{equation}
r_V=\kappa_V f_V |A_{D^+\pi^+V}|\,,
 \label{eq:rVdef}   
\end{equation}
and where all contributions apart from the $\rho^0(770)$, $\omega(782)$, and $\phi(1020)$, which we denote hereafter as $\rho$, $\omega$, and $\phi$ for brevity, 
are parametrized as an integral over
the spectral density $\rho_{exc}(s)$ of continuum and excited 
states with invariant masses above 
the threshold $s_{th} > m_\phi^2$.

The relation (\ref{eq:dispAuns}) deserves several comments.
First of all, we note that, due to the $\sim 1/q^2$ asymptotics of the amplitude ${\cal A}^{(D^+ \pi^+\gamma^*)}(q^2)$, there is no need for subtractions. Nevertheless, in our numerical analysis, we will also use the 
once-subtracted form of the same dispersion relation\footnote{\,Hadronic dispersion relations similar to (\ref{eq:dispA})  were 
used \cite{Khodjamirian:2010vf} in the analysis 
of nonlocal effects in $B\to K^{(*)}\ell^+\ell^-$  decays,
see also \cite{Khodjamirian:2012rm,Bobeth:2017vxj,Gubernari:2020eft}.}
:
\begin{eqnarray}
{\cal A}^{(D^+ \pi^+\gamma^*)}(q^2)
&&={\cal A}^{(D^+ \pi^+\gamma^*)}(q_0^2)
+
(q^2-q_0^2) \Bigg[ \sum_{V=\rho,\omega, \phi } \frac{r_V\, e^{i \varphi_V} }{(m_V^2-q_0^2)(m_V^2-q^2-
i\sqrt{q^2}\Gamma_V(q^2))}
\nonumber\\
&&+\int_{s_{\rm th}}^{\infty} ds
\frac{\rho_{exc}(s)}{(s-q_0^2)(s-q^2-i\epsilon)}\Bigg]\,.
\label{eq:dispA}
\end{eqnarray}
The subtraction constant at a certain spacelike point, $q_0^2<0$ 
will be  calculated from LCSR. In the subtracted version
(\ref{eq:dispA}), the
large-$s$ part of the integral over $\rho_{exc}(s)$ 
is additionally suppressed, making the dispersion relation less dependent on the details of the hadronic
spectral density  at~$s\gg s_{th}$.

Secondly, in the resonance part of the dispersion relation (\ref{eq:dispAuns}), we use a simple Breit-Wigner form for
each vector meson and specify their widths as follows:
\begin{equation}
\Gamma_\rho(q^2)= \frac{\sqrt{q^2}}{m_\rho}
\left( \frac{\beta_\pi(q^2)}{\beta_\pi(m_\rho^2)}\right)^3 \Gamma_\rho^{tot}\theta(q^2-4m_\pi^2)\,,~~
\sqrt{q^2}\Gamma_{\omega(\phi)}(q^2)=m_{\omega(\phi)}\Gamma^{tot}_{\omega(\phi)}\theta(q^2-9m_\pi^2)\,,
\label{eq:width}    
\end{equation}
where $\beta_\pi(q^2)=\sqrt{1-4m_\pi^2/q^2}$.
In particular, for the $\rho$-meson, we adopt the energy-dependent width which
effectively accounts for the continuum dipion states coupled to $\rho$.
A detailed derivation can be found, e.g., in \cite{Bruch:2004py}.
For the narrow resonances $\omega$ and $\phi$, we retain constant widths, demanding that 
they vanish below the threshold of  the $\pi^+\pi^-\pi^0$ state, which is 
the lightest isospin-zero hadronic state in the channel of the 
e.m. current. With the adopted resonance ansatz, we assume 
that  all dominant hadronic states with (invariant) masses  up to $\sim 1$ GeV are taken into account in the hadronic spectral density. 
A possible improvement of this rather simple model of the hadronic spectral density at $s\lesssim 1 $ GeV$^2$ could be achieved by a dedicated dispersive and coupled-channel analysis, which is beyond the scope of the present paper. Our choice of the threshold  $s_{\rm th}$ for the heavier hadronic states in the dispersion relation (\ref{eq:dispA}) will be discussed below, in Sect.\ref{sect:num}. 
Note also that the threshold behaviour adopted in (\ref{eq:width}) ensures that in the spacelike region,
all widths in the denominators of the pole terms in (\ref{eq:dispAuns}) and (\ref{eq:dispA}) vanish.

As a final comment, we emphasize  
that the amplitude ${\cal A}^{(D^+ \pi^+\gamma^*)}(q^2)$ is a complex-valued quantity also at $q^2<0$. The reason is that Im~${\cal A}^{(D^+ \pi^+\gamma^*)}(q^2<0) \neq 0$ is generated
in the channel of the timelike $(p+q)^2$ variable.
For the $D^+ \to \pi^+ \gamma^*$ transition, this variable is fixed
at $(p+q)^2=m_{D^+}^2$. In the absence of an  external momentum
transfer in the weak vertex of this transition, there 
are intermediate  on-shell hadronic states with thresholds
located at $(p+q)^2<m_{D^+}^2$. Via the unitarity condition, these states generate singularities of the $D^+ \to \pi^+ \gamma^*$ amplitude, forming its imaginary part. 
 Note that a quark-hadron duality counterpart of these singularities appears in LCSR 
 (\ref{eq:lcsr}) at $q^2<0$, after the analytic continuation $P^2\to m_{D^+}^2$. Indeed, 
 according to our calculation presented in Appendix~\ref{subapp:A4}, imaginary part emerges in the diagrams (c) and (d) already at LO. At NLO, these effects will also be present in all other diagrams, including both annihilation and loop topologies~\footnote{An emergence of imaginary part at spacelike
 momentum transfer 
 takes place also in the nonlocal form factors of the $B\to K^{(*)}\gamma^*$
 transitions, see \cite{Khodjamirian:2010vf, Khodjamirian:2012rm} for more details.}.
Accordingly, (\ref{eq:dispAuns}) and (\ref{eq:dispA}) should be understood 
as a complex combination of two separate dispersion relations 
for the real and imaginary parts of  ${\cal A}^{(D^+ \pi^+\gamma^*)}$. Merging them together implies that all terms in (\ref{eq:dispAuns}) and (\ref{eq:dispA}) are complex valued.
This is reflected by the generic phases 
attributed to  each resonance term,  and by treating the 
spectral density $\rho_{exc}(s)$ as  a complex valued function. Note that the phases $\phi_V$  in the dispersion relations 
(\ref{eq:dispAuns}) and (\ref{eq:dispA}) are the only unknown parameters in the vector meson terms, all other parameters (absolute values of the residues, the meson masses and widths) are fixed by the 
available data, while also using (\ref{eq:fVexp}) and  (\ref{eq:DpiVampl}).

Modelling the  $D^+\to\pi^+\ell^+\ell^-$ hadronic matrix elements with a 
superposition of the $\rho,\omega$ and $\phi$ meson contributions  was already done before in the literature (see e.g.,~\cite{deBoer:2015boa}) 
to describe the long-distance part of this process. Clearly, to leave the relative phases between the $\rho,\omega $ and $\phi$ terms arbitrary induces 
a very large uncertainty of the resulting
$D^+ \to \pi^+ \ell^+\ell^-$ width. 
Here, we use the LCSR calculation
to constrain the phases between resonance contributions, which leads to a substantial 
reduction of  uncertainties.

Another new element in this paper is that we retain and quantitatively assess, again with the help of LCSR, the integral 
over hadronic spectral density $\rho_{exc}(s)$ in 
(\ref{eq:dispAuns}) and (\ref{eq:dispA}), containing a sum over hadronic states heavier than $\phi$.
It is evident that to simply ignore these states, as it was done
in the three-resonance models before, would violate the
dispersion relation. Albeit being subdominant in the low $s$ region, the
integral over $\rho_{exc}$ induces a   "nonresonant background" there. 
A relevant example is provided by the dispersion relation for the charged pion e.m. form factor. 
At zero momentum transfer, its value that is  fixed by the pion electric charge,
can only be reproduced if the  $\rho$-meson term in that dispersion relation is complemented by
the sum  of excited $\rho$ resonances (for more details, see e.g. \cite{Bruch:2004py}).

In the dispersion relation (\ref{eq:dispAuns}) and in its subtracted version (\ref{eq:dispA}), the contributions included in the 
generic spectral density $\rho_{exc}(s)$ are of numerous origin.
Apart from  excited vector mesons and hadronic continuum  with  light-quark content, 
the vector charmonia ($J/\psi,\psi(2S),...$), and open charm-anticharm states ($D\bar{D}$,...) also 
contribute to $\rho_{exc}(s)$ at $s\geq m_{J/\psi}^2$.  
Altogether, the function $\rho_{exc}(s)$ involves  too many hadronic degrees of freedom to be  
fitted by a few parameters.
However, if we only aim at obtaining the  $D^+\to \pi^+\gamma^*$ amplitude
below $q^2=s_{th}$, i.e. for the lower part of the $q^2$ spectrum, it is sufficient to use a simple parameterization for the dispersion integral involving $\rho_{exc}(s)$. 

To this end, we concentrate  on the subtracted dispersion relation (\ref{eq:dispA}),
which is less sensitive to the large-$s$ behaviour of the hadronic spectral density.
We notice that both imaginary and real parts of the integral over $\rho_{exc}(s)$
are  functions of $q^2$ with no singularities at  $q^2<s_{th}$.
This enables us to use the usual expansion in powers of $z$, so we transform
\begin{equation}
 q^2\to z(q^2)=\frac{\sqrt{s_{th}-q^2}-
 \sqrt{s_{th}}}{\sqrt{s_{th}-q^2}+\sqrt{s_{th}}}\, , 
 \label{eq:ztransf}    
 \end{equation}
where we  adopt the simplest version of this transformation.
The  $z$-transformation  maps
 a certain spacelike interval  $-|q_{1}^2| \leq q^2 \leq -|q^2_{2}|$ , where the LCSR is applicable, onto the interval 
 $z(q_{2}^2)<z<z(q_{1}^2)$ on the positive real axis of the $z$-plane. In particular, we can chose $q_{1}^2$ and $q^2_{2}$ in such a way that 
$|z|\ll 1$ in this interval, justifying a truncation of the expansion in powers of $z$. 

We want to extrapolate the model to the physical region of $D^+ \to \pi^+ \ell^+ \ell^-$,  
so we need to continue our expressions to positive values of $q^2$. From the definition
of $z$ in (\ref{eq:ztransf}), we infer that we have $|z| = 1$ for $q^2 \ge s_{th}$, and hence the $z$ expansion will not converge for such values of $q^2$. This means, in turn, that 
we can trust the $z$-expansion only up to values $q^2 \le s_{max}$ where $|z(s_{max})| \ll 1$. 
The practical choices for $q_{1}^2, q^2_{2} $  and $s_{max}$ will be specified in the next section.

Using the fact that $|z| \ll 1$ in the regions of interest,
the  integral in (\ref{eq:dispA}), 
which is a smooth function of $z$ at $q^2<s_{max}$,  
is approximated by a truncated Taylor expansion around $z=0$~\footnote{ A similar $z$-expansion was
used in analysing nonlocal form factors 
for the $B\to K^{((*)}\ell^+\ell^-$ decays
in  \cite{Bobeth:2017vxj, Gubernari:2020eft} .}:

\begin{equation}
 (q^2-q_0^2)\int\limits_{s_{th}}^{\infty} ds
\frac{\rho_{exc}(s)}{(s-q_0^2)(s-q^2)} 
 =\sum\limits_{k=0}^K \alpha_k [z(q^2)]^k=
 \sum\limits_{k=1}^K \alpha_k\big( [z(q^2)]^k -[z(q_0^2)]^k\big).
 \label{eq:zexp}   
\end{equation}
We remind that the function $\rho_{exc}(s)$ 
in (\ref{eq:zexp}), hence, also
the coefficients $\alpha_k$, are complex valued,
so that 
$\alpha_k= \mbox{Re}\,(\alpha_k) + i\,\mbox{Im}\,(\alpha_k)$. In addition, since at $q^2=q_0^2$
the l.h.s. of (\ref{eq:zexp}) vanishes, there is a constraint  
on the sum on r.h.s., which we implement by expressing the zeroth order
coefficient $\alpha_0$ via the sum over all remaining coefficients. That explains the r.h.s. of (\ref{eq:zexp}).

After replacing  the integral over the spectral density $\rho_{exc}(s)$ with the truncated $z$-expansion (\ref{eq:zexp}),
the once-subtracted dispersion relation (\ref{eq:dispA}) 
for the $D^+\to\pi^+\gamma^*$ amplitude takes  the form:
\begin{eqnarray}
{\cal A}_{(D^+ \pi^+\gamma^*)}^{(disp-z)}(q^2)
=&&{\cal A}_{(D^+ \pi^+\gamma^*)}(q_0^2)
+(q^2-q_0^2)\sum_{V=\rho,\omega,\phi} 
\frac{r_V e^{i \varphi_V} }{(m_V^2-q_0^2)(m_V^2-q^2-
i \sqrt{q^2}\Gamma_V(q^2))}
\nonumber\\
+&&
\sum\limits_{k=1}^K \alpha_k\big( [z(q^2)]^k -[z(q_0^2)]^k\big)\,. 
\label{eq:dispZ}
\end{eqnarray}
This specifies a realistic model for the hadronic dispersion relation. The subtraction point of the dispersion relation is taken at sufficiently negative $q_0^2$, and thus the constant term ${\cal A}^{(D^+ \pi^+\gamma^*)}(q_0^2<0)$ can be calculated from LCSR, while the absolute values of the residues of the poles as well as the widths of the resonances are taken from data.

The parameters of the  model (\ref{eq:dispZ}) are the phases of the residues and the coefficients of the $z$-expansion. We determine  these parameters, continuing the expression (\ref{eq:dispZ}) to negative $q^2$ values and 
fitting it in the chosen interval $-|q_1^2|<q^2<-|q_2^2|$ to the amplitude computed via LCSR.
The details of the numerical fit procedure are described in the next section.
With all  parameters of the hadronic model (\ref{eq:dispZ}) determined, we will 
be able to predict the differential 
branching fraction  of the $D^+\to\pi^+\ell^+\ell^-$ decay 
 in the  region $4m_\ell^2<q^2<s_{max}$.
 
 We emphasize that, applying the $z$-expansion, we  do not 
 need  to model the details of the hadronic spectral density $\rho_{exc}(s)$ above $s_{th}$. 
The price for that is that the region  $q^2>s_{max}$ 
for $D^+\to\pi^+\ell^+\ell^-$ remains inaccessible. 
As an alternative approach, in the next section, an explicit model of hadronic states,
 approximating the spectral density $\rho_{exc}(s)$ 
 in the region $s>s_{th}$ by 
 excited vector mesons will be introduced. It will allow us to extend the prediction for the 
 $D^+\to\pi^+\ell^+\ell^-$  width to the whole physical
 region of this decay, however, at the price of a much higher systematic  uncertainty, which is hard to quantify.

\section{Numerical analysis}
\label{sect:num}
\subsection{Input parameters}
We start our numerical analysis by calculating the 
OPE spectral density entering the LCSR (\ref{eq:lcsr}) for the $D^+\to\pi^+\gamma^*$ amplitude. Since next-to-leading perturbative corrections to the correlation function
are not taken into account, we choose an appropriate renormalization scale,
orienting ourselves at previous analyses of similar correlation functions with the pion DAs and $D$-meson interpolating current. A useful example is provided by LCSRs for the $D\to\pi$ 
form factors. In the most advanced analysis
\cite{Khodjamirian:2009ys} of these LCSRs, the NLO, $O(\alpha_s)$ corrections to the correlation function were taken into account and turned out to be 
reasonably small at the  scale close to $\mu=1.5$ GeV. Here we  
adopt the same scale for the pion DA, for the virtual quark masses, and
also for the Wilson coefficients
in the effective Hamiltonian \cite{deBoer:2016dcg}.
The numerical values of these coefficients at the scale $\mu=1.5 $ GeV in the  next-to-next-to-leading-logarithmic order are taken from
\cite{Feldmann:2017izn}:
\begin{equation}
C_1=1.033, ~~ C_2=-0.647\,, ~~C_9=-0.445.
\end{equation}
The remaining  coefficients $C_{3-8}$ are smaller than 0.1 and we neglect them, having in mind that contributions of the corresponding operators also contain a very small factor $\lambda_b$.

In Table \ref{tab:inpPDG}, we collect the input parameters
that are  taken from the literature.
\begin{table}[h]
  \centering
\begin{tabular}{|c|ll|}
\hline
 & & \\[-4mm]
 & $G_F=1.1664 \cdot 10^{-5} $ GeV$^2$, ~~~
 $\alpha_{em}=1/137$ & \\
&  &\\[-4mm]
Electroweak parameters &$\lambda_d=V_{ud}V^*_{cd}=-(0.2152 \pm 0.004)$\, &\\ 
&  &\\[-4mm]
& $|\lambda_b|=|V_{ub}V^*_{cb}|=(1.57\pm 0.13 ) \times 10^{-4} $ &\\
&  &\\[-4mm]
\hline
&  &\\[-4mm]
 & $\overline{m}_c(\overline{m}_c)=1.2730\pm 0.0046$ GeV  &\\[-2mm]
Quark masses &  &\\[-2mm]
& $\overline{m}_s(2~ \mbox{GeV})=93.5\pm 0.8 $ MeV & \\
 &  &\\[-4mm]
 \hline
& &\\[-4mm]
 & $m_{\pi^\pm}= 139.57$ MeV, 
 ~~~ $f_\pi=(130.2 \pm 1.2)$ MeV & \\
Pion parameters &&\\[-2mm]
 &$a_2^\pi( 1.0~\mbox{GeV})= 0.28 \pm 0.05 $,
 ~~~ $a_4^\pi( 1.0~\mbox{GeV})= 0.19 \pm 0.06 $  \cite{Cheng:2020vwr} &  \\
&&\\[-4mm] 
 \hline
 &&\\[-4mm] 
& $m_{K^0}= 497.61$ MeV, 
 ~~~ $f_K=(155.7\pm0.3)$ MeV & \\
 Kaon parameters
 &&\\[-2mm]
 & $a_1^K( 1.0~\mbox{GeV})= 0.10 \pm 0.04$\,\cite{Chetyrkin:2007vm}\,,~~
 $a_2^K( 1.0~\mbox{GeV})= 0.25\pm 0.15$ \cite{Ball:2006wn} &\\
&&\\[-2mm]
\hline
&& \\[-4mm]
 & $m_{D^\pm}=1869.7$  MeV, 
~~~~~~~$m_{D_s}=1968.3 $ MeV & \\
 && \\[-4mm]
$D_{(s)}$-meson parameters & $f_D=(212.0\pm 0.7)$ MeV, ~~ $f_{D_s}=(249.9\pm 0.5)$ MeV  \cite{FLAG24} &\\
&  & \\[-4mm]
& $\tau_{D^{\pm}}=1.033\pm 0.005$ ps,~~~
 $\tau_{D_s}= 0.5012 \pm 0.0022$ ps & \\[2mm]
\hline
\end{tabular}
\caption{External input parameters taken from \cite{PDG}, unless stated otherwise.} 
\label{tab:inpPDG}
\end{table}
We use the $\overline{MS}$ masses of the $c$ and $s$ quarks and neglect the $d$-quark mass in the OPE diagrams. Note that the expressions for the loop topology diagrams are only applied in the spacelike region, at $q^2\leq -1.0$ GeV$^2$. Therefore,
retaining $m_d\neq 0$  in the loop integral (\ref{eq:dsloop}) would yield a tiny effect less than 0.2 \% 
on the value of this integral. 
For the pion twist-2 DA we use the ansatz (\ref{eq:pionDAmod})  with two Gegenbauer 
polynomials. Their coefficients (Gegenbauer moments)
at a reference scale $\mu_0=1.0$ GeV quoted 
in Table~\ref{tab:inpPDG} are taken from \cite{Cheng:2020vwr} and are based  on the fit of LCSRs for the pion e.m. form factors to experimental data.
In Table \ref{tab:inpLCSR}, additional input needed for the LCSR 
(\ref{eq:lcsr}) is presented, including for convenience also the quark masses
and Gegenbauer moments at the adopted scale.
\begin{table}[h]
    \centering
    \begin{tabular}{|l|l|}
    \hline
    Parameter&Value \\
    \hline
& $m_c(1.5~ \mbox{GeV})=1.195 \pm 0.006$ GeV   \\[-2mm]
Quark masses &\\[-2mm]
&$m_s(1.5 ~\mbox{GeV})=102\pm 0.9 $ MeV \\
 \hline
 \hline
 Borel parameter & $M^2=(2.0\pm 0.5)$ GeV$^2$  \cite{Khodjamirian:2009ys} \\
 & $s_0^D=(5.5\pm 0.5)$ GeV$^2$ 
\\[-2mm]
duality threshold &\\[-2mm]
 & $s_0^{D_s}= ( 6.5\pm 1.0 ) $ GeV$^2$ \\
\hline
&$a_2^\pi( 1.5~\mbox{GeV})= 0.23\pm 0.04 $ \\
&$a_4^\pi( 1.5~\mbox{GeV})= 0.14 \pm 0.04 $\\[-2mm]
Gegenbauer moments &\\[-2mm]
& $a_1^K( 1.5~\mbox{GeV})=0.09\pm 0.035$,\\ ~~ 
&$a_2^K( 1.5~\mbox{GeV})=0.21\pm 0.12 $  \\
\hline
\end{tabular}
\caption{Parameters used in the numerical analysis of  LCSRs  }
\label{tab:inpLCSR}
\end{table}

Our choice of
the Borel parameter interval in Table \ref{tab:inpLCSR} deserves a comment. The annihilation diagrams in Fig.~\ref{fig:lcsr_ann} that are dominating in the OPE for the correlation function (\ref{eq:corr}), contain a $c\bar{d}$ three- and two-point loop and a vacuum-to-pion hadronic matrix element. Therefore, it is 
conceivable to use a range of Borel parameter typical for 
the two-point or three-point QCD sum rules based on local OPE,
rather than a range specific for the LCSRs for the $D^+\to\pi^+$ 
form factors. The latter LCSRs have similarity only with the 
loop-topology diagrams in Fig.~\ref{fig:lcsr_loop}. Their total 
contribution is, however, very small. Based on this argument, in our numerical analysis we use the same interval as in the two-point QCD sum rule for the 
$D$-meson decay constant (see e.g., \cite{Khodjamirian:2020mlb} for a recent use of these sum rules).

Concerning the duality threshold $s_0^D$, we emphasize that this parameter is specific for each 
QCD sum rule. To determine its value, we use the following well-known procedure.  
Differentiating the LCSR (\ref{eq:lcsr}) over $-1/M^2$ and dividing the derivative sum rule by the original one, we obtain an equality for the $D$-meson mass squared:
\begin{eqnarray}
m_D^2=\frac{\int\limits_{m_c^2}^{s_0^D} \!ds \,s \, e^{-s/M^2}
\mbox{Im} F^{(OPE)}(s,q^2,m_D^2)}{
\int\limits_{m_c^2}^{s_0^D} \!ds \,  e^{-s/M^2}
\mbox{Im} F^{(OPE)}(s,q^2,m_D^2)
}\,.
\label{eq:lcsrs0}
\end{eqnarray}
We found that the $D$-meson mass calculated from the above equation 
at $M^2=2.0 \pm 0.5 $ GeV$^2$ deviates from its measured value by  less than  1 \% 
in the whole region of our interest, $-1.0~\mbox{GeV}^2>q^2>-10.0~\mbox{GeV}^2$, if we take 
$s_0^D$ between 5.0 and 6.0 GeV$^2$. That explains our choice for the interval of the threshold parameter quoted in Table~\ref{tab:inpLCSR}.
As an additional test, we estimated the share of the states above $D$ meson in the duality approximation and found that the integral from $s_0^D$ to infinity taken over the integrand on r.h.s. of  the LCSR (\ref{eq:lcsr}) does  not exceed 30\% of the total integral from $m_c^2$ to infinity, a usual criterion for a QCD sum rule. Although the choice of $s^D_0$ is partly correlated with $M^2$, we will conservatively treat them as uncorrelated parameters. 

Furthermore, in Table~\ref{tab:Vpar}, we present the parameters related to the 
resonance part of the hadronic dispersion relations
(\ref{eq:dispAuns}) and (\ref{eq:dispA}). Here we collect together
the masses, leptonic and total widths of vector mesons,
together with the measured branching fractions of $D\to \pi V$ decays, all taken from
\cite{PDG}. The decay constants and nonleptonic amplitudes, also included in this Table, are calculated
from, respectively, (\ref{eq:fVexp}) and (\ref{eq:DpiVampl}). 

\begin{table}[h]
\centering
\begin{tabular}{|c|c|c|c|}
\hline
&&&\\[-3mm]
 Vector meson $V$ & $\rho^0(770)$ &$\omega(782)$&$\phi(1020)$  \\[2mm]
\hline
\hline
&&&\\[-3mm]
$m_V$ (MeV)& 775.26 $\pm$ 0.23 & 782.66 $\pm$ 0.13 & 1019.46 $\pm$ 0.02  
\\[2mm]
 $\Gamma_V^{tot}$(MeV)  & 147.4 $\pm$ 0.8 & 8.68 $\pm$ 0.13& 4.249 $\pm$ 0.013 \\[2mm]
 \hline
 \hline
 &&&\\[-3mm]
  $BR(V\to \mu^+\mu^-)$&$(4.55 \pm 0.28)\!\times\! 10^{-5} $&
 $(7.4 \pm 1.8)\!\times\! 10^{-5} $ &
 $(2.85 \pm 0.19)\!\times\! 10^{-4} $\\[2mm]
 $|f_V|$ (MeV)&  $219.85\pm 1.31 $ &$201.09\pm 1.12$& $ 228.12\pm 0.57$
 \\[2mm]
\hline 
\hline
&&&\\[-3mm]
$BR(D^+\to \pi^+ V)$ & $(8.4 \pm 0.8)\!\times\! 10^{-4}$&  $(2.8 \pm 0.6)\!\times\! 10^{-4}$ &$(5.7 \pm 0.14)\!\times\! 10^{-3}$\\[2mm]
$|A_{D^+ \pi^+ V }|$ (MeV)  & $24.33\pm 1.25 $  &$14.14\pm 1.54  $ & $81.72\pm 1.83$ \\[2mm]
 \hline
 &&&\\[-3mm]
$r_V=\kappa_V f_V|A_{D^+ \pi^+ V }|$  &$3.783\pm 0.195$&
$0.670 \pm 0.073 $ &$-(6.214 \pm 0.140)$
\\[2mm]
(in $10^{-3}~\mbox{GeV}^2$) 
&&&\\
\hline
&&&\\[-3mm] 
 $BR(D^+\to \pi^+V)_{V\to \mu^+\mu^-}$&$(3.82 \pm 0.43)\!\times\! 10^{-8}$ & $(2.1 \pm 0.7)\!\times\! 10^{-8}$&$(1.62 \pm 0.12)\!\times\! 10^{-6}$\\[2mm] 
 \hline
\hline
\end{tabular}
\caption{
The  vector mesons parameters taken from \cite{PDG}
 and the residues in the resonance part of the dispersion relations (\ref{eq:dispAuns})  and (\ref{eq:dispA}). The decay constants $f_V$ are determined from (\ref{eq:fVexp}) using the more accurately measured \cite{PDG} $V\to e^+e^-$ widths\,; the amplitudes of $D^+\to \pi^+ V$ decays are calculated from (\ref{eq:DpiVampl}).
}
 \label{tab:Vpar}
 \end{table}
To choose an optimal value for the lower limit $s_{th}$ of the integral over the spectral 
density $\rho_{exc}(s)$, we assume that 
in the region above $\phi$-meson, $s\gtrsim 1.0$ GeV$^2$,  
and below the upper limit of the  
$D^+ \to\pi^+ \ell^+\ell^-$ physical region $s\leq(m_D-m_\pi)^2\simeq 3.0 $ GeV$^2$, the sum of hadronic intermediate states is dominated by the contributions associated with the radially excited vector resonances. 
We tacitly assume  that continuum hadronic states such as $\pi^+\pi^-$  play 
a secondary role, with their contributions  partially absorbed in the energy-dependent total widths
of resonance terms. This conjecture  allows us to  shift $s_{th}$ 
up to the vicinity of the lowest excited vector resonances.
According to \cite{PDG}, these are 
$\rho(1450)\equiv \rho'$   and   $\omega(1420)\equiv \omega'$. Furthermore, we notice a
strong suppression  of the $\omega(783)$ pole term 
with respect to the one of $\rho(770)$
in the dispersion relation (\ref{eq:dispA}).  The main reason for this suppression is 
the ratio of the normalization factors $k_\omega/k_\rho=1/3$, which is the same also for $\omega'$ and $\rho'$. Hence, we neglect
the $\omega'$ resonance contribution, assuming that the spectral density 
$\rho^{exc}(s)$ starts 
from the mass squared of $\rho'$, shifted downwards by the half of its total width.
Taking from \cite{PDG} the mass and width of $\rho'$ (see also Table~\ref{tab:Vexc} below),
we have :
\begin{equation}
s_{th}= (m_{\rho'}-\Gamma_{\rho'}/2)^2=
1.60\pm 0.15\, \mbox{GeV}^2\,.
\label{eq:sth}
\end{equation}
The parameter $s_{th}$ enters the 
$z$-transformation (\ref{eq:ztransf}). 
In addition, 
as explained in the previous section, we select a maximal 
value of the lepton-pair 
invariant mass squared  in the $D\to \pi\ell^+\ell^-$ decay, $s_{max}=1.2$ GeV$^2<s_{th}$
accessible when using the $z$-expansion (\ref{eq:zexp}). 
With this choice,
the region  
$4m_\ell^2< q^2<s_{max}$   ($\ell=e,\mu$) is mapped onto the interval
$0 \gtrsim z>-0.33$ of sufficiently
small $|z|$. On its turn, the spacelike region,
to be used for the fit to LCSR,
is chosen as 
\begin{equation}
-8.0~\mbox{GeV}^2<q^2< -3.0~\mbox{GeV}^2\,, 
\label{eq:q2region}
\end{equation}
and is  mapped onto $0.42>z>0.26$. This choice 
justifies a truncated $z$-expansion with three terms,
so that in our numerical analysis we take $K=2$ in 
(\ref{eq:zexp}). To complete the choice of parameters, we fix 
the subtraction point at $q_0^2=-2.0$ GeV$^2$.

\subsection{LCSR results for $D^+\to \pi^+\gamma^*$}
\begin{figure}[t]
\centering
\begin{subfigure}{0.49\textwidth}
\includegraphics[scale=0.5]{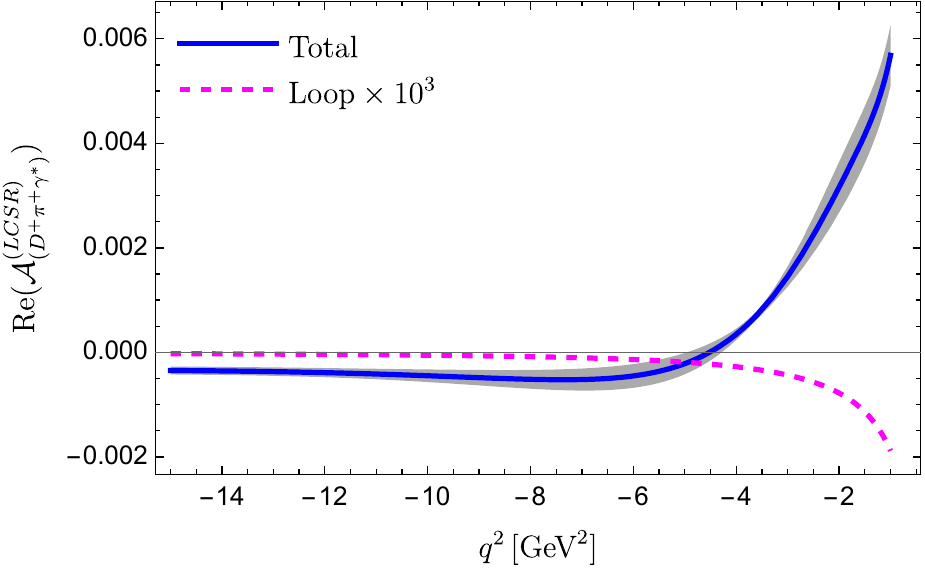}
\end{subfigure}
 \begin{subfigure}{0.49\textwidth}
 \includegraphics[scale=0.51]{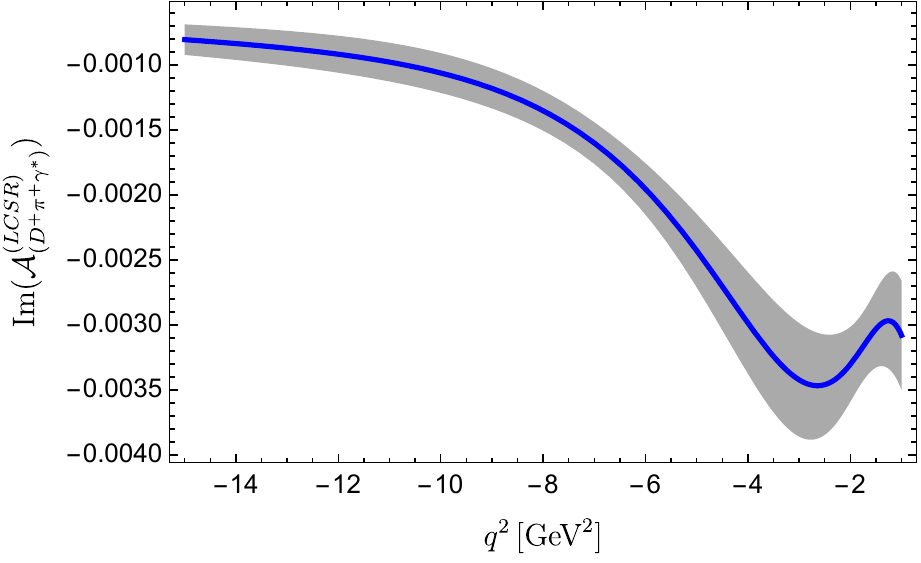}
 \end{subfigure}\\
  \begin{subfigure}{0.7\textwidth}
  \centering
 \includegraphics[scale=0.6]{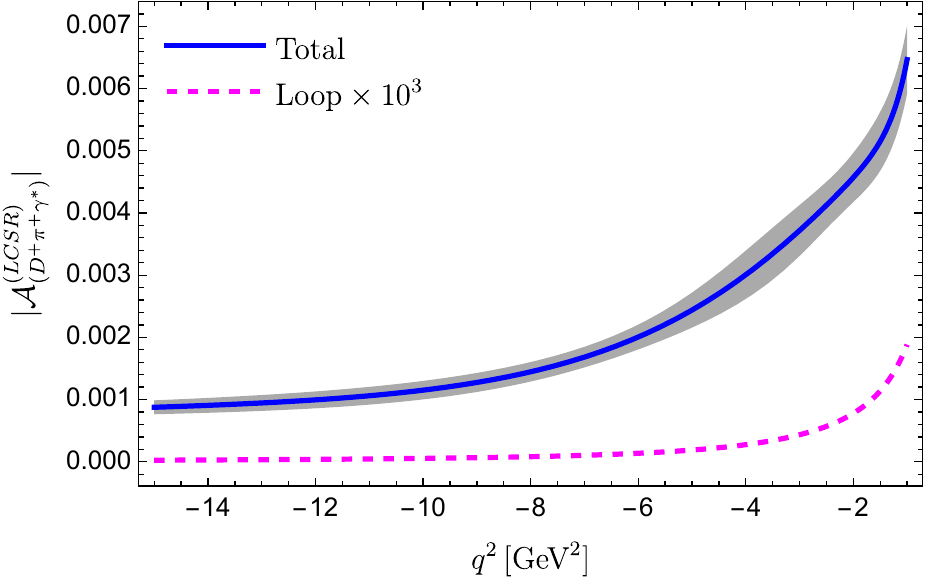}
 \end{subfigure}
 \caption{
  The $D^+\to \pi^+ \gamma^\ast$ 
 amplitude obtained from LCSR at $q^2<0$. The real part, imaginary part and the absolute value are shown with solid lines on the upper-left, upper-right and lower panels, respectively. Grey bands represent the parametric uncertainties, as explained in the text. 
 For comparison, the rescaled contribution of the loop topology diagrams is shown by the dotted line. A tiny difference between the total amplitude and annihilation topology contribution is not visible on these plots. }
\label{Fig:LCSRnumerics}
\end{figure}
Using the input parameters specified above, we calculate the $D^+\to \pi^+\gamma^*$ amplitude at spacelike $q^2$ from the LCSR  (\ref{eq:lcsr}). 
The numerical values for a broader
region $-15.0 ~\mbox{GeV}^2< q^2< -1.0 ~\mbox{GeV}^2$ are collected  in  ancillary data files submitted
with this paper. In Fig.~\ref{Fig:LCSRnumerics}, we plot  -- as a function of~$q^2$ -- the  real and imaginary parts of this amplitude, together with its absolute value. In the same figure, we plot the  sum of the $d$- and $s$-loop diagram contributions (rescaled for visibility). As can be seen from these plots, the overall effect of the loop topology at $q^2<0$ is tiny,
more than three orders of magnitude smaller than the contribution of annihilation topology. The reason for this strong suppression is twofold.
First, the smallness  of the combined
Wilson coefficient  for the loop topology 
(see (\ref{eq:C12comb})  with respect to $C_1$ for the annihilation topology, and second, the effective GIM cancellation between the $s$ and $d$ loop topologies.

The uncertainties of the LCSR results presented in the ancillary files and shown in  Fig.~\ref{Fig:LCSRnumerics} are calculated using the standard error-propagation formula, taking into account   
the uncertainties of  the quark masses $m_c$ and $m_s$,
the decay constants $f_\pi$, $f_D$ and  the Gegenbauer moments $a_2^\pi$, $a_4^\pi$ (see Table~\ref{tab:inpPDG}), as well as the variations within adopted intervals of
the Borel mass and duality threshold in LCSR (see Table ~\ref{tab:inpLCSR}). 
Note that  all these individual uncertainties are treated as uncorrelated.
\begin{figure}
\centering
\includegraphics[scale=0.6]{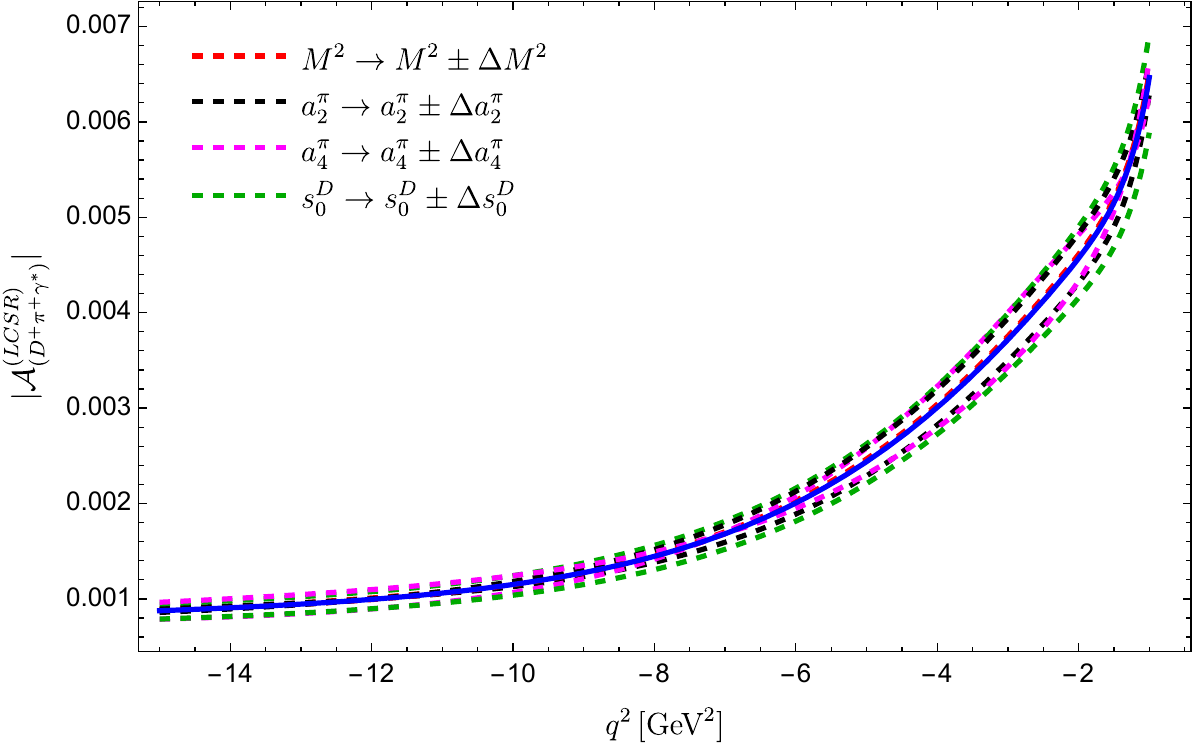}
     \caption{
      The absolute value of the $D^+\to \pi^+ \gamma^\ast$ amplitude calculated from LCSR at the central values of  input parameters (solid line). Dashed lines show the impact of the numerically most significant uncertainties induced by a separate variation of the LCSR parameters within their adopted intervals.
       }
\label{fig:LCSRuncert}
\end{figure}
For completeness, in Fig.~\ref{fig:LCSRuncert} we separately plot the most significant uncertainties that are due to the variations of the parameters $M^2,s_0^D$, $a_2^\pi$ and $a_4^\pi$. In particular, the relatively strong dependence of LCSR on $a_4^\pi$ indicates the sensitivity of 
our sum rule to detailed shape of the pion twist-2 DA.

\subsection{Fit of the Dispersion Relation to the Light-Cone  Sum Rule}
\label{subsect:fit}
The numerical values of the 
$D\to\pi\gamma^*$ amplitude at $q^2<0$, evaluated from LCSR are then fitted 
to the once-subtracted dispersion relation (\ref{eq:dispZ}), where a $z$-polynomial replaces the integral over the spectral density 
$\rho_{exc}$. To this end, we numerically minimized the following expression: 
\begin{eqnarray}
\sum_{q^2<0}\bigg|{\cal A}_{(D^+ \pi^+\gamma^*)}^{(disp-z)}(q^2)
-{\cal A}_{(D^+ \pi^+\gamma^*)}^{(LCSR)}(q^2)\bigg|^2
=min\,,
\label{eq:fitZ}
\end{eqnarray}
as a function of the three phases $\varphi_V$ ($V=\rho,\omega,\phi$) and the two complex-valued coefficients of the $z$-expansion, $\mbox{Re}\,(\alpha_k) +i \mbox{Im}\,(\alpha_k)$, $(k=1,2)$. 
The sum in (\ref{eq:fitZ}) goes over
a set of discrete points for negative $q^2$ where the LCSR calculation is valid. In our numerical analysis, we chose 11 values of $q^2$ taken with the interval $0.5~ \mbox{GeV}^2$ within  the region $-8.0~\mbox{GeV}^2< q^2 <-3.0 ~\mbox{GeV}^2$. 
In this region, simultaneously, the LCSR is valid and $z$-values are still reasonably small, to justify a truncated $z$-expansion. Furthermore, we use  $q_0^2 =-2.0$ GeV$^2$ as the subtraction point.

The minimization procedure was performed employing the \textbf{NMinimize} routine of Mathematica \cite{Mathematica}, with  
Method $\rightarrow$ \{"RandomSearch", "SearchPoints" $\rightarrow$ 2000, 
"RandomSeed" $\rightarrow$ 2\}.
At the adopted central values of all input parameters, the 
minimal value of the sum (\ref{eq:fitZ}) is achieved at \\
\begin{eqnarray}
&&\varphi_\rho= 1.809
, ~~\varphi_\omega=1.804,~~ \varphi_\phi= 
-1.339 , ~~~\mbox{(in radians)}\;,
\nonumber\\      
&&
\alpha_{1}=
(-7.858 + 0.727\,i)\times 10^{-2}
,~~ 
\alpha_{2}=(9.548 +2.184\,i)\times 10^{-2}\,.
\label{eq:zfitparam}
\end{eqnarray}
At this stage of our numerical analysis, we refrain from 
attributing uncertainties
to the above phases and $z$-expansion coefficients. The reason is that  terms at different $q^2$ in the sum (\ref{eq:fitZ}) are strongly correlated and vary concertedly if  a certain input parameter in the LCSR or in the
hadronic dispersion relation is varied within an adopted interval.
Instead of a full uncertainty analysis~\footnote{This would demand a dedicated analysis involving 
Bayesian inference, which is beyond the scope of this paper and can be performed in the future, employing our LCSR results.}, below, in subsection~\ref{subsect:bin} we investigate the propagation of the input uncertainties onto the decay 
branching fractions integrated over certain intervals (bins) in $q^2$.

In Fig.~\ref{fig:amplfit}, the LCSR result and the 
dispersion-relation model with z-expansion for the $D\to\pi^+\gamma^*$ amplitude are compared. 
We also show the eleven points on which the fit is based, including the parametric uncertainties from the sum rule as error bars.
In the whole fit region, the deviation between them does not exceed $10 \%$,
which is a reasonable agreement given the precision of the LCSR calculation.

The phases obtained in (\ref{eq:zfitparam}) from the minimization procedure 
deserve a comment. The almost equal phases of the $\rho$- and $\omega$-pole terms in the dispersion relation are generally expected, since these vector mesons have close masses and the intermediate $D^+\to \rho^0\pi^+$ and $D^+\to \omega\pi^0$ amplitudes entering the residues of these terms are described by the same quark topologies. More important is that the phase
difference between the $\rho$ and $\phi$ terms is found to be close to $\pi$. This enforces a constructive interference of the two main vector-meson terms in the dispersion relation, almost compensating for the relative minus sign between their normalization coefficients $k_\rho$ and $k_\phi$. 
Varying separate input parameters around central values reveals
that, while individual phases are changing, the pattern of phase differences remains almost intact. This important feature is a manifestation of the constraints put on the hadronic dispersion relation 
by LCSR.

With  the central values of the fit parameters presented in 
 (\ref{eq:zfitparam}), we 
can now use the dispersion relation (\ref{eq:dispZ}) 
for the $D^+\to\pi^+\gamma^*$ amplitude in the lower part
\begin{equation}
4m_\ell^2\leq q^2\leq s_{max}=1.2~\mbox{GeV}^2 \,,
\label{eq:rsmax}
\end{equation}
of the physical region for $D^+\to\pi^+\ell^+\ell^-$. The upper limit in the above interval is chosen, as already
explained, to guarantee a reasonable approximation by the truncated $z$-expansion. 
Substituting this amplitude in (\ref{eq:dBR}), we finally 
obtain our prediction for the differential width of the $D^+\to\pi^+\ell^+\ell^-$ decay in the region
(\ref{eq:rsmax}). It is plotted (for the muon  mode) in Fig. \ref{fig:BRfit}. 

\subsection{Extended resonance model}
\label{subsect:reson}

As an alternative to the once-subtracted dispersion relation (\ref{eq:dispZ}),
where the integral over the hadronic spectral density $\rho_{exc}$ was replaced
by a general $z$-expansion, we consider here an explicit model of  this density.
This model will be used for the unsubtracted dispersion relation 
(\ref{eq:dispAuns}) which is equally valid, due to the asymptotic behaviour of the $D^+\to \pi^+\gamma^\ast$ amplitude revealed from LCSR. With this model, we will be able to reach the whole physical region 
of the $D^+\to\pi^+\ell^+\ell^-$ decay.

Since we deal here with the channel of e.m. quark current, it is conceivable to adopt a model similar to the ones 
used for the timelike pion or kaon e.m. form factors in the regions above  
$\rho,\omega$ or $\phi$. Following this analogy, we assume that the spectral density $\rho_{exc}(s)$ is saturated by  the contributions of radially excited vector mesons. According to \cite{PDG}, in the region $s_{max}\leq s\leq (m_D-m_\pi)^2$, that is, above $\phi$ and below the kinematical threshold of the $D^+\to\pi^+ \ell^+\ell^-$ decay, the established vector resonances are
$\rho'\equiv \rho(1450)$, $\omega'\equiv \omega(1420)$, 
$\omega''\equiv \omega(1650)$, $\phi'\equiv\phi(1680)$
and $\rho''\equiv \rho(1700)$,and the latter resonance is located close to this threshold. Furthermore, 
guided  by the smallness of $\omega$ vs $\rho$
residues in the dispersion relation, 
we neglect the $\omega'$ and $\omega''$ contributions, anticipating that they 
are  much smaller than, respectively,  the $\rho'$ and $\rho''$ ones.
This assumption is also supported by the fact that
the decays $D^+\to \omega{'}\pi^+,\omega^{''}\pi^+$ have not been seen 
at the same level as the observed $D^+\to \rho'\pi^+,\rho''\pi^+$ decays. 

To summarize, we include in the spectral density $\rho_{exc}(s)$ the two resonances $\rho'$ and $\phi'$  in the conventional Breit-Wigner form 
with, as usual,  energy-dependent widths. The rest of this density,
as explained below, is replaced by an effective pole with the mass and width 
of the $\rho''$ resonance: $m_{\rho''}=1720\pm 20$ MeV, $\Gamma^{tot}_{\rho''}=250\pm 100$ MeV. The dispersion relation 
(\ref{eq:dispAuns}) is reduced 
to  the following sum over resonance contributions:
\begin{eqnarray}
&&{\cal A}_{(D^+ \pi^+\gamma^*)}^{(disp-res)}(q^2)
=
\sum_{V=\rho,\omega, \phi } \frac{r_Ve^{i \varphi_V} }{m_V^2-q^2-
i\sqrt{q^2}\Gamma_V(q^2)}
\nonumber\\
&&+\sum_{V'=\rho', \phi' } \frac{r_{V'} e^{i \varphi_{V'}} }{m_{V'}^2-q^2-
i\sqrt{q^2}\Gamma_{V'}(q^2)}
+\frac{r_{eff}}{m_{\rho''}^2-q^2 -
i\sqrt{q^2}\Gamma_{\rho''}(q^2)
}
\,,
\label{eq:dispAexc}
\end{eqnarray}
where the residues $r_{V}$  are already defined in (\ref{eq:rVdef}) and their
numerical values are given in Table~\ref{tab:Vpar}. The residues $r_{\rho'}$ and $r_{\phi'}$ given in Table~\ref{tab:Vexc} are obtained differently, invoking additional data. Their determination is described in detail in Appendix~\ref{app:Vexc}.
Note that the extended resonance model (\ref{eq:dispAexc}) 
is also consistent with the $\sim 1/q^2$ asymptotics of the
$D^+\to \pi^+\gamma^*$ amplitude.
\begin{table}[b]
\centering
\begin{tabular}{|c|c|c|}
\hline
&&\\[-3mm]
 Resonance $V'$ & $\rho'\equiv\rho(1450)$ &$\phi'\equiv \phi(1680)$  \\[2mm]
\hline
$m_{V'}$ (MeV) & $1465\pm 25$ &  $1680\pm 20$\\
$\Gamma^{tot}_{V'}$ (MeV)& $400 \pm 60$ & $150 \pm 50$ \\
\hline
&&\\[-3mm]
$BR(D^+\to \pi^+V')_{V'\to \pi^+\pi^-}$ 
& $(1.8\pm 0.5)\times 10^{-4}$ & --
\\
$BR(D^+\to \pi^+V')_{V'\to K^+K^-}$&-- &$(4.9^{+4.0}_{-1.9})\times 10^{-5}$\\
\hline
&&\\[-3mm]
$|r_{V'}|=|\kappa_V f_{V'}A_{D^+ \pi^+ V' }|$ (GeV$^2$) &
$(9.64 ^{+2.92}_{-4.76})\times 10^{-3}$ & $(11.87 ^{+4.74}_{-3.79} )\times 10^{-3}$\\[2mm]
\hline
&&\\[-3mm]
$|f_{V'}|$ (MeV) & 
$140^{+15}_{-35}$  &
$|f_{\phi'}|=|f_{\rho'}|$ \\[2mm]
\hline
\end{tabular}
\caption{Parameters of the $\rho'$ and $\phi'$ resonances 
in the dispersion relation. The residues are estimated  in Appendix~\ref{app:Vexc}.}
\label{tab:Vexc}
\end{table}

The last term in (\ref{eq:dispAexc}) can be understood in the following way.
In the spectral density $\rho^{exc}(s)$ integrated above $\phi'$, we separate the contribution of the next 
resonance  $\rho''$, and transform the integral as follows:
\begin{eqnarray}
\int\limits_{\{> m_{\phi'}^2\}}^{\infty}\!\!\!\! ds\,\frac{\rho_{exc}(s)}{s-q^2} =
\frac{r_{\rho"}}{m_{\rho''}^2-q^2}+ \int\limits_{\{> m_{\rho''}^2\}}^{\infty}\!\!\!\!ds\,\frac{\rho_{exc}(s)}{s-q^2}
\nonumber\\
=
\frac{1}{m_{\rho''}^2-q^2}\Bigg[r_{\rho''}+ (m_{\rho"}^2-q^2)\int\limits_{\{> m_{\rho''}^2\}}^{\infty}\!\!\!\!ds\,\frac{\rho_{exc}(s)}{s-q^2}\Bigg]\,.
\label{eq:effres}
\end{eqnarray}
The expression in brackets on r.h.s., being without singularities at 
$q^2\lesssim m_{\rho''}^2 $, is approximated  by its constant ($q^2$-independent) part
denoted as  $r_{eff}$.
\begin{figure}[t]
\centering
\includegraphics[scale=0.7]{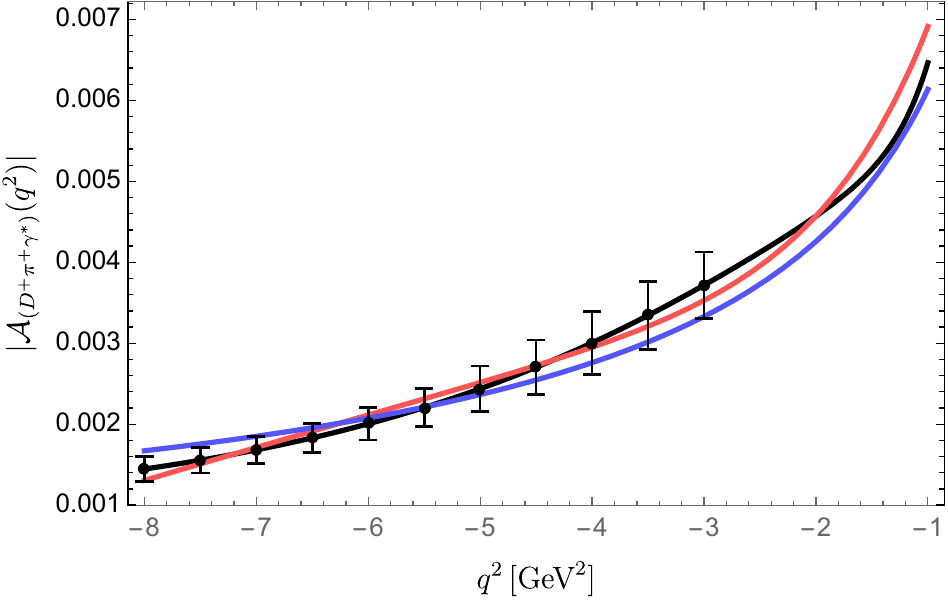}
\caption{
The red (blue) line is the absolute value of the  amplitude 
${\cal A}_{(D^+ \pi^+\gamma^*)}^{(disp-z)}(q^2)$
$\big({\cal A}_{(D^+ \pi^+\gamma^*)}^{(disp-res)}(q^2)\big)$  obtained from the fit to the LCSR result
(black line and points) at the central input. The parametric uncertainties of the LCSR amplitude are shown with error bars.}
\label{fig:amplfit}
\end{figure}

We fitted the resonance model (\ref{eq:dispAexc})
to LCSR, minimizing the following sum
\begin{eqnarray}
\sum_{q^2<0}\bigg|{\cal A}_{(D^+ \pi^+\gamma^*)}^{(disp-res)}(q^2)
-{\cal A}_{(D^+ \pi^+\gamma^*)}^{(LCSR)}(q^2)\bigg|^2
=min\,,
\label{eq:fitres}
\end{eqnarray}
where the range of $q^2$ and the procedure of minimization are the same as for 
the sum in (\ref{eq:fitZ}). The fit returns the unknown 
phases of the $\rho$, $\omega$, $\phi$, $\rho'$ and $\phi'$ resonance terms, as well as the 
modulus and phase of the effective pole residue $r_{eff}$.  
For the central input, we obtain:
\begin{eqnarray}
 &&  \varphi_\rho= 5.944 \,,~~\varphi_\omega= 5.944 \,, ~~\varphi_\phi=2.797 \,,\nonumber\\
 &&\varphi_{\rho'}= 5.929 \,,~~\varphi_{\phi'}= 5.925\,,
\nonumber\\
&&|r_{\rm{eff}}|= 3.538 \times 10^{-2}\,, ~~
Arg[r_{\rm{eff}}]= 3.312\,. 
\label{eq:RRfit}
\end{eqnarray}

In Fig.~\ref{fig:amplfit}, we plot the central value of the $D^+\to \pi^+\gamma^*$ amplitude in the
extended resonance model in the 
fit region of negative $q^2$, comparing it with the LCSR calculation and with the fit obtained for the truncated $z$-expansion.
 We note that in the fit region, the 
deviations between the curves are less than 15\%. In fact,  the fit results for 
the extended resonance model are consistent with the ones for  
the model with $z$ parametrization. Most importantly, the relative phase differences between $\rho$, $\omega$ and $\phi$ come out the same: 
$\phi_\rho - \phi_\omega \sim 0$ and $\phi_\rho - \phi_\Phi \sim \pi$, while different overall phases, 
being unphysical, are due to the different modelling of the hadronic spectral function.

\begin{figure}[h]
\centering
\includegraphics[scale=0.8]{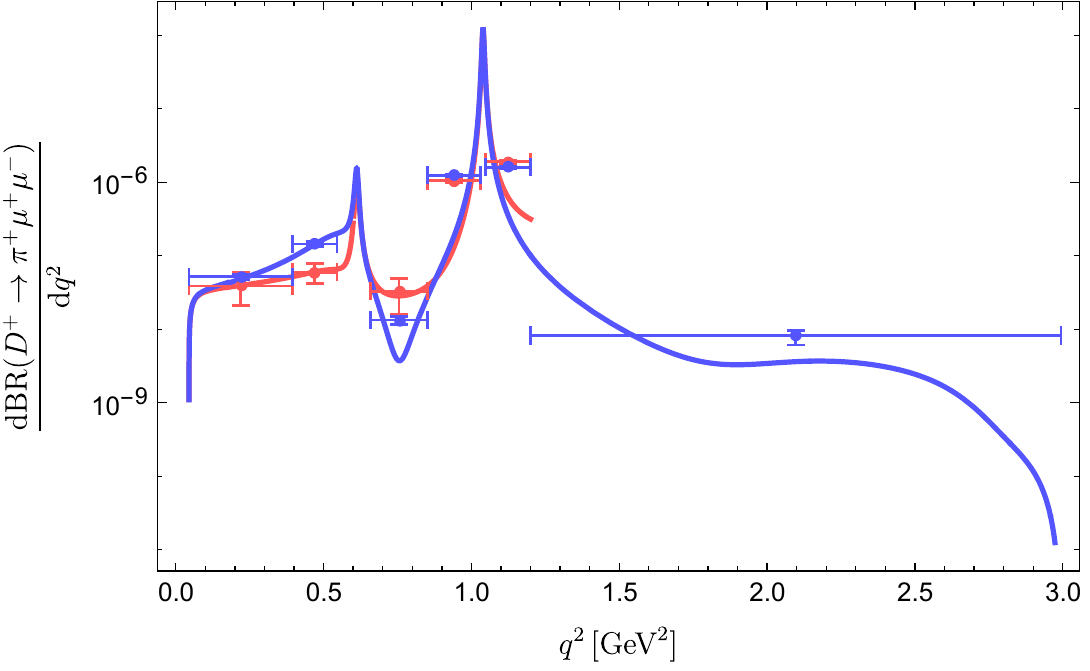}
\caption{
Differential branching fraction of 
$D^+\to\pi^+\mu^+\mu^-$. The orange (blue) line is obtained 
using the $D^+\to\pi^+\gamma^*$ amplitude  given by the dispersion relation  with $z$-expansion
(with extended resonance model) and fitted to LCSR.  
The bins defined in (\ref{eq:bin})  
are shown with the points, horizontal and vertical bars corresponding to, respectively, their centers, sizes 
and uncertainties. }  
\label{fig:BRfit}
\end{figure}

We use the fitted parameters (\ref{eq:RRfit}) to extrapolate 
the extended resonance model into the whole physical region of positive $q^2$, up to $q^2= (m_D - m_\pi)^2$. In Fig~\ref{fig:BRfit}, we show the resulting differential rate for the muonic mode $D^+ \to \pi^+ \mu^+ \mu^-$. 

For the extended resonance model, staying  within our fitting procedure,
it is also difficult to  estimate the uncertainties of the fit parameters in (\ref{eq:RRfit}), 
since the points chosen for the fit are highly correlated. Instead, we will perform an error analysis for $q^2$-bins, discussed in the next subsection. 

\subsection{Binned branching fraction of  $D^+\to \pi^+ \ell^+\ell^-$ }
\label{subsect:bin}

Experimental searches for $D^+\to \pi^+ \ell^+\ell^-$ and similar 
decays are 
usually performed by vetoing the regions of the lepton-pair invariant 
mass near $\rho,\omega,\phi$-resonances. For example, the
latest upper bound obtained by the LHCb collaboration in \cite{LHCb:2020car} 
and quoted in Table~\ref{tab:bins}.
covers a sum of two $q^2$-regions, one below $\rho$ and one above $\phi$.
It is therefore more informative to cast our results on the differential branching 
fraction of this decay into a set of binned branching fractions. 
A bin is defined as
\begin{equation}
\Delta {\cal B}_{(D^+ \to \pi^+ \mu^+ \mu^-)}
(q^2_{ min},q^2_{ max})\equiv
\frac{1}{(q_{max}^2-q_{min}^2)}\int\limits_{q_{min}^2}^{q_{max}^2} d s \,\frac{d \mbox{BR} (D^+ \to \pi^+ \mu^+ \mu^-)}{d s}\, , 
\label{eq:bin}
\end{equation}
so that $\overline{q^2} =(q^2_{max} + q^2_{min})/2$ corresponds to the center of the bin. 

As shown in Table \ref{tab:bins}, we divide the full $q^2$ region for this decay into six bins,
excluding only the two intervals around $\rho,\omega$ and  $\phi$.
For definiteness, we use the total widths of $\rho$ and $\phi$ mesons as a measure 
of  the distance between the bin limits and  resonance positions. We emphasize that, contrary to naive expectation, 
isolating the branching fraction bins from the dominant
$\rho,\omega$ and $\phi$ resonances, we do not 
exclude their contributions.  
The  resonance terms, being a part of the hadronic dispersion relation, contribute to the decay amplitude in the whole 
physical region including the $q^2$-regions chosen
for the bins. The same is true for the excited 
state terms  of the dispersion relations, these terms
also contribute to the regions around $\rho$, $\omega$, and $\phi$ resonances.

In Table~\ref{tab:bins}, we present the  decay branching fractions integrated over the selected bins for both versions of our $D^+\to \pi^+ \gamma^*$  amplitude, the $z$-expanded dispersion relation  and the extended resonance model.  
For completeness and comparison, we separately present the same quantities 
for the two bins  selected in the LHCb experiment \cite{LHCb:2020car}.
In Fig.~\ref{fig:BRfit},  the values of the $q^2$-bins are plotted.

The use of  bins instead of the point-by-point differential branching fraction has another advantage in our approach.
To estimate the total error of a bin, 
caused by the uncertainties of input parameters, one does not 
need to perform a full statistical analysis with correlations 
between  different $q^2$ points.
Instead, we use the following procedure. Inserting central values of all input parameters into minimization equation (\ref{eq:fitZ}), we already obtained the 
corresponding values of fit parameters given in (\ref{eq:zfitparam}). First, we use these
parameters to obtain the central value of a definite $q^2$-bin. 
They are all presented in Table \ref{tab:bins}.
After that,  taking one by one the N input parameters  at their maximal and minimal values, we perform 2N new fits to LCSR. For each input parameter, we obtain two individual sets of  the fit parameters and calculate two corresponding values for the same bins, yielding an individual uncertainty caused by the chosen input parameter. Finally, 
collecting all individual uncertainties calculated in this way, we add them in quadrature. This brings us to the 
uncertainties of $q^2$-bins plotted in 
Fig.~\ref{fig:BRfit} and quoted in Table~\ref{tab:bins}. 

Importantly, the input parameters
used in LCSR and dispersion relation are not correlated with each other, simply because they
originate either from independent lattice-QCD or sum-rule analyses 
or from independent measurements. In more detail, the error budget for the bins includes the uncertainties due of the LCSR parameters $a_2^\pi$, $a_4^\pi$, $M^2$, and $s_0^D$, 
while the intervals of remaining  parameters listed in Tables \ref{tab:inpPDG} and \ref{tab:inpLCSR} have a negligible impact on uncertainties in the LCSR results at $q^2<0$. We also include uncertainties from 
the residue factors $r_\rho$, $r_\omega$, and $r_\phi$. Their intervals are mainly determined by the errors of the measured $D \to \pi V$  widths quoted  in Table~\ref{tab:Vpar}. In addition, in the $z$-expanded dispersion relation we varied: the threshold $s_{th}$ within the interval quoted in (\ref{eq:sth})
and the subtraction point $q_0^2$  between $-1.5~\mbox{GeV}^2$ and $-2.5~\mbox{GeV}^2$. 
On its turn, in the extended resonance model, we varied 
the residue factors $r_{\rho'}$, $r_{\phi'}$ (see Table~\ref{tab:Vexc})  and the 
width $\Gamma_{\rho''}$. 
Note that, conservatively, we neglect possible correlations between separate bin uncertainties. 
\begin{table}[t]
\centering
\begin{tabular}{|c|c|c|c|c|}
\hline
&&&\\[-4mm]
Bin & $q^2_{\rm min}$ & $q^2_{\rm max}$ &
$(q^2_{\rm max}-q^2_{\rm min})\,\Delta{\cal B}^{(disp-z)}$
&$(q^2_{\rm max}-q^2_{\rm min})\,\Delta{\cal B}^{(disp-res)}$\\[1mm]
\hline
&&&&\\[-4mm]
I& $4m_\mu^2$ & $(m_\rho-\Gamma_\rho)^2$  & $1.36^{+0.70}_{-0.63}$
 &$1.81^{+0.14}_{-0.17} $\\[1mm]
\hline
&&&&\\[-4mm]
II& 
$(m_\rho-\Gamma_\rho)^2$ & $(m_\rho-\Gamma_\rho/4)^2$  & $0.90^{+0.29}_{-0.27}$
 &$2.19^{+0.17}_{-0.18} $\\[1mm]
\hline
&&&&\\[-4mm]
III& 
$(m_\rho+\Gamma_\rho/4)^2$ & $(m_\rho+\Gamma_\rho)^2$  & $0.63 ^{+0.31}_{-0.32}$
 &$0.26 ^{+0.03}_{-0.03} $ \\[1mm]
\hline
&&&&\\[-4mm]
IV& $(m_\rho+\Gamma_\rho)^2$ & $(m_\phi-\Gamma_\phi)^2 $
   & $18.88 ^{+1.39}_{-1.40}$ &$22.27 ^{+1.08}_{-1.09}$\\[1mm]
\hline 
&&&&\\[-4mm]
V &$(m_\phi+\Gamma_\phi)^2$  & $1.2~\mbox{GeV}^2$
  & $28.93 ^{+1.70}_{-2.16}$ & $24.27 ^{+1.12}_{-1.13} $ \\[1mm]
\hline
&&&&\\[-4mm]
VI & $1.2~\mbox{GeV}^2$  & $(m_{D^+}-m_{\pi^+})^2$
 & -- &$1.48 ^{+0.88}_{-0.39}$\\[1mm]
\hline
\hline
&&&&\\[-4mm]
LHCb & $4m_\mu^2$ & $(0.525~\mbox{GeV})^2$  & $0.81 ^{+0.43}_{-0.38} $ &$0.93 ^{+0.08}_{-0.10}$ \\
regions & $(1.250~\mbox{GeV})^2 $ & $(m_{D^+}-m_{\pi^+})^2$ & 
  --& $0.40 ^{+0.29}_{-0.25} $ \\[1mm]
\hline
\end{tabular}
\caption{ Branching fraction of $D^+\to\pi^+\mu^+\mu^-$ decay integrated over bin sizes according to (\ref{eq:bin}) (in the units $10^{-8}$). }
\label{tab:bins}
\end{table}

The results for the integrated 
branching fraction presented in Table~\ref{tab:bins} 
can be directly compared  with the future 
measurements of this observable. 
Note that our choice of individual bin sizes can be easily modified and adjusted to 
experimental cuts, as it is done for the LHCb  regions.
Importantly,  the predicted total branching fraction over the two LHCb regions
amounts to $\left(1.33^{+0.17}_{-0.24}\right)\times 10^{-8}$, which is of the same order of magnitude 
as the current upper bound 
$6.7 \times 10^{-8}$ quoted in 
Table~\ref{tab:modes}. Hence, we expect that in the future more accurate measurements of $D^+\to \pi^+\ell^+\ell^-$ will be able to probe our predictions.
We also remind that both Bin I and the low LHCb region in Table~\ref{tab:bins} exclude 
intermediate  $\eta$ and $\eta'$ mesons
in $D\to \pi\ell^+\ell^-$, their 
effects being of higher order in $\alpha_{em}$.
Since both mesons are very narrow, their contributions to the measured $D^+\to\pi^+\ell^+\ell^-$ partial width can be simply subtracted, estimating  each of them with
the  product of measurable branching fractions, $BR(D^+\to \pi^+ \eta^{(')})\times 
BR(\eta^{(')}\to \ell^+\ell^-)$.

\subsection{Contribution of the operator $O_9$}
\label{subsect:O9}

Up to here, our numerical calculation was performed  in the GIM limit, at $\lambda_b=0$. Switching on $\lambda_b\neq 0$,   
adds to the $D^+\to \pi^+\ell^+\ell^-$ amplitude, among other very small effects,  the contribution (\ref{eq:ampl3}) of the $O_9$ operator.
Here, we estimate this contribution
in the form of an effective addition to the 
$D^+\to \pi^+ \gamma^*$ amplitude,
as given in (\ref{eq:O9add}). The only hadronic input needed for this estimate is the vector form factor of $D^+\to\pi^+$ transition, which in the isospin symmetry limit, is equal to the very accurately calculated and measured form factor 
$f^+_{D\pi}(q^2)$  of the semileptonic 
$D^0\to \pi^-\ell^+\nu_\ell$ decay. For our estimate, we use the currently most advanced lattice QCD results for this form factor presented  in \cite{FermilabLattice:2022gku} in a form of $z$-expansion
(see eq.(5.24) and Table XIII there). We checked that the LCSR calculation of the $f^+_{D\pi}(q^2)$ obtained much earlier in \cite{Khodjamirian:2009ys} agrees with this input within uncertainties. 
Inserting the $D\to \pi$ form factor values and the 
SM value for $C_9 (\mu = 1.5 \, {\rm GeV})$ in (\ref{eq:O9add}), we find that an 
effective addition to the amplitude ${\cal A}^{(D^+ \pi^+ \gamma^*)}$ 
 due to  $O_9$ lies in the interval from $1.3\times 10^{-6}$ to $6.3\times 10^{-6}$,
at  $0<q^2<(m_D-m_\pi)^2$, that is, at least three orders 
of magnitude smaller than the main part of the $D^+\to \pi^+\gamma^*$ amplitude induced by $O_1$ and $O_2$. 
We note that current global fits for the FCNC $B$~decays indicate an anomaly in $C_9$, implying an increase of this Wilson coefficient by a factor up to four. If one assumes that this anomaly is due to BSM physics, which has a similar size also in the charm sector, that will still be far from sufficient to overcome the enormous long-distance dominance in 
$D^+\to \pi^+\ell^+\ell^-$.

\section{$D^+_s\to\pi^+\ell^+\ell^-$ and $D^0\to \bar{K}^0\ell^+\ell^-$ decays}
As realized above, the short-distance $c \to u \ell^+ \ell^-$ transition  has a negligible contribution 
to the decay $D^+ \to \pi^+ \ell^+ \ell^-$ in the SM.
Moreover, the diagrams corresponding to the $d$- and $s$-loop topologies suffer from a strong cancellation, so that the SM amplitude for this SCS decay is dominated by the diagrams with the weak annihilation 
and virtual photon emission.
Based on this observation, it is interesting to study the CF decay modes $D_s^+ \to \pi^+ \ell^+\ell^-$ and 
$D^0\to \bar{K}^0\ell^+\ell^-$, which are determined, respectively, by the $D_s^+ \to \pi^+ \gamma^*$ and 
$D^0\to \bar{K}^0\gamma^*$ nonlocal amplitudes.
As already mentioned, the latter amplitudes are totally described by 
the annihilation topology diagrams, similar to the ones for $D^+\to\pi^+\gamma^*$. 
To this end, the CF modes $D_s^+ \to \pi^+ \ell^+\ell^-$ and $D^0\to \bar{K}^0\ell^+\ell^-$ 
can serve as control channels constraining the SCS decays of interest. 

The method developed in this paper can be straightforwardly applied to the CF modes. Note that the $D^+_s\to \pi^+\gamma^*$  amplitude is especially interesting, 
because in the $U$-spin symmetry limit it is  
equal to the $D^+ \to \pi^+ \gamma^*$ amplitude. We will assess the effect of this symmetry violation at a quantitative level, taking into account 
the mass of $s$-quark in the LCSR correlation function. The $D^0\to \bar{K}^0\gamma^*$  amplitude is, on the contrary,  not related to $D^+\to \pi^+\gamma^*$, being determined by the second independent $U$-spin invariant amplitude  in (\ref{eq:HME3}). 

The LCSRs for ${\cal A}_{(D_s^+\pi^+\gamma^*)}(q^2)$ and ${\cal A}_{(D^0 \bar{K}^0\gamma^*)}(q^2)$  
at spacelike $q^2$ are obtained, respectively,  from the correlation functions 
(\ref{eq:corrDs}) and (\ref{eq:corrD0}). 
The final expressions for these sum rules are very similar to 
(\ref{eq:lcsr}) and we will not present them here. Note only that for the $D_s$ mode, the following replacements should be done:
$m_D\to m_{D_s}$, $f_D\to f_{D_s}$, $s_0^D\to s_0^{D_s}$, whereas for the $D^0$ meson we use the same parameters as for $D^+$ in the adopted isospin symmetry limit.
All these input parameters are given in Tables~\ref{tab:inpPDG} and
\ref{tab:inpLCSR}. The respective OPE spectral densities, 
$\mbox{Im} F^{(OPE)}_{(D_s \pi^+\gamma^*)}$ and 
$\mbox{Im} F^{(OPE)}_{(D^0 \bar{K}^0\gamma^*)}$ 
are computed from the annihilation  topology diagrams, 
obtained from the ones in Fig.~\ref{fig:lcsr_ann},
replacing  the light-quark flavours accordingly. 
Details of this computation are given in Appendix~\ref{app:CF}.
\begin{table}[h]
\centering
{\small
\begin{tabular}{|c|c|c|c|}
\hline
&&&\\[-3mm]
 Vector meson $V$ & $\rho^0(770)$ &$\omega(782)$&$\phi(1020)$  \\[2mm]
\hline
&&&\\[-3mm]
$BR(D^+_s\to \pi^+V)$&$(1.12\pm 0.17)\times 10^{-4}$ &
$(1.92\pm 0.30)\times 10^{-3}$ & $(4.5\pm 0.4)\%$\\[2mm]
$|A_{D_s \pi^+ V }|$ (MeV)&
2.59 $\pm$ 0.20 & 10.79 $\pm$ 0.85 & 64.86 $\pm$ 2.91\\[2mm]
\hline
&&&\\[-3mm]
$r_V^{(D_s \pi^+)}=\kappa_V f_V|A_{D_s \pi^+ V }|$  &
$0.403 \pm 0.031$&
$0.512 \pm 0.040 $&
$-(4.932 \pm 0.222)$
\\[2mm]
(in $10^{-3}~\mbox{GeV}^2 $) &&&\\
\hline
&&&\\[-3mm]
$BR(D^+_s\to \pi^+V)_{V\to \mu^+\mu^-}$ & $(5.096  \pm 0.835)\times 10^{-9}$  & 
$(1.421 \pm 0.411\times 10^{-7}$ & 
$(1.282 \pm 0.142 )\times 10^{-5}$\\[2mm]
\hline
\hline
&&&\\[-3mm]
$BR(D^0\to \bar{K}^0 V)$&$1.26^{+0.12}_{-0.16} ~ \%$&
$2.32\pm 0.08~ \%$ & $0.83\pm 0.05 ~\%$\\[2mm]
$|A_{D^0 \bar{K}^0 V}|$ (MeV)&
$41.13 \pm 2.62$  & $56.29\pm 1.03 $  &$49.10\pm 1.40$\\[2mm]
\hline
&&&\\[-3mm]
$r_V^{(D^0 \bar{K}^0)}=\kappa_V f_V|A_{D^0 \bar{K}^0 V }|$&
 $6.39 \pm  0.41$ & 
 $2.67\pm 0.05$ &
$-(3.73 \pm 0.11)$\\[2mm]
(in $10^{-3}~\mbox{GeV}^2 $) &&&\\
\hline
&&&\\[-3mm]
$BR(D^0 \to \bar{K}^0 V )_{V\to \mu^+\mu^-}$ &  $(5.73 \pm 0.81 )\times 10^{-7}$ & 
$1.72\pm 0.42  )\times 10^{-6}$ & $(2.36 \pm 0.20)\times 10^{-6}$\\[2mm]
\hline
\end{tabular}
}
\caption{ 
The  data on $D_{s}^+\to \pi^+ V$ and $ D^0\to \bar{K}^0 V$ decays ($V=\rho^0,\omega,\phi$) from \cite{PDG}
 and the corresponding residues used 
 in the resonance part of the hadronic model analogous to (\ref{eq:dispZ}). }
 \label{tab:CFVpar}
 \end{table}

\begin{figure}
\centering
\begin{subfigure}{0.49\textwidth}
  \includegraphics[scale=0.5]{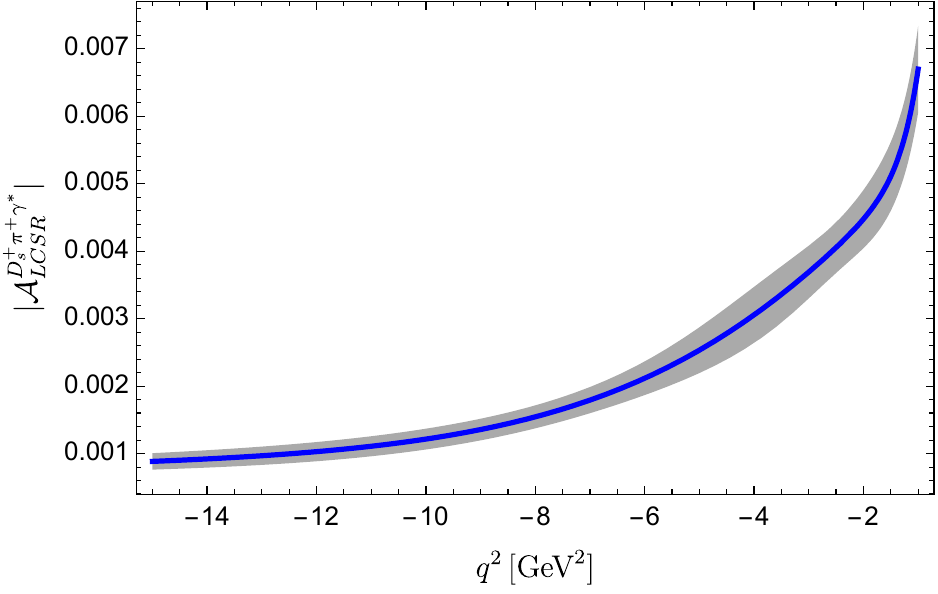} 
  \caption{}
\end{subfigure}
\begin{subfigure}{0.49\textwidth}
  \includegraphics[scale=0.535]{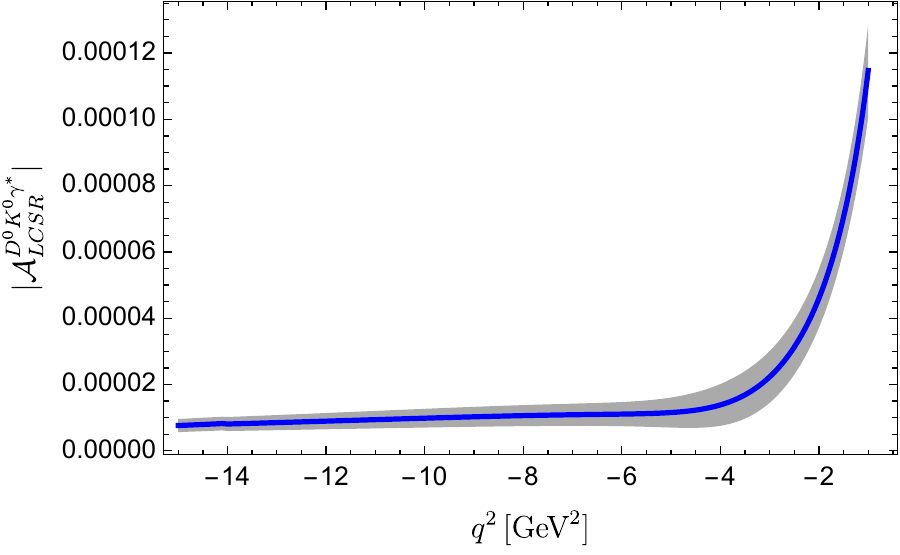}  
  \caption{}
\end{subfigure}
\caption{
The absolute values of the 
(a) $D_{s}^+\to \pi^+ \gamma^\ast$ and 
(b) $D^0\to K^0 \gamma^\ast$
amplitudes obtained from LCSRs. Grey bands represent the parametric uncertainties.}
\label{fig:CFLCSR}
\end{figure}

The numerical results for the amplitudes of $D_s^+\to \pi^+ \gamma^*$ and 
$D^0\to \bar{K}^0\gamma^*$  at $q^2<0$ obtained from LCSRs are presented in the ancillary files 
submitted with this paper.
The absolute values of these amplitudes are plotted in Fig.~\ref{fig:CFLCSR}. Our results reveal a small but visible 
$SU(3)_{fl}$ ($U$-spin) symmetry violation due to the $O(m_s)$ contributions taken into account 
in LCSR for $D_s^+\to \pi^+ \gamma$. The ratio of absolute values 
 $|\mathcal{A}^{(D_s^+\to \pi^+ \gamma^\ast)}_{(LCSR)}(q^2)|/
 |\mathcal{A}^{(D^+\to \pi^+ \gamma^\ast)}_{(LCSR)}(q^2)|$ varies by about $\pm 7\% $
 around unit at $-1.0<q^2<-15$ GeV$^2$. In the ratios of real and imaginary parts taken separately, this effect is more pronounced. 
Note also a suppression  of the $D^0\to \bar{K}^0\gamma^*$ amplitude 
 which is due to the fact that it is proportional to a
 numerically small combination of Wilson coefficients $C_{1,2}$ 
 (see Appendix~\ref{app:CF} for details).

As a next step, we
consider once-subtracted hadronic dispersion relations
similar to (\ref{eq:dispA}) for the
$D_s^+ \to \pi^+ \gamma^*$ and $D^0\to \bar{K}^0\gamma^*$ amplitudes,
modelling them with the  $\rho^0,\omega$ and $\phi$ 
contributions complemented by $z$-expansion for the heavier state contributions, as in (\ref{eq:dispZ}).
To obtain the residues $r_V$, we need the amplitudes of the weak 
$D_s^+ \to \pi^+ V$  and $D^0\to \bar{K}^0 V$ decays
($V=\rho^0,\omega,\phi$). Their absolute values  
are calculated taking the branching fractions 
from \cite{PDG}, and using the formula similar to (\ref{eq:DpiVampl}),
where the CKM factor $|V_{cs}V_{ud}^*|$ replaces $\lambda_d$.
The decay constants of vector mesons 
were already given in  Table~\ref{tab:Vpar}.
The measured branching fractions of weak decays 
and the resulting residues 
are given in Table~\ref{tab:CFVpar}. Note that the $D^0\to \bar{K^0} V$ branching fractions quoted in this Table are obtained
multiplying the measured $BR(D^0\to K_S V)$ by a factor 2.

Applying then a minimization procedure similar to (\ref{eq:fitZ}), we fit these dispersion relations to the LCSRs  for both transition amplitudes. 
The  fit region at $q^2<0$, the threshold $s_{th}$ entering the $z$ definition, and the subtraction point remain the same.
The results of the fit at the central input are: 
\begin{itemize}
    
\item for the $D^+_s\to \pi^+ \gamma^*$ transition:
\begin{eqnarray}
&&\varphi_\rho = 2.154\,, ~~\varphi_\omega = 2.148 \,,~~ 
\varphi_\phi = -0.997, 
\nonumber\\
&& \alpha_1= (-7.121 - 1.697 i)\times 10^{-2},~~
\alpha_2= (8.092 + 4.799 i)\times 10^{-2},~~
\label{eq:resDspifit}    
\end{eqnarray}

\item 
for the $D^0\to \bar{K}^0 \gamma^*$ transition:
\begin{eqnarray}
&&\varphi_\rho = -0.008 \,, ~~\varphi_\omega =  3.139 \,,~~ 
\varphi_\phi = -0.015 , 
\nonumber\\
&& \alpha_1= ( 0.261 + 0.038 i )\times 10^{-2},~~
\alpha_2= (- 0.313 - 0.046 i)\times 10^{-2},~~
\label{eq:resD0Kfit}    
\end{eqnarray}
\end{itemize}
The goodness of the fit for $D^+_s\to \pi^+ \gamma^*$,
reflected by up to $7\%$ deviations of the hadronic model from LCSR 
in the fit region, $-8.0$ GeV$^2 \leq q^2 \leq -3.0$ GeV$^2$, is at the same level as for $D^+\to \pi^+ \gamma^*$ (with deviations up to $10\%$), but turns out worse for the $D^0\to \bar{K}^0 \gamma^*$ (with deviations  up to $25\%$).

\begin{figure}[t]
\centering
\begin{subfigure}{0.49\textwidth}
  \includegraphics[scale=0.45]{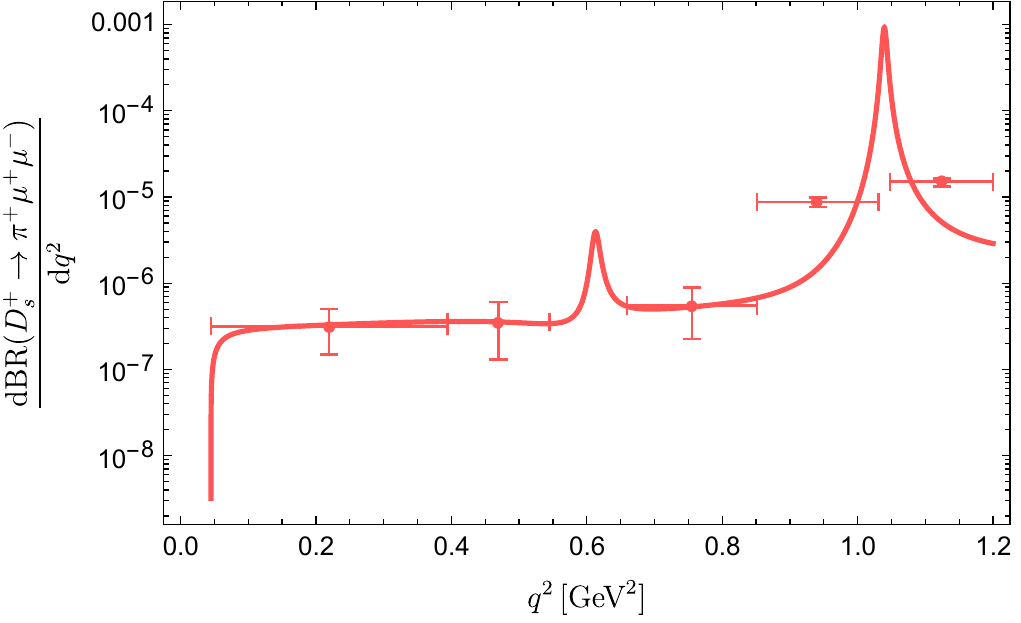}
  \caption{}
\end{subfigure}
\begin{subfigure}{0.49\textwidth}
  \includegraphics[scale=0.4]{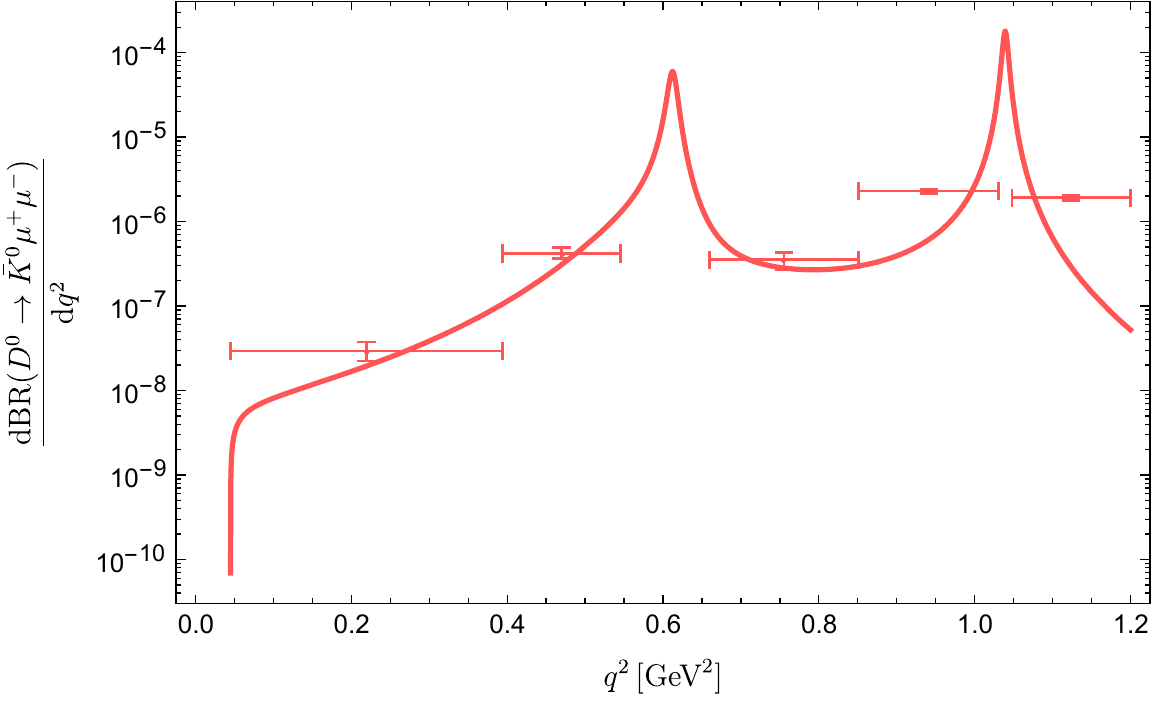}
  \caption{}
\end{subfigure}
\caption{
The differential branching fraction
of (a) $D_{s}^+\to \pi^+ \ell^+\ell^-$ and (b) $D^{0}\to K^0 \ell^+\ell^-$. }
\label{fig:CFBR}
\end{figure}

With these fit parameters, we calculate the nonlocal 
$D_s^+\to \pi^+ \gamma^*$ and $D^0\to \bar{K}^0\gamma^*$
amplitudes in the timelike region $q^2<s_{max}$. 
The resulting differential branching fractions for
the $D_{s}^+\to \pi^+ \ell^+\ell^-$ and 
$D^0\to \bar{K}^0 \ell^+\ell^-$ decays
are plotted in Fig.~\ref{fig:CFBR} (a) and (b), respectively.
In the same figures, we display the 
$q^2$-bins with estimated uncertainties. The branching fractions integrated over these bins are 
presented in  Table~\ref{tab:CFbins}. 
We notice that the predicted $D^+_{s}\to \pi^+ \ell^+\ell^-$ branching fraction integrated over the lower part of the LHCb region 
is only around 20-60\%  of the current upper limit for this observable presented in Table~\ref{tab:modes}. The $D^0\to \bar{K}^0 \ell^+\ell^-$ branching fraction 
in the same region is, on the contrary, far below 
the measured limit. 
%
%
%
%
\begin{table}[t]
\centering
\begin{tabular}{|c|c|c|c|c|}
\hline
&&&\\[-4mm]
Bin & $q^2_{\rm min}$ & $q^2_{\rm max}$ &
$(q^2_{\rm max}-q^2_{\rm min})\,\Delta{\cal B}^{(disp-z)}_{D_s\to \pi^+ \mu^+ \mu^-}$
&$(q^2_{\rm max}-q^2_{\rm min})\,\Delta{\cal B}^{(disp-z)}
_{D^0\to \bar{K}^0 \mu^+ \mu^-}$\\[1mm]
\hline
&&&&\\[-4mm]
I& $4m_\mu^2$ & $(m_\rho-\Gamma_\rho)^2$  & $ 1.10^{+0.66}_{-0.58} $ 
 & $0.10^{+0.03}_{-0.02}$\\[1mm]
\hline
&&&&\\[-4mm]
II& 
$(m_\rho-\Gamma_\rho)^2$ & $(m_\rho-\Gamma_\rho/4)^2$  & $ 0.53^{+0.38}_{-0.34}$
 &$0.63^{+0.12}_{-0.08} $\\[1mm]
\hline
&&&&\\[-4mm]
III& 
$(m_\rho+\Gamma_\rho/4)^2$ & $(m_\rho+\Gamma_\rho)^2$  & $ 1.06^{+0.67}_{-0.63} $ 
 &$0.68^{+0.14}_{-0.16} $ \\[1mm]
\hline
&&&&\\[-4mm]
IV& $(m_\rho+\Gamma_\rho)^2$ & $(m_\phi-\Gamma_\phi)^2 $
   & $ 15.69 ^{+1.93}_{-1.92}  $ & $4.11^{+0.25}_{-0.25}$\\[1mm]
\hline 
&&&&\\[-4mm]
V &$(m_\phi+\Gamma_\phi)^2$  & $1.2 ~\mbox{GeV}^2$
  & $ 22.88 ^{+2.24}_{-2.64} $ & $2.91^{+0.19}_{-0.20}$\\[1mm]
\hline
&&&&\\[-4mm]
\hline
&&&&\\[-4mm]
LHCb & $4m_\mu^2$ & $(0.525~\mbox{GeV})^2$  & $ 0.69^{+0.39}_{-0.35}   $ & $0.03^{+0.01}_{-0.01}$\\
\hline
\end{tabular}
\caption{ Branching fractions of the
$D_s^+\to\pi^+\mu^+\mu^-$ and $D^0\to\bar{K}^0 \mu^+\mu^-$ decays integrated over bin sizes  (in units $10^{-7}$) 
}
\label{tab:CFbins}
\end{table}
To our knowledge, there are no data available for
the weak decays $D_s\to \pi^+ V'$ and $D^0\to \bar{K}^0 V'$, where 
$V'=\rho',\phi'$. Hence, an extended  resonance model, similar to the one introduced for the $D^+\to \pi^+ \gamma^*$ amplitudes, cannot yet be completely fixed, and its study has  to be postponed to the future.

Having at hand the width of the CF decay  $D_s^+ \to \pi^+ \mu^+ \mu^-$, we can readily test  the accuracy of the 
$U$-spin symmetry relation between this and the $D^+ \to \pi^+ \mu^+ \mu^-$ decays.
As already mentioned, we expect that the symmetry  relation in (\ref{eq:rel_widths}) is 
satisfied for the integrated widths. The total $q^2$-region accessible from our analysis 
of both decays  corresponds to a bin with $q^2_{min}=4m_\mu^2$ and  $q^2_{max}=1.2$ GeV$^2$
and includes all three lowest resonances.
The $U$-spin relation for this region yields: 
\begin{eqnarray}
\int\limits_{4m_{\mu}^2}^{1.2\,GeV^2} \!\!\!\!\!\!dq^2 \,\frac{dBR(D_s^+ \to \pi^+ \mu^+ \mu^-)}{dq^2}
=\frac{|V_{cs}|^2}{|V_{cd}|^2}\frac{\tau_{D_s}}{\tau_{D}} 
\!\!\!\!\int\limits_{4m_\mu^2}^{1.2\,GeV^2}\!\!\!\!\!\!\!dq^2\,
\!\left[\frac{p_{(D_s^+\pi^+\gamma^*)}(q^2)}{p_{(D^+\pi^+\gamma^*)}(q^2)}\right]^3\frac{dBR(D^+ \to \pi^+ \mu^+ \mu^-)}{dq^2},
\label{eq:rel_BR_U} 
\end{eqnarray}
where  the kinematical correction to the widths is taken into account, according to (\ref{eq:dBR}). 
Using  on both sides our hadronic model with $z$-expansion and taking the values  (\ref{eq:zfitparam}) and (\ref{eq:resDspifit}) for  the fitted  parameters, we obtain, respectively, 
$1.31\times 10^{-5}$ for l.h.s. and $2.09 \times 10^{-5}$ for r.h.s. of the above equation. The observed 
difference corresponds to a 
$\sim 20\%$ violation of $U$-spin symmetry at the amplitude level. This is a typical size of  $SU(3)_{fl}$ symmetry  violation in the hadronic matrix elements, e.g., the difference between decay constants $f_{D_s}$ and $f_D$ (see Table~\ref{tab:inpPDG}) is at the same level. We also tested the relations analogous to (\ref{eq:rel_BR_U}) for all separate bins listed in Table~\ref{tab:CFbins} and found that the amount of the $U$-spin violation at amplitude level varies between 15 \% in the Bin I and the LHCb region and 30\% in Bin II, the one adjacent to the $\rho$-resonance.

\section{Discussion}
\label{sect:concl}
In this paper, we suggest a new method to analyse the 
rare semileptonic $D\to P \ell^+\ell^-$ decays, concentrating ourselves on the most important  mode $D^+\to \pi^+ \ell^+\ell^-$.
The decays of this type have attracted a lot of attention, since 
they can proceed at short distances via the FCNC  
$c \to u \ell^+ \ell^-$  transition  
that could be modified by new physics effects induced by yet unknown heavy particles. 
At the same time, within SM, the overwhelming mechanism of the $D^+\to \pi^+ \ell^+\ell^-$ decay  is not a 
short-distance FCNC transition, but a genuine long-distance process, in which a weak SCS hadronic transition is combined with an e.m. emission of a lepton  pair via a virtual photon. We obtained the underlying  $D\to\pi\gamma^*$ transition amplitude, which can be 
identified as a typical nonlocal  form factor. 

The method we used in this paper relies on an LCSR calculation in the unphysical region of the dilepton invariant mass $q^2$. For 
    sufficiently large $-q^2$, we  calculated 
the $D\to\pi\gamma^*$ amplitude, 
starting from a correlation function which contains a perturbative QCD part convoluted  with the nonperturbative light-cone distribution amplitudes of the pion.
The dominant contribution to LCSR stems from the annihilation topology diagrams, whereas the $d$- and $s$-loop topology diagrams largely cancel each other, which is an evident manifestation of the GIM mechanism.

The other major element of our method is the hadronic dispersion relation used to extrapolate the  $D\to\pi\gamma^*$ amplitude  to positive values of $q^2$ in the physical region of the $D\to\pi\ell^+\ell^-$ decay. However, this extrapolation requires to model the spectral density in the dispersion relation. We probed two different models, both containing 
the lowest-lying vector resonances  $V=\rho,\omega,\phi$, but treating the heavier hadronic states differently. 
Some properties of this spectral density are known, in particular the masses, widths and decay constants of 
$V$ mesons, as well as the $D\to \pi V$ amplitudes and all of them were used as model inputs. Using two different parameterizations for the 
spectral density above the 
$\rho,\omega,\phi$ resonances, we fitted the modelled hadronic dispersion relation to LCSR, and determined the relative phases of these resonances
together with parameters of the models describing the heavier hadronic states.
Essentially, in this paper, we refined 
and extended to charm decays the method of\, ``LCSR assisted" hadronic models that were previously applied to 
nonlocal form factors in $B\to P \ell^+\ell^-$ decays.

The main result of our investigation is  a prediction for the $q^2$ spectrum of $D^+\to \pi^+ \ell^+\ell^-$. 
At least for values of $q^2 \le 1.2$ GeV$^2$, we obtain a
consistent picture from different models of dispersion relation, including also predictions for $q^2$-bins with estimates of uncertainties. 
To demonstrate the universality of our method, we derived LCSRs 
and obtained $q^2$-spectra and bins also for the Cabibbo-favored decay modes, $D_s^+\to \pi^+ \ell^+\ell^-$ and  $D^0\to \bar{K}^0 \ell^+\ell^-$.

It is instructive to compare 
our  method and results with the most recent detailed study of $D^+\to \pi^+ \ell^+\ell^-$ in \cite{Bharucha:2020eup}. In that paper, both
important aspects: the dominance of weak annihilation contributions and the role of the vector mesons beyond the lowest ones were put forward.
However, there are several major differences with respect to our approach.
First of all, we use the QCD-based LCSRs in the spacelike region of $q^2$, 
so that weak annihilation and loop  topologies reveal themselves at the level of OPE diagrams in the correlation function.
In contrast, the approach used in \cite{Bharucha:2020eup}
provides an essentially  partonic description
of $D^+\to \pi^+ \gamma^*$ amplitude, with quark-topology contributions calculated directly at positive $q^2$, applying QCDf
\cite{Feldmann:2017izn}, \cite{Beneke:2001at}
at small $q^2$ 
and  OPE \cite{Beylich:2011aq} at large $q^2$, both approaches taken from the description of
$B$-meson FCNC decays.
Secondly, our LCSRs are derived  at finite $c$-quark mass, 
whereas in \cite{Bharucha:2020eup} the heavy quark limit and $D$-meson light-cone DAs are used, so that 
there is an issue of possible  large $O(1/m_c)$ corrections.
Furthermore, in \cite{Bharucha:2020eup}, a completely different 
treatment of vector resonances, as compared to our 
approach described above, is applied. The vector-meson resonances  are installed in the $D\to \pi \gamma^*$ loops following the parton-hadron duality ansatz suggested in \cite{Feldmann:2017izn}. In addition, in \cite{Bharucha:2020eup}, individual vector-resonance (V) contributions  were fitted from $e^+e^-$ data, assuming a factorizable approximation for $D\to \pi V$ decays, but then confronted with a necessity to adjust the $\phi$-resonance residue by a large inflating factor to match the data on the $D\to \pi \phi$ decay. Also in that study, unknown phases
of separate $V$-terms remain the main source of uncertainty. 
The numerical prediction for the low $q^2$-bin of  $D^+\to \pi^+\ell^+\ell^-$ in \cite{Bharucha:2020eup} is in good agreement with our result in the $z$-expansion model for the same bin, the latter having a smaller uncertainty.
The large $q^2$-bin predicted in \cite{Bharucha:2020eup} is smaller than our prediction of the extended resonance model, there is only a  marginal agreement, if one takes into account  large uncertainties.

In many other previous  
studies of $D\to \pi \ell^+\ell^- $ and related decays 
(see e.g.,~\cite{Fajfer:2005ke,deBoer:2015boa,Sanchez:2022nsh})
a three-resonance model 
with $\rho^0,\omega,\phi$, was used to describe the long-distance 
$D\to \pi\gamma^*$ amplitude. Note that such a model simply violates the general unitarity relation 
for this amplitude because the heavier intermediate states with vector-meson quantum numbers are ignored. 
Moreover,  a negative relative sign  between $\rho$ and $\omega$ terms is fixed in these models, based on the assumption that $\rho^0$ and $\omega$ mesons originate predominantly from the $\bar{d}d$ state which has opposite signs in their quark content.
However, this assumption neglects the dominant 
weak annihilation topology, where the emission of a virtual photon can occur not only  from $\bar{d}d$ state but also  from $\bar{u}u $ state, (cf. the diagram with a virtual photon emission from the $u$-quark in Fig.\ref{fig:lcsr_ann}).
 Note that our fit
of LCSRs to the modelled dispersion relations predicts almost equal phases for  $\rho$ and $\omega$. 

We emphasize that our study in this paper is the first and, to a large extent, an exploratory attempt to apply the LCSR-assisted method to $D_{(s)}\to P \ell^+\ell^-$   decays. Naturally, then, there is still a considerable room for further improvements and refinements. 
First of all, since our computation of LCSRs
was done at LO and with the lowest twist-2 accuracy for the pion and kaon DAs, further upgrades of light-cone OPE, including NLO and higher twist terms are desirable, albeit technically demanding. The NLO perturbative gluon exchanges 
(both factorizable and nonfactorizable ones)
added to the annihilation and loop diagrams  are not expected 
to produce large corrections, as we know, e.g., from studies of LCSRs for the $D\to\pi,K$ form factors.
More important could be soft-gluon nonfactorizable corrections involving three-particles pion or kaon DAs. 
Another avenue of improvements concerns the hadronic dispersion relation in the photon virtuality, where a simple Breit-Wigner ansatz for lowest-lying  resonance contributions can be replaced by a less model-dependent and 
more accurate dispersive analysis of the same kind as the ones used nowadays for the  local
pion and kaon form factors.

Last but not least, a 
more detailed study of uncertainties and correlations emerging in 
the fitting procedure could eventually reduce the uncertainties in the differential
branching fractions. In this respect, invoking Bayesian methods, and in parallel,  probing other than $z$-expansion parameterizations for heavier states, may prove useful.

On the experimental side, it is desirable 
to have more accurate data on the nonleptonic decays of the type $D_{(s)}\to \pi(K) V'$, where $V'$ are the excited vector meson
resonances. Their amplitudes are necessary 
as inputs in the extended hadronic models of the dispersion relation.

The method suggested in this paper can also be applied
to other modes of $c\to u \ell^+\ell^-$ transitions such as $D\to \pi^+\pi^- \ell^+ \ell^-$, applying LCSRs with dipion DAs. Here, theory 
can already be challenged  by the existing measurement
\cite{LHCb:2017uns}
of this decay channel by the LHCb collaboration.

We conclude that the situation for the $D\to\pi\ell^+\ell^-$ decay and other 
SCS rare decays of this type listed in Table \ref{tab:modes} is vastly different 
from e.g. $B \to K^{(*)} \ell^+ \ell^-$ decays: It will be almost impossible to trace the short-distance part of FCNC transitions in these decays, unless BSM effects enhance the SM Wilson coefficients by orders of magnitude. 
Still, as our study in this paper shows, the 
$D_{(s)}\to P \ell^+\ell^-$  decays offer a very useful testing ground for understanding the hadronic dynamics of nonlocal heavy-to-light form factors. 

\section*{Acknowledgements}
This research was supported by the Deutsche Forschungsgemeinschaft (DFG, German Research Foundation) under grant 396021762 - TRR 257 “Particle Physics Phenomenology after the Higgs Discovery''.

\appendix

\section{Annihilation topology diagrams for $D^+\to \pi^+\gamma^*$}
\label{app:diagA}

Here we present in detail the calculation of the annihilation diagrams 
in Fig.~\ref{fig:lcsr_ann} contributing to 
the LO of the correlation function (\ref{eq:corr}) for the $D^+\to \pi^+\gamma^*$ transition. Note that in this case, only the operator $O_1^d$
contributes.

\subsection{Expressions for the diagrams in terms of loop integrals}

Contracting the quark fields into the propagators
and isolating the vacuum-to-pion hadronic matrix element,
we obtain the initial 
expressions \footnote{ \,Our convention for covariant derivative in QED is $D_\mu= \partial_\mu-ie Q_fA_\mu$, where $Q_f=-1,2/3,-1/3$, respectively,  for a lepton, up-type quark, down-type quark and $e=\sqrt{4\pi\alpha_{em}}$
is the e.m. coupling. The $\varepsilon$-tensor  is defined as:
$\varepsilon_{0123}\!\!=\!\!-\varepsilon^{0123}\!\! =\!\! 1$,
$\mbox{Tr} (\gamma^\mu \gamma^\nu \gamma^\rho\gamma^\lambda\gamma^5)\!\! =\!
4i\varepsilon^{\mu\nu\rho\lambda}$\,.
}.
For the diagrams (a) and (b), where the virtual photon is emitted, respectively, from the 
initial $d$-quark and $c$-quark line, we have
\footnote{\,Since we aim at the imaginary part of the loop diagrams,
it is sufficient to consider them at $D=4$.}
:
\begin{eqnarray} 
&& {\cal F}_\mu^{(a)}(p,q,k)=\frac{3i}{4}C_1Q_d(m_c+m_d)(p-k)_\rho f_\pi
\nonumber\\
&&\times \int\frac{d^4f}{(2\pi)^4}
\mbox{Tr}\Bigg[ \frac{(\slashed{f} + \slashed{p}+ \slashed{q}+m_c)}{(f+p+q)^2-m_c^2}
\gamma_5\frac{(\slashed{f} +m_d)}{f^2-m_d^2}\gamma_\mu 
\frac{(\slashed{f} + \slashed{q}+m_d)}{(f+q)^2-m_d^2} \gamma_\rho(1-\gamma_5)\Bigg]\,,
\label{eq:diagainit}
\end{eqnarray}
and
\begin{eqnarray} 
&&{\cal F}_\mu^{(b)}(p,q,k)=\frac{3i}{4}C_1 Q_c(m_c+m_d)(p-k)_\rho f_\pi
\nonumber\\
&&\times \int\frac{d^4f}{(2\pi)^4}
\mbox{Tr}\Bigg[ \frac{(\slashed{f} + \slashed{p}+ \slashed{q}+m_c)}{(f+p+q)^2-m_c^2}
\gamma_5\frac{(\slashed{f} +m_d)}{f^2-m_d^2}\gamma_\rho(1-\gamma_5) 
\frac{(\slashed{f} + \slashed{p}+m_c)}{(f+p)^2-m_c^2} \gamma_\mu\Bigg]\,.
\label{eq:diagbinit}
\end{eqnarray}
In both diagrams, the vacuum-to-pion matrix element is reduced to the pion decay constant:
\begin{equation}
\langle \pi^+(p-k)|\bar{u} \gamma_\rho \gamma_5 d|0\rangle=-if_\pi(p-k)_\rho\,.
\end{equation}

In the diagrams (c) and (d) the virtual photon is emitted  from the final-state $u$- and $d$-quark lines, respectively.
In this case, the emerging nonlocal vacuum-to-pion matrix element 
is expressed via the pion DAs with growing twist. 
We use the standard definition:
\begin{eqnarray}
 \langle \pi^+(p-k)|\bar{u}_\omega^i(x_1) 
d^j_\xi(x_2)|0\rangle
  &=& \frac{i\delta^{ij}}{12}f_\pi 
\int_0^1 du~e^{iu (p-k)\cdot x_1+ i\bar{u} (p-k)\cdot x_2   }
 [(\slashed{p}-\slashed{k})\gamma_5]_{\xi\omega} \varphi_\pi(u)
\nonumber\\
&+&\dots\,,
\label{eq:tw2}
\end{eqnarray}
where $\bar{u}=1-u$,  and explicitly shown is only the term with the lowest twist-2  DA that will be needed. 

In the twist-2 approximation, the initial expressions for the diagrams (c) and (d) are:
\begin{eqnarray} 
&&{\cal F}_\mu^{(c)}(p,q,k)=\frac{3i}{16}C_1 Q_u(m_c+m_d)f_\pi
\int\frac{d^4f}{(2\pi)^4}
\mbox{Tr}\Bigg[ \frac{(\slashed{f} +m_d)}{f^2-m_d^2}\gamma_\rho(1-\gamma_5)
\frac{(\slashed{f} + \slashed{p}+\slashed{q}+m_c)}{(f+p+q)^2-m_c^2} 
\gamma_5 \Bigg]
\nonumber\\
&&\times\int\limits_0^1 du \frac{\varphi_\pi(u)}{(q+u(p-k))^2-m_u^2}
\mbox{Tr} \bigg[\gamma^\rho(1-\gamma_5)(\slashed{p}-\slashed{k})\gamma_5\gamma_\mu
(u(\slashed{p}-\slashed{k})+\slashed{q}+m_u)\bigg]\,,
\label{eq:diagcinit}
\end{eqnarray}
and
\begin{eqnarray} 
&&{\cal F}_\mu^{(d)}(p,q,k)=\frac{3i}{16}C_1 Q_d(m_c+m_d)f_\pi
\int\frac{d^4f}{(2\pi)^4}
\mbox{Tr}\Bigg[ \frac{(\slashed{f} +m_d)}{f^2-m_d^2}\gamma_\rho(1-\gamma_5)
\frac{(\slashed{f} + \slashed{p}+\slashed{q}+m_c)}{(f+p+q)^2-m_c^2} 
\gamma_5 \Bigg]
\nonumber\\
&&\times\int\limits_0^1 du \frac{\varphi_\pi(u)}{(q+\bar{u}(p-k))^2-m_d^2}
\mbox{Tr} \bigg[\gamma^\rho(1-\gamma_5)
(-\bar{u}(\slashed{p}-\slashed{k})-\slashed{q}+m_d) \gamma_\mu  (\slashed{p}-\slashed{k})\gamma_5 \bigg]\,.
\label{eq:diagdinit}
\end{eqnarray}
After taking traces, we omit the kinematical
structures proportional to $k_\mu$, also the one 
with $\epsilon$ tensor, since they are 
not relevant for the sum rule for the $D^+ \to \pi^+ \gamma^\ast$ amplitude. 

According to our chosen
configuration of invariant variables,
in all diagrams we keep $k^2=0$, $p^2=0$ and $(p-k)^2=m_\pi^2$.
In addition, we
put the  $u$ and $d$ quark masses, as well as the pion mass to zero, i.e.
we adopt the chiral limit, which is a usual approximation for the LCSRs with pion DAs.
Note that in this limit, the twist-3 terms in the expansion of the vacuum-to pion matrix element (\ref{eq:tw2}) 
do not contribute to the correlation function, being proportional to 
$\mu_\pi m_u \sim \mu_\pi m_d \sim m_\pi^2$, where $\mu_\pi=m_\pi^2/(m_u+m_d)$ is the normalization factor of 
the pion twist-3 DAs. The next, twist-4 terms do contribute, but they 
are power suppressed with respect to the twist-2 term, typically as $\Lambda_{QCD}^2/m_c^2$ 
and therefore neglected 
(cf. the twist expansion in the LCSRs for the pion e.m. form factor 
\cite{Braun:1999uj}). Finally, we obtain:
\begin{eqnarray} 
{\cal F}_\mu^{(a)}(p,q,k)=(3iC_1Q_d m_c^2 f_\pi)\!
\int\frac{d^4f}{(2\pi)^4 [(f+p+q)^2-m_c^2]f^2(f+q)^2 }
\nonumber\\
\times\Big[-2f_\mu (f\cdot (p-k))-f_\mu (q\cdot(p-k)) - q_\mu(f\cdot(p-k))+p_\mu (f^2 +(f\cdot q))\Big]
\,,
\label{eq:diagaTr}
\end{eqnarray}
\begin{eqnarray} 
{\cal F}_\mu^{(b)}(p,q,k)=(3iC_1Q_c m_c^2 f_\pi)\!
\int\frac{d^4f}{(2\pi)^4 [(f+p+q)^2-m_c^2]f^2[(f+p)^2-m_c^2] }
\nonumber\\
\times\Big(-2f_\mu (f\cdot (p-k))+f_\mu (q\cdot (p-k)) -q_\mu(f\cdot(p-k))
-p_\mu (2(f\cdot (p-k)) +(f\cdot q))\Big]\,,
\label{eq:diagbTr}
\end{eqnarray}
\begin{eqnarray} 
&&{\cal F}_\mu^{(c)}(p,q,k)=3iC_1 Q_u m_c^2f_\pi
\int\limits_0^1 du \frac{\varphi_\pi(u)}{(q+u(p-k))^2}
\nonumber\\
&&\times \int\frac{d^4f}{(2\pi)^4}
\frac{\Big[ f_\mu(q\cdot(p-k))+f_\mu(p-k)^2u
-p_\mu(f\cdot q)-
2p_\mu(f\cdot(p-k))u \Big]}{[(f+p+q)^2-m_c^2 ]\,f^2}
\,,
\label{eq:diagcTr}
\end{eqnarray}
and
\begin{eqnarray} 
&&{\cal F}_\mu^{(d)}(p,q,k)=3iC_1 Q_d m_c^2f_\pi
\int\limits_0^1 du \frac{\varphi_\pi(u)}{(q+\bar{u}(p-k))^2}
\nonumber\\
&&\times \int\frac{d^4f}{(2\pi)^4}
\frac{\Big[ -f_\mu(q\cdot(p-k))-f_\mu(p-k)^2\bar{u}
+p_\mu(f\cdot q)+ 2p_\mu(f\cdot(p-k))\bar{u}
\Big]}{[(f+p+q)^2-m_c^2 ]\,f^2}
\,.
\label{eq:diagdTr}
\end{eqnarray}
We notice that the above expression for 
${\cal F}_\mu^{(d)}$, after transforming
the integration variable $u \to \bar{u}$, coincides 
with the expression (\ref{eq:diagcTr}) for ${\cal F}_\mu^{(c)}$, up to a replacement
$Q_u\to -Q_d$. Here we use that in the isospin symmetry limit, the pion DA is symmetric: 
$$\varphi_\pi(u)=\varphi_\pi(\bar{u}).$$
This allows us to obtain a compact expression 
for the sum of two diagrams:
\begin{eqnarray} 
&&{\cal F}_\mu^{(c)\oplus(d)}(p,q,k)=3iC_1 (Q_u-Q_d) m_c^2f_\pi
\int\limits_0^1 du \frac{\varphi_\pi(u)}{(q+u(p-k))^2}
\nonumber\\
&&\times \int\frac{d^4f}{(2\pi)^4}
\frac{\Big[ f_\mu(q\cdot(p-k))+f_\mu(p-k)^2 u
-p_\mu(f\cdot q)- 2p_\mu(f\cdot(p-k))u
\Big]}{[(f+p+q)^2-m_c^2 ]\,f^2}\,.
\label{eq:diagcdTr}
\end{eqnarray}
Furthermore, for the pion twist-2 DA, we adopt the usual 
expansion  in Gegenbauer polynomials, retaining the first two
nonasymptotic terms:
\begin{eqnarray}
\varphi_\pi(u,\mu)=6u(1-u)\bigg[1+a_2(\mu)C_2^{3/2}(u-\bar{u})+a_4(\mu)C_4^{3/2}(u-\bar{u})\bigg] \,,
\label{eq:pionDAmod}
\end{eqnarray}
(see e.g., \cite{Cheng:2020vwr} for more details concerning this choice).

\subsection{Scalar integrals and dispersion representations} 
\label{app:scaldisp}

Our goal is to transform the expressions for the annihilation diagrams presented in the previous subsection into a form of dispersion relation in the variable $(p+q)^2$. To this end, we use 
Cutkosky rules to calculate the imaginary part 
of the three- and two-point loop diagrams 
\footnote{a formula for Cutkosky rules applied to a general Feynman amplitude 
can be found, e.g.,
in \cite{Itzykson:1980rh}.}.
This technique was used recently in \cite{Bordone:2022drp} to obtain sum rules for the $B_c\to J/\psi$ form factors. 

In this subsection, we collect 
dispersion forms of scalar (traceless) Feynman 
integrals needed for the annihilation diagrams.
Deriving these forms, we neglect the terms in the Feynman integrals 
which do not develop imaginary part, 
i.e. constant or polynomial terms in $(p+q)^2$, 
because they all vanish
after the Borel transform in $(p+q)^2$, which will be performed to obtain the sum rule.

For the three-point
scalar integral needed for the diagram (a), we obtain (at $p^2=0$):
\begin{eqnarray}
I_{cdd}((p+q)^2,q^2)&&\equiv 
i\int \frac{d^4f}{(2\pi)^4} 
\frac{1}{\big[(f+p+q)^2-m_c^2\big]f^2(f+q)^2}
\nonumber\\
&&= -\frac{1}{16\pi^2}\int\limits_{m_c^2}^{\infty}
\frac{ds}{[s-(p+q)^2](s-q^2)}
\log\Bigg[\frac{-q^2m_c^2}{s(s-m_c^2-q^2)}\Bigg]\,.
\label{eq:Icdd}
\end{eqnarray}
Hereafter, the indices such as $cdd$ in the above integral indicate the flavour content of the 
quark loop.
Note that the same four-dimensional integral can be calculated with a standard  Feynman paramatrization, yielding 
at $p^2=0$:
\begin{eqnarray}
I_{cdd}((p+q)^2,q^2)= \frac{1}{16\pi^2}\int_0^1 dx \int_0^1 dy \,
\frac{1}{m_c^2y-q^2x(1-y)-(p+q)^2xy}\,.
\nonumber
\end{eqnarray}
 The Borel transform of this integral in $(p+q)^2$ 
 is easily performed, 
 and we checked that it coincides numerically with the
 Borel transformed dispersion form (\ref{eq:Icdd}). 

For the two-point integrals, we have:
\begin{eqnarray}
I_{cd}((p+q)^2)\equiv 
i\int \frac{d^4f}{(2\pi)^4} 
\frac{1}{\big[(f+p+q)^2-m_c^2\big]f^2}
=-\frac{1}{16\pi^2}\int\limits_{m_c^2}^{\infty}
\frac{ds (s-m_c^2)}{[s-(p+q)^2]s}\,,
\label{eq:Icd}
\end{eqnarray}
\begin{eqnarray}
I_{cdfq}((p+q)^2,q^2)\equiv 
i\int \frac{d^4f}{(2\pi)^4} 
\frac{(f\cdot q)}{\big[(f+p+q)^2-m_c^2\big]f^2}
=\frac{1}{64\pi^2}\int\limits_{m_c^2}^{\infty}
\frac{ds (s-m_c^2)^2(s+q^2)}{[s-(p+q)^2]s^2}\,.
\label{eq:Icdfq}
\end{eqnarray}
For the diagram (b), we need an additional three-point scalar integral
\begin{eqnarray}
I_{cdc}((p+q)^2,q^2)\equiv &&
i\int \frac{d^4f}{(2\pi)^4} 
\frac{1}{\big[(f+p+q)^2-m_c^2\big]f^2\big[(f+p)^2-m_c^2\big]}
\nonumber\\
= &&-\frac{1}{16\pi^2}\int\limits_{m_c^2}^{\infty}
\frac{ds}{[s-(p+q)^2](s-q^2)}\log\frac{m_c^2s}{s(s-q^2)+m_c^2q^2}
\,,
\label{eq:Icdc}
\end{eqnarray}
and the two-point integral
\begin{eqnarray}
I_{cdfp}((p+q)^2,q^2)\equiv 
i\int \frac{d^4f}{(2\pi)^4} 
\frac{(f\cdot p)}{\big[(f+p+q)^2-m_c^2\big]f^2}
=
\frac{1}{64\pi^2}\int\limits_{m_c^2}^{\infty}
ds \frac{(s-m_c^2)^2(s-q^2)}{[s-(p+q)^2]s^2}\,.
\label{eq:Icdfp}
\end{eqnarray}

Finally,
for the combined expression of the
diagrams (c) and (d), we will only need the two-point scalar integral $I_{cd}$ given in (\ref{eq:Icd}).

\subsection{Reduction of vector and tensor integrals}
\label{sec:Areduction}
The expressions (\ref{eq:diagaTr}) - (\ref{eq:diagcdTr}) contain integrals with one or two powers of the loop four-momentum $f$ in the numerator. Here we reduce them to combinations of scalar integrals multiplied by 
Lorentz-covariant structures in terms of external four-momenta.  

We start from the vector integral originating from the three-point loop:
\begin{eqnarray}
I^\alpha_{cdd}(p,q)\equiv 
i\int \frac{d^4f}{(2\pi)^4} 
\frac{f^\alpha}{\big[(f+p+q)^2-m_c^2\big]f^2(f+q)^2}\,.
\label{eq:intfalpha}
\end{eqnarray}
Multiplying this integral by $p_\alpha$ and $q_\alpha$, we 
express the results in terms of scalar integrals 
(\ref{eq:Icdd}) and (\ref{eq:Icd}):
\begin{eqnarray}
&&p_\alpha I_{cdd}^\alpha= -\frac12\big[I_{cd}+(s-m_c^2-q^2)I_{cdd}\big]\,,
\nonumber\\
&&q_\alpha I_{cdd}^\alpha= \frac12\big[I_{cd}-q^2I_{cdd}\big]\,,
\label{eq:pIqI}
\end{eqnarray}
where we omit the contributions  which do not develop Im part in $(p+q)^2=s$ at $s>m_c^2$.
For brevity, in this subsection, we do not display  the 
dependence of scalar integrals on kinematical invariants
and replace the variable $(p+q)^2$ by $s$ in the coefficients
of the scalar integrals,
anticipating that these coefficients will enter the integrands of the dispersion
integrals.

To proceed, we  expand the integral (\ref{eq:intfalpha}) 
in two independent momenta $p$ and $q$:
\begin{eqnarray}
I_{cdd}^\alpha= p^\alpha ~I^{(p)}_{cdd}+ q^\alpha ~I^{(q)}_{cdd}\,.
\label{eq:Icddal_exp}
\end{eqnarray}
Multiplying  both parts 
of this expansion by these momenta, we solve the system of two linear equations for the  coefficients in the above expansion, 
expressing them via scalar
integrals:
\begin{eqnarray}
I^{(p)}_{cdd}= &&
\frac{
    (s+\text{q2})}{(s-\text{q2})^2}\,\text{Icd}
    +\frac{\text{q2} (-2
    \text{mc2}+s-\text{q2})}{(s-\text{q2})^2}\,
    \text{Icdd}\,,
\nonumber\\
I^{(q)}_{cdd}= &&
-\frac{\text{Icd}}{s-\text{q2}}-
    \left(1-\frac{\text{mc2}}{s-\text{q2}}\right)\!
    \text{Icdd}\,.
\end{eqnarray}
A similar method works for the tensor integral:
\begin{eqnarray}
I^{\alpha\beta}_{cdd}(p,q)\equiv 
i\int \frac{d^4f}{(2\pi)^4} 
\frac{f^\alpha f^\beta}{\big[(f+p+q)^2-m_c^2\big]f^2(f+q)^2}.
\label{eq:intfalphabeta}
\end{eqnarray}
We use an expansion in all possible 
symmetric combinations of four-momenta, including also $g_{\alpha\beta}$:
\begin{eqnarray}
I_{cdd}^{\alpha\beta}= \big(p^\alpha p^\beta) ~I^{(pp)}_{cdd}+
\big(q^\alpha q^\beta)~I^{(qq)}_{cdd} + 
(q^\alpha p^\beta+q^\beta p^\alpha)I^{(qp)}_{cdd}
+g^{\alpha\beta}\,I^{(g)}_{cdd}
\,,
\label{eq:Icddalbet_exp}
\end{eqnarray}
and use the following equalities:
\begin{eqnarray}
&& g_{\mu\nu} I_{cdd}^{\mu\nu} =0 \,,
\nonumber\\
&&p_{\mu}p_{\nu} I_{cdd}^{\mu\nu} = (m_c^2-s+q^2)^2 \frac{I_{cdd}}{4}-(2m_c^2-2s+q^2)\frac{I_{cd}}{4}+\frac{I_{cdfq}}{2}\,,
\nonumber\\
&& q_{\mu}q_{\nu} I_{cdd}^{\mu\nu}  =\frac{(q^2)^2}{4}I_{cdd}-\frac{q^2}{4}I_{cd}+\frac{I_{cdfq}}{2}\,,
\nonumber\\
&&p_{\mu}q_{\nu}  I_{cdd}^{\mu\nu}  = -q^2(m_c^2-s+q^2)\frac{I_{cdd}}{4}+(m_c^2-s+q^2)\frac{I_{cd}}{4}-\frac{I_{cdfq}}{2}\,.
\end{eqnarray}
We obtain the following results for the coefficients in 
(\ref{eq:Icddalbet_exp}) expressed via the three scalar integrals:
\begin{eqnarray}
I^{(pp)}_{cdd}
=&&-\frac{\text{q2} \left(6 \text{mc2} (s+\text{q2})-5s^2-2s\text{q2}+(\text{q2})^2\right)}{(s-\text{q2})^4}\text{Icd} 
\nonumber\\    
  &&  +\frac{
    (\text{q2})^2 \left(6 (\text{mc2})^2+6 \text{mc2}
    (\text{q2}-s)+(s-\text{q2})^2\right)}{(s-
    \text{q2})^4}\text{Icdd} 
  \nonumber\\  
&&+\frac{2 \left(s^2+4 \text{pplq2}\text{q2}
+(\text{q2})^2\right)}{(\text{pplq2}-\text{q2})^4}\text{Icdfq}\,; 
\label{eq:I_pp}
\end{eqnarray}
\begin{eqnarray}
I^{(qq)}_{cdd} =
\frac{ (2
    \text{pplq2}-2 \text{mc2}   -\text{q2})}{(\text{pplq2}-\text{q2})^2}\text{Icd}
+\frac{(\text{mc2}-\text{pplq2}+\text{q2})^2}{(\text{pplq2}
    -\text{q2})^2}\text{Icdd}
+\frac{2}{(\text{pplq2}-\text{q2})^2} \text{Icdfq}\, ;
\label{eq:I_qq}
\end{eqnarray}

\begin{eqnarray}
I^{(qp)}_{cdd}=
&&\frac{2 \text{mc2} (\text{pplq2}+2 \text{q2})-2
    s^2-2 \text{pplq2}
    \text{q2}+(\text{q2})^2}{(\text{pplq2}-\text{q2})^3}\text{Icd} 
\nonumber\\    
&&    +\frac{
    \text{q2} (\text{mc2}-\text{pplq2}+\text{q2}) (3
    \text{mc2}-\text{pplq2}+\text{q2})}{(\text{q2}-\text{pplq2})^3}
    \text{Icdd}
-\frac{2 (2 \text{pplq2}+\text{q2})}{(\text{pplq2}-\text{q2})^3
}\text{Icdfq}\,;
\label{eq:I_qp}
\end{eqnarray}
\begin{eqnarray}
I^{(g)}_{cdd}= \frac{ s^2-\text{mc2}
    (\text{pplq2}+\text{q2})}{2
    (\text{pplq2}-\text{q2})^2}\,\text{Icd}
   +\frac{ \text{mc2} \text{q2}
    (\text{mc2}-\text{pplq2}+\text{q2})}{2
    (\text{pplq2}-\text{q2})^2}\,\text{Icdd}
+\frac{\text{pplq2}}{(\text{pplq2}-\text{q2})^2}
\,\text{Icdfq}
\,.
\label{eq:I_g}
\end{eqnarray}
For diagram (b), the vector integral 
\begin{eqnarray}
I^\alpha_{cdc}(p,q)\equiv 
i\int \frac{d^4f}{(2\pi)^4} 
\frac{f^\alpha}{\big[(f+p+q)^2-m_c^2\big]f^2\big[(f+p)^2-m_c^2\big]}\,,
\label{eq:intfalphab}
\end{eqnarray}
originating from the triangle loop with two $c$-quark lines
is also  parameterised as
\begin{eqnarray}
I_{cdc}^\alpha= p^\alpha ~I^{(p)}_{cdc}+ q^\alpha ~I^{(q)}_{cdc}\,,
\label{eq:Icdcal_exp}
\end{eqnarray}
where the coefficients are obtained, using the same method as  
for the diagram (a):
\begin{eqnarray}
 &&I^{(p)}_{cdc}=-\frac{\text{pplq2}+\text{q2}}{(\text{pplq2}-
 \text{q2})^2}\text{Icd}
 +\frac{\text{pplq2} (\text{q2}-\text{pplq2})-2 \text{mc2}
    \text{q2}}{(\text{pplq2}-\text{q2})^2}\text{Icdc}\,, 
\nonumber\\
&&I^{(q)}_{cdc}=
\frac{\text{Icd}}{\text{pplq2}-\text{q2}}+\frac{
\text{mc2}}{\text{pplq2}-\text{q2}}\text{Icdc}\,.
\end{eqnarray}
Turning to  the tensor integral for diagram (b):
\begin{eqnarray}
I^{\alpha\beta}_{cdc}(p,q)\equiv 
i\int \frac{d^4f}{(2\pi)^4} 
\frac{f^\alpha f^\beta}{\big[(f+p+q)^2-m_c^2\big]f^2\big[(f+p)^2-m_c^2\big]}\,,
\label{eq:intfalphabetab}
\end{eqnarray}
we expand it similar to (\ref{eq:Icddalbet_exp}), 
\begin{eqnarray}
I_{cdc}^{\alpha\beta}= \big(p^\alpha p^\beta) ~I^{(pp)}_{cdc}+
\big(q^\alpha q^\beta)~I^{(qq)}_{cdc} + 
(q^\alpha p^\beta+q^\beta p^\alpha)I^{(qp)}_{cdc}
+g^{\alpha\beta}\,I^{(g)}_{cdc}
\,,
\label{eq:Icdcalbet_exp}
\end{eqnarray}
where:
\begin{eqnarray}
I^{(pp)}_{cdc}=&& \frac{m_c^2\left(5(q^2)^2+2q^2s-s^2\right)+
2s\left(-2(q^2)^2+q^2s+s^2\right)}{(s-q^2)^4} I_{cd}
\nonumber\\ &&+\frac{6(q^2)^2 m_c^4+ 6 q^2 s m_c^2(s-q^2)+s^2(s-q^2)^2}{(s-q^2)^4}I_{cdc}
+\frac{2\left((q^2)^2+4q^2s+s^2\right)}{(s-q^2)^4}
I_{cdfp}\,;
\label{Ippcdc}
\end{eqnarray}

\begin{eqnarray}
I^{(qq)}_{cdc}=&& \frac{m_c^2}{(s-q^2)^2}I_{cd}
+\frac{m_c^4}{(s-q^2)^2}I_{cdc}  + 
\frac{2}{(s-q^2)^2}I_{cdfp}\,:
\label{eq:Iqqcdc}
\end{eqnarray}

\begin{eqnarray}
I^{(qp)}_{cdc}=\frac{3q^2m_c^2+2s(s-q^2)}{(q^2-s)^3}I_{cd}
+\frac{2sm_c^2(s-q^2)+3q^2m_c^4}{(q^2-s)^3}I_{cdc}
+\frac{2(q^2+2s)}{(q^2-s)^3}I_{cdfp}\,;
\end{eqnarray}
\begin{eqnarray}
I^{(g)}_{cdc}=&& \frac{q^2m_c^2+s(s-q^2)}{2(s-q^2)^2}
I_{cd}
+\frac{s m_c^2(s-q^2)+q^2m_c^4}{2(s-q^2)^2}I_{cdc}
+\frac{s}{(s-q^2)^2}I_{cdfp}\,.
\label{eq:Igcdc}
\end{eqnarray}
In order to derive the above equations we used:
\begin{eqnarray}
&&  g_{\mu\nu} I_{cdc}^{\mu\nu} =0\,,
\nonumber\\ 
&&p_{\mu}p_{\nu} I_{cdc}^{\mu\nu}  = \frac{1}{4}\big[m_c^2 I_{cd} + m_c^4 I_{cdc}+2I_{cdfp}\big]\,,
\nonumber\\
&&q_{\mu}q_{\nu}  I_{cdc}^{\mu\nu}  = \frac{1}{4}\big[(2s -m_c^2)I_{cd}+ s^2I_{cdc}+2I_{cdfp}\big]\,,
\nonumber\\
&&p_{\mu}q_{\nu} I_{cdc}^{\mu\nu}  = \frac{1}{4}\big[-m_c^2sI_{cdc}-s I_{cd}
-2I_{cdfp}\big]\,.
\nonumber
\end{eqnarray}
In addition, for the diagrams (c) and (d), we need a two-point loop vector integral:
\begin{eqnarray}
I^\alpha_{cd}(p,q)\equiv 
i\int \frac{d^4f}{(2\pi)^4} 
\frac{f^\alpha}{\big[(f+p+q)^2-m_c^2\big]f^2}\,,
\label{eq:intfalphacd}
\end{eqnarray}
which depends on the external momentum $p+q$  and is easily reduced to
\begin{eqnarray}
I^\alpha_{cd}=-(p+q)^\alpha\, \frac{s-m_c^2}{2s}I_{cd}\,.
\label{eq:intfalphacd1}
\end{eqnarray}

\subsection{Resulting expressions for the annihilation diagrams}
\label{subapp:A4}
The full expression  for the diagram (a) expressed in terms of the integrals
defined in the previous subsections reads:
\begin{eqnarray} 
&&{\cal F}^{(a)\mu}(p,q,k)=(3C_1Q_d m_c^2 f_\pi)
\nonumber\\
&&\times\Big[-2(p-k)_\beta\,I_{cdd}^{\mu\beta}
- (q\cdot(p-k))\, I_{cdd}^{\mu}
- q^\mu (p-k)_\beta\, I_{cdd}^{\beta} + p^\mu (I_{cdd\,\beta}^{\beta} +
 q_\beta\, I_{cdd}^{\beta})\Big]
\,.
\label{eq:diagaInt1}
\end{eqnarray}
Retaining only the integrals contributing to the imaginary part in $(p+q)^2=s$, and reducing them to scalar integrals, we obtain for the invariant amplitude
multiplying $(-q^2)p_\mu$ and thus (according to (\ref{eq:expan})) relevant for our sum rule
\footnote{Note that apart from the  conditions $p^2=k^2=0$ and $(p-k)^2=m_\pi^2$, we also use the relation  
$  q\cdot k= \frac12\Big((p+q)^2-P^2+m_\pi^2\Big)$
following from the definition $P^2=(p+q-k)^2$.}:
\begin{eqnarray}
F^{(a)}((p+q)^2,q^2,P^2)=&&\frac{(3C_1Q_d m_c^2 f_\pi)}{q^2}
\Big[m_\pi^2\,I_{cdd}^{(pp)}+ \big(P^2-m_\pi^2-q^2\big)\,I_{cdd}^{(qp)}+2\,I_{cdd}^{(g)}
\nonumber\\
&&+\frac12\big(P^2-m_\pi^2-(p+q)^2\big)I_{cdd}^{(p)} -q^2\,I_{cdd}^{(q)}\Big]\,.
\label{eq:diagaIntfin}
\end{eqnarray}
We then use the expressions for $I_{cdd}^{(pp,qp,g,p,q)}$ 
given in Section \ref{sec:Areduction} in terms of 
scalar integrals. In these expressions, at $q^2<0$ it is possible to replace 
the variable $(p+q)^2 \to s$, and the same can be done in the above equation.
After that, representing the whole amplitude in the dispersion form 
is straightforward, having at hand the scalar
integrals written in the  dispersion form  in Section~\ref{app:scaldisp}. We finally obtain the  spectral density
\begin{eqnarray}
\frac1\pi \mbox{Im} F^{(a)}(s,q^2,P^2)
=&&(3C_1Q_d m_c^2 f_\pi)\Bigg\{
\frac{-\rho_{cd}(s)}{2 q^2(\text{pplq2}-\text{q2})^4}
\Big[4 \text{mc2} \text{mpi}^2 \left(s^2+4
    \text{pplq2} \text{q2}+(\text{q2})^2\right)
\nonumber\\    
&&+2 \text{mc2}
    (\text{pplq2}-\text{q2}) \left(-2 \text{P2} (\text{pplq2}+2
    \text{q2})+\text{pplq2}^2+2 \text{pplq2} \text{q2}+3
    (\text{q2})^2\right)-\left(s^3 \left(3 \text{mpi}^2-3
    \text{P2}+\text{pplq2}\right)\right)
\nonumber\\    
 &&   +s^2 \text{q2} \left(-11
    \text{mpi}^2+\text{P2}-3 \text{pplq2}\right)+(\text{q2})^3
    \left(\text{mpi}^2+\text{P2}+\text{pplq2}\right)+\text{pplq2} (\text{q2})^2
    \left(\text{mpi}^2-5 \text{P2}+3 \text{pplq2}\right)\Big]
\nonumber\\    
 &&   -\frac{\rho_{cdd}(s,q^2)}{2(\text{pplq2}-\text{q2})^4}
 \Big[-2
    m_c^4 \left(3 \text{mpi}^2
    (\text{pplq2}+\text{q2})-(\text{pplq2}-\text{q2}) (3
    \text{P2}-\text{pplq2}-2 \text{q2})\right)
\nonumber\\    
    &&+2 \text{mc2}
    (\text{pplq2}-\text{q2}) \left(3 \text{mpi}^2
    (\text{pplq2}+\text{q2})-(\text{pplq2}-\text{q2}) (3
    \text{P2}-\text{pplq2}-2 \text{q2})\right)
    \nonumber\\
   && +(\text{pplq2}-\text{q2})^2
    \left((\text{P2}-\text{pplq2}) (\text{pplq2}-\text{q2})-\text{mpi}^2
    (\text{pplq2}+\text{q2})\right)\Big]
\nonumber\\    
   && -\frac{2\rho_{cdfq}(s,q^2)}{q^2(\text{pplq2}-\text{q2})^4} \Big[-3 \text{mpi}^2
    \text{pplq2} (\text{pplq2}+\text{q2})+\text{P2} (\text{pplq2}-\text{q2}) (2
    \text{pplq2}+\text{q2})-\text{pplq2}^3+(\text{q2})^3\Big]\Bigg\}\,,
\label{eq:ImFa}
\end{eqnarray}
where we isolate the parts originating from the scalar integrals
\begin{eqnarray}
&&\rho_{cd}(s)
=-\frac{1}{16\pi^2}\frac{(s-m_c^2)}{s}\,;
\nonumber\\
&&\rho_{cdd}(s,q^2)=-\frac{1}{16\pi^2}
\frac{1}{(s-q^2)}
\log\Bigg[\frac{-q^2m_c^2}{s(-q^2+s-m_c^2)}\Bigg]\,;
\nonumber\\
&&\rho_{cdfq}(s,q^2)
=\frac{1}{64\pi^2}
\frac{ (s-m_c^2)^2(s+q^2)}{s^2}\,.
\nonumber
\end{eqnarray}
For the diagram (b) we obtain, with a similar procedure:
\begin{eqnarray}
F^{(b)}((p+q)^2,q^2,P^2)=&&\frac{(3C_1Q_c\, m_c^2 f_\pi)}{q^2}
\Big[m_\pi^2\,I_{cdc}^{(pp)}+ \big(P^2-m_\pi^2-q^2\big)\,I_{cdc}^{(qp)}+2\,I_{cdc}^{(g)}
\nonumber\\
&&+\frac12\big((p+q)^2 -P^2+3m_\pi^2\big)I_{cdc}^{(p)} +\,I_{cdc}^{(q)}(P^2-m_\pi^2)\Big]\,,
\label{eq:diagbIntfin}
\end{eqnarray}
with the  spectral density
\begin{eqnarray}
\frac1\pi \mbox{Im} F^{(b)}(s,q^2,P^2)
=&&3C_1 Q_c m_c^2f_\pi\Big[
\frac{ \rho_{cd}(s) }{2q^2 (s-\text{q2})^4}
\Big[2m_c^2\big\{q^2(s-q^2)\left(s+2q^2-3P^2\right)
\nonumber\\
&& -m_\pi^2\left(s^2-5q^2s-2(q^2)^2\right)\big\}
    \nonumber\\
&&  +(s-q^2)\left\{(s-P^2)(s^2-(q^2)^2)+m_\pi^2\left(3s^2+8q^2s+(q^2)^2\right)\right\} \Big]
     \nonumber\\
 &&   +\frac{\rho_{cdc}(s,q^2)}{2q^2(s-\text{q2})^4}
    \Big[2q^2m_c^4\left\{3m_\pi^2(s+q^2)+(s-q^2)\left(s+2q^2-3P^2\right)\right\}
    \nonumber\\
   && +2m_c^2(s-q^2)\left\{s(s-P^2)(s-q^2)+m_\pi^2(s+q^2)(s+2q^2)\right\} \nonumber\\
 &&-s(s-q^2)^2\left\{m_\pi^2(s-3q^2)+(s-P^2)(s-q^2)\right\}\Big]
    \nonumber\\
 &&   +\frac{2 \rho_{cdfp}(s,q^2)}{q^2(s-\text{q2})^4}  \Big[3m_\pi^2s(s+q^2)-P^2\left(2s^2-q^2s-(q^2)^2\right)\!-\!(q^2)^3\!+\!s^3
 \Big]\Big]\,,
\label{eq:ImFb}
\end{eqnarray}
where
\begin{eqnarray}
&&\rho_{cdc}(s,q^2)=-\frac{1}{16\pi^2}
\frac{1}{(s-q^2)}
\log\frac{m_c^2s}{s(s-q^2)+m_c^2q^2
}\,,
\nonumber\\
&&\rho_{cdfp}(s,q^2)
=\frac{1}{64\pi^2}
\frac{ (s-m_c^2)^2(s-q^2)}{s^2}\,.
\nonumber
\end{eqnarray}
Turning to the sum of diagrams (c) and (d) given by (\ref{eq:diagcdTr}), we notice that the part with the twist-2 pion DA at LO 
of the correlation function, due to the additional four-momentum does not 
generate an additional imaginary part in the variable $(p+q)^2=s$.
The two-point loop part of this diagram is reduced to the integral (\ref{eq:intfalphacd}). 
In the dispersion form, the complete answer for the sum of $(c)$ and $(d)$ contributions to the invariant amplitude reads:

\begin{eqnarray}
&&F^{(c)\oplus(d)}((p+q)^2,q^2,P^2)= -\frac34 C_1 (Q_u-Q_d) m_c^2f_\pi \frac{m_c^2-(p+q)^2}{q^2(p+q)^2}I_{cd}
\nonumber\\
&&\times \int\limits_0^1 du \frac{\varphi_\pi(u)}{u\,P^2 +\bar{u}\,q^2-u\bar{u}m_\pi^2}
  \left[\text{mpi}^2 (2 u-1)+2 \text{P2}
    u+\text{P2}-\text{pplq2}-2 \text{q2} (u+1)\right]\,,
\label{eq:resFcd}
\end{eqnarray}
with the spectral density
\begin{eqnarray}
\frac1\pi \mbox{Im} F^{(c\,\oplus\, d)}(s,q^2,P^2)=&&-\frac34 C_1 (Q_u-Q_d) m_c^2f_\pi \frac{\rho_{cd}(s)}{q^2s} (s-m_c^2)
\nonumber\\
&&\times \int\limits_0^1 du\, \varphi_\pi(u)\frac{m_\pi^2(1-2u)-P^2(2u+1)+2q^2(u+1)+s}{u\,P^2 +\bar{u}\,q^2-u\bar{u}m_\pi^2} \,.
\label{eq:ImFcd}
\end{eqnarray}
Numerical integration of the above spectral function 
in the LCSR (5.10) after analytic continuation $P^2\to m_D^2$  
deserves comments. First of all, we notice that
the integration 
over $s$ with Borel exponent factorizes from the integration 
over $u$ with the pion DA.  Introducing two integrals that are independent of $q^2$: 
$$I_n(M^2)=\int\limits_{m_c^2}^{s_0^D} ds e^{-s/M^2}\rho_{cd}(s)(s-m_c^2)s^{n-1} ~~~ (n=0,1\,),$$ 
we obtain, adopting as before the chiral limit $m_\pi=0$:
\begin{eqnarray}
\frac1\pi \int 
\limits_{m_c^2}^{s_0^D}\!\! ds e^{-s/M^2}
\mbox{Im} F^{(c\,\oplus\, d)}(s,q^2,m_D^2)=
\frac34 C_1 (Q_u-Q_d) m_c^2f_\pi 
\big[I_0(M^2)J_0(q^2)+I_1(M^2)J_1(q^2)\big],
\label{eq:intImFcd}
\end{eqnarray}
where the integrals over $u$ are
\begin{eqnarray}
J_n(q^2)=\frac{1}{(m_D^2-q^2)(-q^2)  }\int\limits_0^1 du\, \varphi_\pi(u)\frac{[-m_D^2(2u+1)+2q^2(u+1)]^{1-n}}{
(u-u_*)}\,, 
\label{eq:intcd}
\end{eqnarray}
with $$~~ u_*\equiv\frac{-q^2}{m_D^2-q^2}\,.$$
The integrand in (\ref{eq:intcd})  has a pole at $u=u_*$ within the integration region, since $0<u_*<1$ at $q^2<0$.
 The origin of this pole is traced to the additional imaginary part emerging after 
 analytic continuation $P^2\to m_D^2$ of the initial integral  
 (\ref{eq:ImFcd}). Considering this integral as a function $I(P^2)$ of $P^2$, we have  
 $\mbox{Im}_{P^2}I(P^2)=1/(2i)[I(P^2+i\epsilon)-I(P^2-i\epsilon)]$. 
  Accordingly, in (\ref{eq:intcd}), we use the formula 
$$\frac1{u-u_*+i\epsilon}= PV\frac1{u-u_*}-i\pi\delta(u-u_*)\,,$$ 
yielding 
\begin{eqnarray}
J_n(q^2)=&&\frac{1}{(m_D^2-q^2)(-q^2)}
\Bigg [ PV\left\{\int\limits_0^1 du\, \varphi_\pi(u)\frac{[-m_D^2(2u+1)+2q^2(u+1)]^{1-n}}{
(u-u_*)}\right \} 
\nonumber
\\
&&-i\pi\varphi_\pi(u_*)\bigg(-m_D^2(2u_*+1)+2q^2(u_*+1)\bigg)^{1-n}
\Bigg]
\,.
\label{eq:intcd1}
\end{eqnarray}

The imaginary part  emerging after the transition $P^2\to m_D^2$   
is interpreted as the duality counterpart of the 
strong phase generated by the lighter than $D$ intermediate on-shell states in $D\to \pi\gamma^*$ amplitude that are present also at $q^2<0$.

Finally, to derive the asymptotic behaviour of the $D^+\to \pi^+\gamma^*$ amplitude at $q^2\to -\infty$,
we calculate in this limit the contributions of separate diagrams. For consistency, the pion DA is also taken in the asymptotic form. 
For the diagrams (a) and (b), we conclude, respectively, from (\ref{eq:ImFa}) and (\ref{eq:ImFb}), that they both  behave $\sim 1/q^2$ with equal coefficients, differing only by the quark charge factors. We obtain the asymptotic limit of the sum of (a) and (b):
\begin{eqnarray}
\Big[{\cal A}^{(D^+ \pi^+\gamma^*)}(q^2\to -\infty)\Big]^{(a)\oplus (b)}
= \frac{1}{\pi m_D^2f_D}\int 
\limits_{m_c^2}^{s_0^D} ds \,e^{(m_D^2-s)/M^2}\mbox{Im} F^{(a)\oplus (b)}(s,q^2\to -\infty,m_D^2)
\nonumber\\
=\left(\frac{1}{q^2}\right)\frac{3C_1(Q_c-Q_d) m_c^2 f_\pi}{32\pi^2 m_D^2f_D }\int 
\limits_{m_c^2}^{s_0^D} ds \,e^{(m_D^2-s)/M^2}\frac{(s-m_c^2)^2}{s^2} 
+O\left(\frac{1}{(q^2)^2}\right)\,.
\label{eq:q2asympt_ab}
\end{eqnarray}
For the sum of the diagrams $(c)$ and $(d)$, the $\sim 1/q^2$ asymptotics 
is provided by the real part of the integral $J_0(q^2)$ in (\ref{eq:intcd1}): 
$$\mbox{Re}[J_0(q^2\to -\infty)]=-\frac{10}{q^2} +O\left(\frac{1}{(q^2)^2}\right),
$$ 
whereas its imaginary part in this limit 
is suppressed by an additional  power of $1/q^2$. 
Also the asymptotics of $J_1(q^2)$ starts at $1/(q^2)^2$.
 Using the above limit, we obtain from (\ref{eq:intImFcd}):
\begin{eqnarray}
\Big[{\cal A}^{(D^+ \pi^+\gamma^*)}(q^2\to -\infty)\Big]^{(c\,\oplus\, d)}
= \frac{1}{\pi m_D^2f_D}\int 
\limits_{m_c^2}^{s_0^D} ds \,e^{(m_D^2-s)/M^2}\mbox{Im} F^{(c)\oplus (d)}(s,q^2\to -\infty,m_D^2)
\nonumber\\
=
\left(\frac{1}{q^2}\right)
\frac{15 C_1(Q_u-Q_d) m_c^2 f_\pi}{32\pi^2 m_D^2f_D }\int 
\limits_{m_c^2}^{s_0^D} ds \,e^{-s/M^2}\frac{(s-m_c^2)^2}{s^2} 
+O\left(\frac{1}{(q^2)^2}\right)\,.
\label{eq:q2asympt_cd}
\end{eqnarray}

\section{Loop topology diagrams for $D^+\to \pi^+\gamma^*$}
\label{app:diagL}
The LO diagrams with the loop topology depicted in Fig.~\ref{fig:lcsr_loop}
differ from the annihilation topology diagrams by their underlying effective operators.
 Instead of $O^d_1$, there are  two four-quark operators which we denote as
\begin{equation}
O^{{\cal D}{\cal D}}\equiv \big(\bar{{\cal D}}_L\gamma_\rho {\cal D}_L\big)
\big(\bar{u}_L\gamma_\rho c_L\big)\,,
~~~~{\cal D}=d,s \,.
\label{eq:qqoper}
\end{equation}
Their common Wilson coefficient 
\begin{equation}
\frac13\left(C_1+\frac43 C_2\right)
\label{eq:C12comb}
\end{equation}
is obtained by applying Fierz transformations 
to the combination $C_1\,O_1^{{\cal D}}+ C_2\,O_2^{{\cal D}}   $ entering the 
effective Hamiltonian.
First, we use 
$$O_2^{\cal D}=\frac12\big(O^{{\cal D}{\cal D}}-\frac13 O_1^{\cal D}\big)\,,$$  
and then 
$$O_1^{\cal D}=\frac 13 O^{{\cal D}{\cal D}} +2
\left(\bar{{\cal D}}_L\gamma_\rho t^a{\cal D}_L\right)\big(\bar{u}_L\gamma_\rho t^ac_L\big)\,,$$
noticing that the product of colour-octet operators does not contribute at LO.

Substituting the operators (\ref{eq:qqoper}) and the coefficient (\ref{eq:C12comb}) in the correlation function (\ref{eq:corr}),
we obtain for the sum of $d$- and $s$-loop diagrams the following 
factorized expression written in a compact form: 
\begin{eqnarray}
{\cal F}^{(L)}_\mu(p,q,k)=
\frac13\left(C_1+\frac43 C_2\right)\Big[Q_d\Pi^d_{\mu\rho}(q)-
Q_s\Pi^s_{\mu\rho}(q)\Big ] G^\rho(p,q,k)\,, 
\label{eq:loopF}
\end{eqnarray}
where
\begin{eqnarray}
\Pi^{{\cal D}}_{\mu\rho}(q)=\frac{i}{2}
\int \! d^4x \, e^{iq\cdot x}
\langle 0| T\big\{ \left(\bar{{\cal D}}(x)\gamma_\mu {\cal D}(x)\right) 
\left(\bar{{\cal D}}(0)\gamma_\rho {\cal D}(0)
\right)\big\}|0\rangle\,, ~~ {\cal D}=d,s\,,
\label{eq:Pivac}
\end{eqnarray}
and
\begin{eqnarray}
\nonumber\\
G_\rho(p,q,k)=i\int \! d^4y \, e^{-i(p+q)\cdot y}\langle \pi^+(p-k)|T\Big\{
\big(\bar{u}_L(0)\gamma_\rho c_L(0)\big)
j_5^D(y)\Big\}|0\rangle\,.
\label{eq:Tcorr}
\end{eqnarray}
Note that in (\ref{eq:loopF}), a complete GIM cancellation does not take place due to the mass difference 
between $s$ and $d$ quarks, which is a calculable effect in our approach.

The two-point correlation function (\ref{eq:Pivac}) yields after integration 
in $D\neq 4$ dimensions:
\begin{eqnarray}
&&\Pi^{{\cal D}}_{\mu\rho}(q)=(q_\mu q_\rho-g_{\mu\rho}q^2)\Pi^{{\cal D}}(q^2)\,,
~~~({\cal D}=d,s )
\nonumber\\
&&\Pi^{{\cal D}}(q^2)=\frac{12 \,\Gamma(2-\frac{D}{2}) }{(16\pi^2)^{\frac{D}{4}}}\int\limits_0^1
dx\, x\,(1-x)\big[m_{\cal D}^2-q^2x(1-x)\big]^{\frac{D}{2}-2}.
\label{eq:Pi}
\end{eqnarray}
The difference of $d$ and $s$-quark loops entering (\ref{eq:loopF}) 
is finite, yielding at $D=4$:
\begin{eqnarray}
\Pi^{d}(q^2)-\Pi^{s}(q^2)\equiv \Pi^{(d-s)}(q^2)=\frac{3}{4\pi^2}
\int\limits_0^1 dx\, x\,(1-x) \log\bigg(\frac{m_s^2-q^2x(1-x)}{m_d^2-q^2x(1-x)}\bigg)\,.
\label{eq:dsloop}
\end{eqnarray}

The remaining part of the loop-topology diagrams is reduced to the 
vacuum-to-pion correlation function (\ref{eq:Tcorr}) of the weak effective Hamiltonian and the interpolating current. Note that up to a replacement $u\to d$ 
and at $k\to 0$, this part of the diagrams coincides with 
the correlation function used to obtain the LCSR  for $D\to \pi$ semileptonic form factors. A  detailed  analysis of this sum rule can be found in \cite{Khodjamirian:2009ys}. Hence, the computation follows the same routine.
At LO, after contracting  the $c$-quark fields into a free propagator, we insert the pion DA in the leading twist-2 approximation. For the invariant amplitude multiplying $p_\rho$  
which is relevant for LCSR:
\begin{eqnarray}
G_\rho(p,q,k)= p_\rho G((p+q)^2,q^2,P^2)+ \dots \,, 
\nonumber\\
\end{eqnarray}
we obtain
\begin{eqnarray}
G((p+q)^2,q^2,P^2)=\frac12 f_\pi(m_c+m_d)m_c
\int\limits_0^1 du
\frac{\varphi_\pi(u)}{\bar{u}P^2 -\bar{u}q^2+m_c^2-(p+q)^2}\,.
\label{eq:Tcorr1}
\end{eqnarray}

The contribution of the loop topology diagrams to the relevant Lorentz structure of correlation 
function becomes:
\begin{eqnarray}
{\cal F}^{(L)}_\mu(p,q,k)=-\big[ (p\cdot q) q_\mu -q^2 p_\mu \big]
\frac19\left(C_1+\frac43 C_2\right)\Pi^{(d-s)}(q^2)G((p+q)^2,q^2,P^2)\,, 
\label{eq:loopF2}
\end{eqnarray}
so that the full expression for the invariant amplitude reads:
\begin{eqnarray}
F^{(L)} ((p+q)^2,q^2,P^2) &&=
-\frac{1}{18}\left(C_1+\frac43 C_2\right)\Pi^{(d-s)}(q^2)
\nonumber\\
&&\times f_\pi(m_c+m_d)m_c
\int\limits_0^1 du
\frac{\varphi_\pi(u)}{\bar{u}P^2 -\bar{u}q^2+m_c^2-(p+q)^2}\,. 
\label{eq:loopF3}
\end{eqnarray}

For the $D^+\to \pi^+ \gamma^*$ transition we use this expression at $P^2=m_D^2$ 
and in the chiral limit 
$m_\pi^2=0$, $m_d=0$.
Instead of taking the imaginary part of (\ref{eq:loopF3}) in the variable $(p+q)^2$ directly, it is more convenient to transform the integral over the variable $u$ into a dispersion integral form using the following transformation of the integration variable:
$$ u \to s= \bar{u}(m_D^2-q^2)+m_c^2, ~~~~ u=1-\frac{s-m_c^2}{m_D^2-q^2}$$
After applying duality and Borel transformation 
in $(p+q)^2$, it is also more convenient to return to the integration 
variable $u$. We obtain the final expression for the 
contribution  of the loop-topology diagrams to LCSR: 
\begin{eqnarray}
\frac{1}{\pi}\int\limits_{m_c^2}^{s_0^D} \!ds \,e^{-s/M^2}
\mbox{Im} F^{(L)}(s,q^2,m_D^2)= &&
-\frac{1}{18}\left(C_1+\frac43 C_2\right)
f_\pi(m_c+m_d)m_c\Pi^{(d-s)}(q^2)
\nonumber\\
&&\times \int\limits _{u^D_0}^1 du\, \varphi_\pi(u)
\exp\bigg[-\frac{\bar{u}(m_D^2-q^2)+m_c^2}{M^2}\bigg]\,,
\label{eq:ImFL}
\end{eqnarray}
where $$u^D_0=1-\frac{s_0^D-m_c^2}{m_D^2-q^2}\,.$$

Note that we  derived the loop-topology contribution
to LCSR in the same twist-2 approximation for pion DAs
 as for the annihilation-topology contribution. 
Opposite to the latter, the twist-3 pion DAs 
neglected in (\ref{eq:Tcorr1}) play a larger role
and should be taken into account in future improvements
of this sum rule.
That will, however, not significantly influence the overall accuracy of LCSR, 
since, as our numerical analysis has revealed, the sum of $d-s$ loop-topology diagrams at $q^2<0$ is 
numerically very small with respect to the dominant annihilation-topology diagrams.

\section{Annihilation topology 
diagrams for Cabibbo favoured transitions}
\label{app:CF}

Here, we discuss in detail the modifications in the diagrams
and in the resulting OPE spectral densities, which are necessary to obtain the LCSRs for the two CF modes:

\subsection*{ The $D_s^+\to\pi^+\gamma^*$ transition}

To compute the correlation function (\ref{eq:corrDs}) for  this transition, 
we only have to replace in all  four diagrams in Fig.~\ref{fig:lcsr_ann}
the virtual $d$-quark in the loop by the $s$-quark,
and, correspondingly, replace $m_d\to m_s$ in the initial
expressions for these diagrams, given in the equations 
(\ref{eq:diagainit}),(\ref{eq:diagbinit}),
(\ref{eq:diagcinit}) and (\ref{eq:diagdinit}),
Since we retain a nonvanishing $s$-quark mass in this
case, the OPE spectral functions
of annihilation diagrams will acquire terms with $m_s$. The final 
 expressions replacing 
  (\ref{eq:ImFa}),  (\ref{eq:ImFb}) and
    (\ref{eq:ImFcd}),  are  cumbersome, and we do not 
present them here~\footnote{The Mathematica notebook of these 
 expressions can be provided by the authors on request.}.

\subsection*{ The $D^0\to\bar{K}^0\gamma^*$ transition}

For this transition, we compute the correlation 
function (\ref{eq:corrD0}), which differs from the one for the $D^+\to\pi^+\gamma^*$ transition
in several aspects. First of all, we encounter a different 
combination of Wilson coefficients in the factorizable
approximation of the annihilation topology diagrams. Instead of $C_1$, one has the same linear combination (\ref{eq:C12comb})
of $C_1$ and $C_2$, as in the  loop-topology diagrams. Consequently, there is a strong numerical suppression
of the OPE result. 

Furthermore, the diagrams  for $D^0\to\bar{K}^0\gamma^*$  are obtained, replacing the quark flavours and, correspondingly, the quark charge factors in  the Fig.~\ref{fig:lcsr_ann} diagrams. The loop parts of all annihilation diagrams 
remain massless in the adopted limit $m_u=m_d=0$, hence we can use the expressions 
for the loop integrals obtained in Appendix~\ref{app:diagA}. 
In addition, the charged pion  hadronic matrix elements are replaced by the neutral kaon ones. 
For the diagrams in Fig.~\ref{fig:lcsr_ann}(a) and (b), 
we use the $\bar{K}^0$ decay constant defined as:
\begin{equation}
\langle \bar{K}^0(p-k)|\bar{s} \gamma_\rho \gamma_5 d|0\rangle=-if_{K}(p-k)_\rho\,,
\end{equation}
which is equal to the charged kaon decay constant in the isospin symmetry limit that we adopt.
Finally, the kaon DA of twist-2 replacing the pion DA in the Fig.~\ref{fig:lcsr_ann}(c) and (d) diagrams, 
is defined as:
\begin{eqnarray}
 \langle \bar{K}^0(p-k)|\bar{s}_\omega^i(x_1) 
d^j_\xi(x_2)|0\rangle
  &=& \frac{i\delta^{ij}}{12}f_K 
\int_0^1 du~e^{iu (p-k)\cdot x_1+ i\bar{u} (p-k)\cdot x_2   }
 [(\slashed{p}-\slashed{k})\gamma_5]_{\xi\omega} \varphi_K(u).
\label{eq:tw2K}
\end{eqnarray}
For this DA, we adopt the ansatz with two Gegenbauer moments,  the same approximation as
e.g., in the LCSR for $D\to K$ form factors in \cite{Khodjamirian:2009ys}:
\begin{eqnarray}
\varphi_K(u,\mu)=
\underbrace{a_1^K(\mu)\bigg[6u(1-u)C_1^{3/2}(u-\bar{u})\bigg]}_{\varphi_K^{(asym)}(u)}\
+\underbrace{6u(1-u)\bigg[1+a_2^K(\mu)C_2^{3/2}(u-
\bar{u})\bigg]}_{\varphi_K^{(sym)}(u)}  \,,
\label{eq:KDAmod}
\end{eqnarray}
where we separate the antisymmetric and the symmetric parts with respect to the $u\to\bar{u}$ transformation. 
Note that we only retain the effects of $O(m_s)\sim O(m_K^2)$ in OPE diagrams. In particular,   
in the diagram analogous to Fig.~\ref{fig:lcsr_ann}(d),
we neglect the term $m_s^2$ in the  propagator of the virtual $s$-quark, because this term is much smaller
than a characteristic virtuality. A similar approximation
was used in the LCSR calculation of the kaon e.m. form factor (see e.g. \cite{Bijnens:2002mg}).

The input values for the kaon decay  constant and  
Gegenbauer moments are given in Tables~\ref{tab:inpPDG},
\ref{tab:inpLCSR}. Note that the $SU(3)_{fl}$ 
violation effects are caused by $f_K\neq f_\pi$ and by the appearance of antisymmetry with respect to 
$u \to \bar{u} $ 
in the kaon DA (\ref{eq:pionDAmod}), due  to the first Gegenbauer moment 
$a_1^K\neq 0$.

To summarize, the following replacements are made 
in the expressions (\ref{eq:ImFa}),  (\ref{eq:ImFb}),
 (\ref{eq:ImFcd}) for the OPE spectral density 
 to obtain the corresponding expressions for the 
 $D^0\to \bar{K}^0\gamma^*$ transition: 
\begin{center}
\begin{tabular}{|c|c|}
\hline
 Equation  & Replacement \\
 \hline
(\ref{eq:ImFa}),  (\ref{eq:ImFb}), (\ref{eq:ImFcd}) 
 & $C_1\to \frac{1}{3}\Big(C_1+\frac{4}{3}C_2\Big)$, ~
 $f_\pi\to f_K$,~ $m_\pi\to m_K$ \\  
\hline
(\ref{eq:ImFa})& $Q_d\to Q_u$\\
\hline
(\ref{eq:ImFcd})& $(Q_u-Q_d)\to (Q_s+Q_d) $ ~~$\varphi_\pi(u)\to \varphi_K^{(asym)}(u)$\\
\hline
\end{tabular}
\end{center}
In addition, there are minor modifications in the final expression (\ref{eq:intImFcd}) for the sum of the contributions of the diagrams (c) and (d).
We take into account that the position of the pole in the u-integration region 
is shifted due to the factor proportional to $m_K^2$ that appears in the denominator.
For completeness, we write down the transformed expression
\begin{eqnarray}
\frac1\pi \int 
\limits_{m_c^2}^{s_0^D} ds e^{-s/M^2}
\Big[\mbox{Im} F^{(c\,\oplus\, d)}(s,q^2,m_D^2)\Big]^{(D^0\to \bar{K}^0\gamma^*)}&=&
\frac14 (C_1 +\frac{4}{3}C_2)(Q_s+Q_d) m_c^2f_K
\nonumber\\
&\times&\bigg[I_0(M^2)J_0(q^2)+I_1(M^2)J_1(q^2)\bigg]\,,
\label{eq:intImFcdD0KO}
\end{eqnarray}
\\
where the integrals  $I_{0,1}(M^2)$ are the same as in (\ref{eq:intImFcd}) , and the integrals 
over $u$ become:
\begin{eqnarray}
J_n(q^2)&=&\frac{2}{(-q^2)}\int\limits_0^1\!\! du\, 
\frac{\varphi_K^{(asym)}(u)\big[-m_D^2(2u+1)+2q^2(u+1)+m_K^2(1-2u)\big]^{1-n}}{(u-u_*)
\big[m_K^2(2u-1)+(m_D^2-q^2)+\sqrt{(m_D^2-q^2-m_K^2)^2-4m_K^2q^2}\big]}, 
\label{eq:intcdD0K0}
\end{eqnarray}
where the position of the pole in the integrand is at 
\begin{equation}
u_*= \frac{-(m_D^2-q^2-m_K^2)+\sqrt{(m_D^2-q^2-m_K^2)^2-4m_K^2q^2}}{2m_K^2}\,.
\nonumber
\end{equation}
Note that at $q^2<0$ the second pole from the expression in brackets in the denominator of (\ref{eq:intcdD0K0}) lies outside the $u$-integration region.

\section{Parameters of the $\rho'$ and $\phi'$ 
resonances}
\label{app:Vexc}
Here we explain how the residues ($r_{\rho'}$ and 
$r_{\phi'}$) of the excited vector mesons 
entering the dispersion relation (\ref{eq:dispAexc}) 
are obtained. 

Considering first  the $\rho'$ residue, we notice that
there are  no separate data on the  branching fraction of the weak decay $D^+\to\pi^+\rho^{\,' 0}$. This channel was most recently 
extracted \cite{LHCb:2022lja} from the Dalitz-plot analysis of the 
three-body $D^+\to\pi^+ \pi^+\pi^-$ decay.
The outcome of that fit taken from \cite{PDG} 
and presented in Table~\ref{tab:Vexc} 
is a product of the $BR(D^+\to\pi^+\rho^{\,' 0})$ and 
$BR(\rho^{\,'0}\to \pi^+\pi^-)$.
The amplitude of the latter decay is determined by 
the invariant $\rho' \pi\pi$ strong coupling:
\begin{eqnarray}
 A\big(\rho^{\,' 0}(q)\to\pi^+(q^+)\pi^-(q^-)\big)=g_{\rho'\pi\pi}~
 \epsilon_{\rho'}\cdot(q^+-q^-)\,,
\label{eq:VpipiA} 
\end{eqnarray}
so that the branching fraction is
\begin{equation}
BR(\rho^{\,'0}\to\pi^+\pi^-)=\frac{m_{\rho'}[\beta_\pi(m_{\rho'}^2)]^{3}}{48\pi
\Gamma_{\rho'}}|g_{\rho'\pi\pi}|^2\,.
\label{eq:VpipiBR}
\end{equation}
An indirect experimental information on
the $g_{\rho'\pi\pi}$ coupling  is available, e.g., 
from the analysis of the pion timelike e.m. form factor. 
This coupling enters the residue  of the $\rho'$ contribution
in the form-factor dispersion relation. Note, however, that the latter residue also contains the decay constant $f_{\rho'}$ defined similar to (\ref{eq:fV}),
but not directly measured, as opposed to $f_\rho$. 
Nevertheless, for our purpose, it is sufficient 
to extract the absolute value of the product $|f_{\rho'}g_{\rho'\pi\pi}|$. To this end, we use the
BaBar collaboration fit \cite{BaBar:2012bdw}  of the measured 
pion form factor  to a multi-$\rho$ resonance model. 
Comparing equation (26) in that paper with the conventional normalization of the $\rho'$ contribution to 
the form factor (see e.g., \cite{Bruch:2004py}), we obtain the following relation: 
\begin{equation}
\frac{|k_\rho f_{\rho'}g_{\rho'\pi\pi}|}{m_{\rho'}}=
\frac{|c_{\rho'}|}{|1+c_{\rho'}+c_{\rho''}+c_{\rho'''}|}\
\equiv {\cal C}_{\rho'}=0.178 
\label{eq:crho}
\end{equation}
where the complex-valued fit parameters $c_{\rho'},c_{\rho''}$, $c_{\rho'''}$ are taken from Table VI in 
\cite{BaBar:2012bdw}.

To proceed further, we use the relations (\ref{eq:DpiVampl}) and (\ref{eq:VpipiBR}), and express  the product of 
the $D\to\pi\rho'$ amplitude and the $\rho'\pi\pi$ strong 
coupling via the product of branching fractions:
\begin{eqnarray}
  |A_{D^+ \pi^+ \rho'} \,g_{\rho'\pi\pi}|=
 \bigg[N_{\rho'}BR(D^+\to \pi^+ \rho^{'\,0}\,)BR( \,\rho^{'\,0}\to\pi^+\pi^-) \bigg]^{1/2}\,,
\label{eq:prod}  
\end{eqnarray}
where a shorthand notation
\begin{eqnarray}
N_{\rho'}\equiv \frac{384\pi^2\Gamma_{\rho'}  }{\tau_{D^+} G_F^2 
|\lambda_d|^2 m_{D^+}^3 m_{\rho'}\lambda ^{3/2}_{D^+ \pi^+\rho'}
[\beta_\pi(m_{\rho'}^2)]^{3}}
\end{eqnarray}
is introduced for a combination of measured quantities and kinematical factors.
The relation (\ref{eq:crho}) extracted from the pion form factor data, allows us 
to exclude the unknown strong coupling from the l.h.s. of 
(\ref{eq:prod}).
We finally obtain for the  residue of the $\rho'$ 
pole in the dispersion relation (\ref{eq:dispAexc}):
\begin{equation}
r_{\rho'}= \frac{(k_{\rho}f_{\rho'})^2}{
{\cal C}_{\rho'}m_{\rho'}}
\bigg[N_{\rho'}BR(D^+\to \pi^+ \rho^{'\,0}\,)BR( \,\rho^{'\,0}\to\pi^+\pi^-)\bigg]^{1/2}
\label{eq:rrho1}    
\end{equation}

It remains to estimate the $\rho'$ decay constant entering the above relation. For that, we  use the conventional  two-point QCD (SVZ) sum rule \cite{Shifman:1978bx} for the $\rho$ meson channel. Since the correlation function for this sum rule is a positive
definite quantity, each  resonance term in the hadronic sum
brings a positive contribution. The  ground-state $\rho$-meson term in this  sum rule is determined  by the  known value of 
the decay constant square, $f_{\rho}^2$. Hence, subtracting this term from the sum rule, 
we can estimate the decay constant squared of the first excited state $\rho'$. To this end,
we use the expression for the Borel-transformed sum rule 
\cite{Shifman:1978bx}
presented in a convenient form in the 
review \cite{Colangelo:2000dp} (see eq.(60) there).
Instead of, as usual, leaving only the lowest $\rho$-meson term,
we retain also the $\rho'$ contribution, approximaitng the sum 
over higher states contributions in the quark-hadron duality  approximation, 
The duality threshold parameter has to be adjusted accordingly.
We obtain:
\begin{eqnarray}
&&f_{\rho'}^2=e^{m_{\rho'}^2/M^2}\Big[\Pi(M^2,s^{(\rho')}_0)- f_\rho^2 e^{-m_\rho^2/M^2}\Big]\,,
\label{eq:frhoSVZ}
\end{eqnarray}
where
\begin{eqnarray}
&&\Pi(M^2,s_0)=
M^2\Big[\frac{1}{4\pi^2}\Big( 1-e^{-s_0/M^2}\Big)\Big(1+\frac{\alpha_s(M)}{\pi}\Big)
\nonumber\\
&&+
\frac{(m_u+m_d)\langle\bar{q}q\rangle}{M^4}+
\frac{1}{12}\frac{\alpha_s/\pi\langle GG \rangle}{M^4}-
\frac{112\pi}{81}\frac{\alpha_s \langle\bar{q}q\rangle ^2}{M^6}\Big]\,.
\label{eq:frhoSVZ1}
\end{eqnarray}
In the above, condensate contributions up to dimension 6 are taken into account, with the usual vacuum-saturation for the four-quark condensate term.
In the perturbative part of this sum rule, the $u$ and $d$ quark masses are neglected, and we assume that the scale of $\alpha_s$ is equal to the Borel parameter $M$. 
Note that the numerators of condensate terms are either scale-independent or their scale dependence is negligible. Since we are dealing with the 
excited state $\rho'$, we adopt a somewhat larger than usual interval $1.0~\mbox{GeV}^2 <M^2<2.0~\mbox{GeV}^2$ of the Borel parameter,
so that QCD coupling varies between 
$\alpha_s(1.0~\mbox{GeV})= 0.474$
and $\alpha_s(1.41~\mbox{GeV})= 0.36$. 
For the quark condensate term  in (\ref{eq:frhoSVZ1}), we use  
the Gell-Mann-Oakes-Renner relation $(m_u+m_d)\langle\bar{q}q\rangle=-\frac12 m_\pi^2 f_\pi^2= -1.651\cdot10^{-4}~\mbox{GeV}^4 $ (neglecting a small error). 
For the gluon condensate density \cite{Shifman:1978bx} we take $\alpha_s/\pi\langle GG \rangle=0.012^{+0.006}_{-0.012} ~\mbox{GeV}^{\,4}$, allowing for a  large uncertainty. Finally, using the above relation and the value $(m_u +m_d)/2=3.49\pm 0.07~\mbox{MeV}$ at the scale $\mu=2~\mbox{GeV}$ from \cite{PDG}, we obtain  $\alpha_s\langle\bar{q}q\rangle^2= (1.499 \pm 0.06)\cdot 10^{-4}~\mbox{GeV}^{\,6}$ for  the four-quark condensate term, where the inessential scale dependence is neglected.
In the chosen Borel-parameter range, the OPE in condensates is well behaved, so that the $d=4,6$ contributions are substantially smaller than the perturbative term.

A numerical analysis of this QCD
sum rule, where  only the ground-state $\rho$ meson contribution is isolated, 
reveals that an optimal choice of the duality threshold $s_0$ for the Borel mass interval of our choice is in the ballpark of the mass squared of the first excited state, that is,  $s_0^{(\rho)}=m_{\rho'}^2$.
With the same token, we choose $s_0^{(\rho')}=m_{\rho''}^2$
in the modified sum rule (\ref{eq:frhoSVZ})
and obtain the interval $f_{\rho'}=105-155$ MeV, 
quoted in Table~\ref{tab:Vexc}.  
The interval  reflects 
 parametric uncertainties, dominated by the intervals of the Borel parameter and of the gluon-condensate density.
With this estimate, we finally obtain from
(\ref{eq:rrho1}), the residue of $\rho'$
given in the same Table.

Turning to the residue of $\phi'$, albeit the charged kaon e.m. form factor 
was measured quite accurately, we could not find 
fits of this form factor in terms of the 
vector resonances. Hence, the combination of parameters ${\cal C}_{\phi'}$ analogous to
${\cal C}_{\rho'}$ in (\ref{eq:crho}) is not available yet. We simply take the relations between the decay constants and strong couplings 
that follow from the valence quark content (neglecting the
$s$- and $u,d$-quark mass difference):
\begin{equation}
f_{\phi'}=f_{\rho'}, 
~~g_{\phi'K^+K^-}=g_{\rho'\pi^+\pi-}/\sqrt{2}\,,
\label{eq:rel}
\end{equation}
and obtain: 
\begin{equation}
{\cal C}_{\phi'} = \frac{|k_\phi f_{\phi'}g_{\phi'K^+K^-}|}{m_{\phi'}} =
\frac{m_{\rho'}}{3 m_{\phi'}} {\cal C}_{\rho'}=0.052 \,.
\label{eq:phi}
\end{equation}
Using the approximation (\ref{eq:rel}), we rely on the fact that it is well obeyed for the lowest vector mesons. Indeed, the difference between $f_\rho $ and $f_\phi$
(see Table~\ref{tab:Vpar}) amounts to about 4\%, whereas calculating the ratio of strong couplings (using for them the formulas analogous to (\ref{eq:VpipiBR})) we observe again a relatively small deviation from the unit: 
$g_{\phi K^+K^-}/(g_{\rho \pi^+\pi-}/\sqrt{2})\simeq 1.08$.

Finally, we obtain  for the $\phi'$-residue 
\begin{equation}
r_{\phi'}= \frac{(k_{\phi}f_{\phi'})^2}{
{\cal C}_{\phi'}m_{\phi'}}
\bigg[N_{\phi'}BR(D^+\to \pi^+ \phi'\,)BR( \,\phi'\to K^+ K^-)\bigg]^{1/2}\,,
\label{eq:rphi1}    
\end{equation}
where all inputs, including  the normalization factor 
\begin{eqnarray}
N_{\phi'}\equiv \frac{384\pi^2\Gamma^{tot}_{\phi'}  }{\tau_{D^+} G_F^2 
|\lambda_d|^2 m_{D^+}^3 m_{\phi'}\lambda ^{3/2}_{D^+ \pi^+\phi'}
[\beta_K(m_{\phi'}^2)]^{3}}
\end{eqnarray}
are specified and yield the numerical value of $r_{\phi'}$ presented in Table~\ref{tab:Vexc}. 
Note that by construction, the minus sign at the coefficient $k_\phi$ is not included in the residue $r_{\phi'}$.

\bibliographystyle{JHEP.bst}
\bibliography{references.bib}

\end{document}